\documentclass[aps,prd,twocolumn,groupedaddress,floatfix,showpacs]{revtex4}

\usepackage[dvips]{graphicx}
\usepackage{amsmath,amssymb}
\usepackage{hyperref}
\usepackage{bm}
\usepackage{longtable}
\usepackage{tabularx}
\usepackage{slashbox}

\newcommand{\norm}[1]{\left| #1 \right|}

\newcommand{\uvec}[1]{\bm{\hat{#1}}}
\newcommand{\dvec}[1]{\dot{\bm{#1}}}
\newcommand{\duvec}[1]{\dot{\bm{\hat{#1}}}}

\usepackage{color}
\DeclareMathAlphabet{\mathpzc}{OT1}{pzc}{m}{it}

\newcommand\Msun{M_{\odot}}
\newcommand\intAll{\int_{-\infty}^{\infty}}
\newcommand{\pd}[2]{\frac{\partial #1}{\partial #2}}
\renewcommand{\vec}[1]{\bm{#1}}

\begin{document}
%\preprint{}

\title{Testing General Relativity with LISA including Spin Precession and Higher Harmonics in the Waveform}

% author. Explanatory text should go in the 's, actual e-mail

\author{C\'{e}dric \surname{Huwyler}}
\email{chuwyler@physik.uzh.ch}
\affiliation{Institut f\"ur Theoretische Physik, Universit\"at Z\"urich,
Winterthurerstrasse 190, 8057 Z\"urich}

\author{Antoine \surname{Klein}}
\affiliation{Institut f\"ur Theoretische Physik, Universit\"at Z\"urich,
Winterthurerstrasse 190, 8057 Z\"urich}

\author{Philippe \surname{Jetzer}}
\affiliation{Institut f\"ur Theoretische Physik, Universit\"at Z\"urich,
Winterthurerstrasse 190, 8057 Z\"urich}


\date{\today}

\begin{abstract}
We compute the accuracy at which a LISA-like space-based gravitational wave detector will be able to observe deviations from General
Relativity in the low frequency approximation. To do so, we introduce six correction parameters that account for modified gravity
in the second post-Newtonian gravitational wave phase for inspiralling supermassive black hole binaries with spin precession on quasi-circular
orbits. Our implementation can be regarded as a subset of the ppE formalism developed by Yunes and Pretorius, being able to investigate
also next-to-leading order effects. 
%The precession 
%of the spins and the angular momentum modulate the gravitational waveform, resulting in additional structure which could reduce correlations in the
%parameter space and increase the detection accuracy of the alternative theory parameters. Also, the use of higher harmonics could create further structure and increase
%the time during which the signal lasts in the frequency window of the detector. 
In order to find error distributions
for the alternative theory parameters, we use the Fisher information formalism and carry out Monte Carlo simulations for 17 different binary black hole mass 
configurations in the range $10^5 \Msun < M < 10^8 \Msun$ with $10^3$ randomly distributed points in the parameter space each, comparing the full (FWF) and restricted (RWF) version of the 
gravitational waveform. We find that the binaries can roughly be separated into two groups: one with low ($\lesssim 10^7 \Msun$) and one with high total masses 
($\gtrsim 10^7 \Msun$). The RWF errors on the alternative theory
parameters are two orders of magnitude higher than the FWF errors for high-mass binaries while almost comparable for low-mass binaries. Due to dilution
of the available information, the accuracy of the binary parameters is reduced by factors of a few, except for the luminosity distance which
is affected more seriously in the high-mass regime. 
As an application and to compare our research with previous work, we compute an optimal lower bound on the graviton Compton wavelength which is increased by a factor of $\sim1.6$ when using the FWF.
\end{abstract}

%\keywords{}

\maketitle


\section{\label{sec:intro}Introduction}

Although General Relativity (GR) has so far passed all experimental and observational tests \cite{will2006}, some unsatisfactorily explained phenomena still remain which 
could be more elegantly described by alternative gravity theories. Among these theories are the proposed inflationary epoch of the universe shortly after the big bang which explains the temperature
homogeneity of the cosmic
microwave background, dark matter which should account for the missing 23\% of the mass in the universe and dark energy introduced as an attempt to drive the observed late
accelerated expansion of the universe. Moreover, attempts to quantize GR or to unify gravitation with the other three fundamental forces
are as yet incomplete. Consequently, several modifications to GR have been proposed. Certain alternative theories work by introducing additional fields to the
Einstein-Hilbert action of GR. Scalar-tensor field theories such as Brans-Dicke theory \cite{bransdicke1961} are candidates for reproducing inflation. 
Modified Newtonian Dynamics (MOND) \cite{bekensteinmilgrom1984} attempts to get rid of dark matter by modifying the $1/r^2$ behavior of the gravitational potential; a relativistic
version introducing scalar and vector fields called Tensor-Vector-Scalar gravity (TeVeS) has also been proposed \cite{bekenstein2004}.
The class of $f(R)$ theories \cite{faraoni2008} modify the Einstein-Hilbert action by replacing the Riemann scalar by a function of it. 
More phenomenological approaches such as Massive Graviton theories \cite{will1998, stavridiswill2009} study the wave propagation of a 'massive' gravitational field.

Since alternatives to GR can be heavily constrained by the observation of Solar System effects and pulsar binaries \cite{will2006}, viable alternative
theory candidates should reduce to GR in the limit of weak fields.
In spacetime regions with strong dynamical gravity, such as binary black holes (BBHs), comparable constraints do not yet exist and should be tested for.
A good review of currently discussed alternatives to GR can be found in the appendix of \cite{yunespretorius2009}.

Among the most popular gravitational wave detectors are laser interferometers. Several ground-based interferometers such as LIGO (USA), Virgo (Italy) and 
GEO600 (Germany) have been built and are already operating, being sensitive to high frequencies between 10 Hz and 1 kHz. Currently LIGO is being upgraded to 
\emph{Advanced LIGO} with a sensitivity ten times better, and is expected to observe several events per year and make gravitational wave detection likely within the next five years. Hence gravitational
waves could finally be observed directly a hundred years after their theoretical prediction by Einstein.

Complementary to ground-based detectors restricted by their short arm-length and seismic noise at low frequencies, the spaceborne, low frequency detector 
eLISA/NGO  (evolved Laser Interferometer Space Antenna / Next Gravitational Wave Observatory) has been proposed, sensitive in a range of $\sim10^{-5}-1$ Hz. 
The mission was originally planned as an ESA/NASA 
collaboration, consisting of three spacecrafts separated by five million km, forming an equilateral triangle of laser arms. In 2011, NASA discontinued 
their participation in the LISA project; the European Space Agency planned to realize the project on their own with a reduced, affordable mission design called
eLISA/NGO \cite{ngoscience}. 
Although not selected as the first large L1 mission, there is a high chance that eLISA/NGO will be selected within the next few
years as an L2 mission.
In this paper we perform calculations for the originally planned LISA-like detector, as this enables us to compare our results to other studies
and also since it is currently unknown with what technical specifications eLISA will fly. 
We will use the term 'LISA' for a classic LISA-like mission throughout this paper.

Among the strongest sources which LISA will detect are supermassive black hole binaries with masses between $10^5 - 10^7 \Msun$. After a long inspiral phase, 
such binaries could merge into one single Kerr black hole which rings down from its excited state by emitting gravitational radiation. Compact binary inspirals produce a
very clean and long-lasting gravitational signal which may be accurately described by harmonics of the orbital phase using the post-Newtonian (PN) formalism. 
Inspiralling BBHs emit gravitational radiation carrying information about binary parameters such as the individual black hole masses and 
spins in its amplitude and phase. By using matched filtering techniques \cite{finn1992, poissonwill1995},
the binary parameters can be extracted from the noisy signal measured by the detector. Alternative gravity theories will also leave their imprints on gravitational waves, 
since they modify the strong-field dynamics of the BBH, resulting in a different orbital phase evolution. Also a possible 'graviton mass' will influence gravitational waves 
on their way to us by making their velocity frequency dependent. Since alternative theories are heavily constrained and LISA is expected to observe signals with very high signal-to-noise ratio (SNR),
a signal from a BBH will be detected with GR waveform templates regardless whether or not GR is true. This could create a fundamental bias \cite{yunespretorius2009} in parameter extraction 
if the signal is fitted with an incorrect GR waveform template, leading to incorrect parameter estimation. To fix this bias, additional parameters controlling deviations from
GR can be introduced. Adding parameters while having the same information from the detectors increases the correlation between the extracted parameters and thus decreases 
the accuracy in the recovered parameter values. 

Previous papers computed bounds which LISA could place on the Brans-Dicke parameter $\omega_{\text{BD}}$ (see e.g. \cite{will1994, scharrewill2002}) or on the graviton 
Compton wavelength $\lambda_g$ (see e.g. \cite{will1998}) using matched filtering. Due to the no hair theorem, for BBHs, scalar field effects in Brans-Dicke 
theory arising from the inner structure of compact objects cannot be distinguished; however, such massive binaries are an excellent environment to test massive gravity effects. 
The effects of 'massive' propagation have been investigated by various authors, considering different source and detector models.
After a first analysis of massive graviton propagation by Will \cite{will1998}, Berti et al. \cite{berti2005} introduced spin parameters and 
spin-orbit/spin-spin couplings, finding a loss of accuracy due to the extra parameters included in the model. Stavridis and Will \cite{stavridiswill2009} considered the 
full precession of the spins and discovered that the resulting phase modulation restores the lost accuracy on $\lambda_g$. Yagi and Tanaka \cite{yagitanaka2010} included
eccentricity to the system and found that the additional structure through both precession and eccentricity increases the measurement accuracy by an
order of magnitude. Arun and Will \cite{arunwill2009} showed that the bounds on $\lambda_g$
are improved by almost an order of magnitude for non-spinning BBHs when using the full waveform (FWF) instead of the restricted waveform (RWF) which takes the 
phase up to full PN order but considers the amplitude only to leading order. Taking higher harmonics into consideration increases the time during which the signal
stays in the frequency window of LISA and shows a richer structure in the gravitational wave, leading to less correlation in the parameter space.
Keppel and Ajith \cite{keppelajith2010} used hybrid 
inspiral-merger-ringdown waveforms and found that they lead to a $\sim10$ times higher accuracy than for inspiral-only waveforms.
Moreover, Berti et al. \cite{bertietal2011} pointed out that the combination of the bounds on $\lambda_g$ from individually observed inspirals in a two-year running time
can again raise the accuracy by an order of magnitude. Tables summarizing lower bounds 
on $\lambda_{g}$ and upper bounds on $\omega_{\text{BD}}$ found by previous works are e.g. provided by \cite{yagitanaka2010, keppelajith2010}. Arun et al. \cite{arunetal2006} re-interpreted 
the matched filtering method and fitted the post-Newtonian coefficients to the waveform instead of the parameters usually extracted from them. They discussed
to what extent LISA will be able to measure deviations from the 3.5PN gravitational wave phase parameters in General Relativity. Yunes and Pretorius \cite{yunespretorius2009} 
generalized this approach to a \emph{parameterized post-Einsteinian (ppE)} formalism which maps different types of alternative theories to the gravitational waveform
of a compact binary merger. Cornish et al. \cite{cornishsampson2011} used Markov Chain Monte Carlo simulations to investigate parameter biases and possible bounds on 
the ppE parameters. 

In this work we parametrize alternative theories by introducing corrections to the post-Newtonian coefficients of the orbital phase for a BBH 
inspiral, including the full 2PN precession of spins and angular momentum. We add higher harmonics to the waveform by considering the full 2PN
amplitude.
We postpone the discussion of eccentric orbits to later work and restrict our calculations 
to quasi-circular orbits. Since matched filtering is far more sensitive to the gravitational wave phase than to the amplitude, we do not consider corrections to
the amplitude of the wave. 
We evaluate the measurement accuracy with which a LISA-like mission will be able to detect such corrections for BBHs.  To estimate the errors on the parameters, we make use of the Fisher information formalism which is legitimate in the limit of high SNR
which LISA will provide.

The organization of this paper is as follows. In Sec. \ref{sec:evolution} we shortly introduce the necessary equations to describe the evolution of the inspiral phase, 
the spins and the angular momentum of a BBH up to 2PN. In Sec. \ref{sec:orbitalphase} we introduce small departures from GR into the post-Newtonian frequency 
evolution equation. We then compute the modified orbital phase evolution in this
scheme, incorporate it into a modified waveform template in Sec. \ref{sec:htildeModifications}, taking the waveform to be the sum of harmonics of the orbital
phase, compute the Fourier transformed waveform including alternative theory parameters and compare it with the ppE formalism in sec. \ref{sec:ppE}. In Sec. \ref{sec:ParameterEstimation} we review the Fisher information 
formalism in order to estimate the errors on the parameters. In sec. \ref{sec:simulations} we explain the details of the Monte Carlo simulations
we carried out. We discuss the resulting error distributions on selected parameters in Sec. \ref{sec:results} to
see to what extent we can measure deviations from the 2PN gravitational wave phase predicted by GR and how strongly the binary parameters are affected by the introduction
of six new parameters to the model. We discuss two representative BBH systems in Secs. \ref{sec:results_lowmass}) and \ref{sec:results_highmass}). 
In sec. \ref{sec:correlations} we have a closer look at correlations between the newly-introduced parameters.
Because systems at higher redshifts experience higher errors, we plot the maximal
redshifts for different upper error limits of the alternative theory parameters in Section \ref{sec:results_redshifts}. As an example, we calculate the resulting optimal 
lower bounds on the Compton wavelength of the graviton in Sec. \ref{sec:massivegraviton}. We summarize our work and discuss possible
extensions in Sec. \ref{sec:conclusion}. 
In Appendix \ref{sec:breakdown} we discuss the breakdown of three approximations used in this work and where the integrations should be stopped.
The expressions we used for the 2.5PN and 3PN frequency evolution are given in appendix \ref{sec:PN}. We give tables with best-case, worst-case and median measurement errors of both the binary and alternative
theory parameters in Appendix \ref{sec:tables}.

\section{\label{sec:evolution}Evolution of Black Hole Binaries with Precessing Spins}

A complete description of the inspiral evolution of two spinning black holes on a quasi-circular orbit with two individual masses $m_{1,2}$ and the corresponding spin vectors
$\vec{S}_{1,2}(t)$ is given by the angular momentum unit vector $\vec{\hat{L}}(t)$, the orbital angular frequency $\omega(t)$ and an initial value for the orbital phase 
$\varphi(t_0)$.
Further characteristics such as the orbital separation can be related to $\omega$ using post-Newtonian expressions. Therefore a quasi-circular BBH inspiral can be described by 12 intrinsic parameters.
In order to relate the binary with a detector, a unit vector $\vec{\hat{n}}$ pointing from the detector to the barycenter, and a luminosity distance $d_L$
between the two
can be introduced, bringing an additional set of 3 extrinsic parameters into play. Thus, to describe a BBH inspiral on quasi-circular orbit, 
15 parameters are required.

Since a description of the motion of such a system with full General Relativity is only possible with numerical methods and at high computational cost, 
an analytic expansion of the Einstein equations in powers of $v/c$ has been studied: the post-Newtonian (PN) formalism. Currently, the equations of motion
for spinning objects are known up to 2.5PN, while spin-spin and spin-orbit coupling terms are only known up to 2PN \cite{blanchet2006}. Therefore we take all the 
relevant expressions up to 2PN, i.e. $\mathcal{O}[(v/c)^4]$ away from leading order.
The evolution equation for the angular frequency of a BBH system is \cite{blanchetbuonanno2006}

\begin{align}
 \label{dxdtGR}
 \frac{dx}{dt} &= \frac{64\nu}{5} \frac{c^3}{GM} x^{5}
\Bigg[ 1 - \left( \frac{743}{336} + \frac{11\nu}{4} \right) x \nonumber\\
 &+ \left( 4\pi -
\frac{1}{12} \beta(113,75) \right) x^{3/2} \\
 &+ \left( \frac{34103}{18144} + \frac{13661\nu}{2016} + \frac{59\nu^2}{18} -
\frac{1}{48} \sigma(247,721)  \right) x^2 \Bigg], \nonumber
\end{align}
where

\begin{equation}
 x \equiv \left( \frac{GM\omega}{c^3}\right)^{2/3}
\end{equation}
is the dimensionless orbital frequency parameter, $M = m_1+m_2$ is the total mass and $\nu = m_1 m_2 / M^2$ is the symmetric mass ratio. The spin-orbit and spin-spin couplings
are given by

\begin{equation}
 \beta(a,b) = \frac{c}{G} \sum_{i=1}^2 \left( \frac{a}{M^2} + \frac{b\nu}{m^2_i}\right) \vec{S}_i \cdot \vec{\hat{L}},
\end{equation}
and

\begin{equation}
 \sigma(a,b) = \frac{c^2}{\nu M^4 G^2} (a \text{ } \vec{S}_1 \cdot \vec{S}_2 - b (\vec{S}_1 \cdot \vec{\hat{L}})(\vec{S}_2 \cdot \vec{\hat{L}})),
\end{equation}
respectively. The precession of $\vec{\hat{L}}$ and $\vec{S}_{1,2}$ induces a time dependence for these couplings, and thus a modulation of the gravitational wave phase. 
The orbit-averaged evolution equations without radiation reaction ($\dvec{L}+\dvec{S_1}+\dvec{S_2} = 0$) at 2PN order are \cite{apostolatos1994}

\small
\begin{eqnarray}
 \dvec{L} &= \frac{G}{c^2} \frac{1}{r^3} \left( \left( 2 + \frac{3
m_2}{2
m_1} \right) \bm{S}_1 + \left( 2 + \frac{3 m_1}{2 m_2} \right) \bm{S}_2
\right) \times \bm{L} \nonumber\\
 &\phantom{=} - \frac{3 G}{2 c^2} \frac{1}{r^3} \left( \left( \bm{S}_2 \cdot
\uvec{L} \right) \bm{S}_1 + \left( \bm{S}_1 \cdot \uvec{L} \right)
\bm{S}_2
\right) \times \uvec{L}, \label{Lhatdot}\\
 \dvec{S}_i &= \frac{G}{c^2} \frac{1}{r^3} \left[ \left( 2 + \frac{3
m_j}{2 m_i} \right) \bm{L} + \frac{1}{2} \bm{S}_j - \frac{3}{2} \left(
\bm{S}_j \cdot \uvec{L} \right) \uvec{L} \right] \times \bm{S}_i,
\label{spinprecession}
\end{eqnarray}
\normalsize
with $i \neq j$ and $i,j \in \{1,2\}$. The orbital separation $r$ and the angular momentum are related to the orbital frequency by the Newtonian relations

\begin{eqnarray}
\label{NewtonianLandr}
 L &= &\mu \left( \frac{G^2M^2}{\omega} \right)^{1/3}, \\
 r &= &\left( \frac{GM}{\omega^2} \right)^{1/3},
\end{eqnarray}
since higher-order corrections would exceed the 2PN order. Eqs. (\ref{dxdtGR}) and (\ref{NewtonianLandr}) enable us to express the evolution equations (\ref{spinprecession})
in terms of the frequency $\omega$:

\begin{align}
\label{spinprecession_omega}
 \frac{d\bm{S}_i}{d\omega} &= \frac{5}{96} \frac{c^3}{G M} \omega^{-2}
\Bigg[
\uvec{L} \times \bm{\Sigma}_i  \nonumber \\
&+ \frac{1}{2L} \left(
\bm{S}_j - 3 \left(
\bm{S}_j \cdot \uvec{L} \right) \uvec{L} \right) \times \bm{S}_i
\Bigg],
\\
 \frac{d\uvec{L}}{d\omega} &= \frac{5}{96} \frac{c^3}{G M} \omega^{-2}
\frac{1}{L}\left[ \bm{\Sigma}_1 + \bm{\Sigma}_2 - \frac{3}{2L} \left(
\bm{\sigma}_1 + \bm{\sigma}_2 \right) \right] \times \uvec{L}
\label{dLhat_df}\\
 &= -\frac{1}{L} \left( \frac{d\bm{S}_1}{d\omega} +
\frac{d\bm{S}_2}{d\omega}
\right),
\nonumber
\end{align}
with
\begin{equation}
 \bm{\Sigma}_i = \left( 2 + \frac{3 m_j}{2 m_i} \right) \bm{S}_i,\\
\end{equation}
and
\begin{equation}
 \bm{\sigma}_i = \left( \bm{S}_j \cdot \uvec{L} \right) \bm{S}_i.
\end{equation}

We express the gravitational wave phase in terms of the ``principal + direction'' \cite{arunbuonannofaye2009} defined as the direction of the vector 
$\vec{\hat{L}} \times \vec{\hat{n}}$. A precession of the angular momentum vector changes the principal + direction. The resulting modulation of the 
gravitational waveform can be expressed by modifying the phase by

\begin{align}
 \delta \varphi &= - \int_t^{t_c} \frac{\uvec{L} \cdot \uvec{n}}{1 - \left(
\uvec{L} \cdot \uvec{n} \right)^2} \left( \uvec{L} \times \uvec{n} \right) \cdot
\duvec{L} \, dt \nonumber\\
 &= \delta\varphi_0 + \int_{\omega_0}^\omega \frac{\uvec{L} \cdot \uvec{n}}{1 -
\left(
\uvec{L} \cdot \uvec{n} \right)^2} \left( \uvec{L} \times \uvec{n} \right) \cdot
\frac{d\uvec{L}}{d\omega} \, d\omega, \label{deltaphiprec}
\end{align} 
where $\omega_0$ is the orbital frequency at time $t_0$,
$\delta \varphi_0 = - \int_{t_0}^{t_c}
(d\delta
\varphi/dt) dt$, and $d\uvec{L}/d\omega$ is given in Eq. \eqref{dLhat_df}.
The resulting 2PN orbital phase is then, expressed in terms of the orbital
angular frequency:
$\phi(\omega) = \varphi(\omega) + \delta \varphi(\omega)$.

A signal observed from a BBH at cosmological distance is redshifted, i.e. the observed frequency is $f_{\text{o}} = f_{\text{e}} / (1+z)$, where $f_e$ is the frequency
of the gravitational waves emitted by the binary. The relation between redshift and luminosity distance in a $\Lambda$CDM cosmology without radiation and with
$\Omega_\Lambda = 0.72$, $\Omega_m=0.28$ and $H_0 = 70.1$ km/s/Mpc \cite{komatsu2009} is

\begin{equation}
\label{dL}
 d_L(z) = (1+z) \frac{c}{H_0} \int_0^z \frac{dz'}{\sqrt{\Omega_m (1+z')^3 + \Omega_\Lambda}}.
\end{equation}
For binaries at cosmological distance, the redshifted signal can be expressed as one coming from a binary with 'redshifted' masses 
$\tilde{m}_{1,2} = (1+z) \, m_{1,2}$ at luminosity distance $d_L(z)$.
Unfortunately, for gravitational wave experiments, it is not possible to disentangle redshift, mass and distance: only two parameters out of these three
can be inferred. Simultaneous observations of electromagnetic counterparts, through which the actual redshift could be measured, could break this correlation and lead to 
interesting astrophysical insights.

\section{\label{sec:orbitalphase}Modifications to the 2PN Orbital Phase}

Matched filtering techniques are more sensitive to the gravitational wave phase than to the amplitude.
The signal from a BBH inspiral can be described as a sum of harmonics of its orbital phase; to find the imprints of alternative gravity theories 
on gravitational waves it is therefore reasonable to look at how the orbital phase evolution of a BBH changes for small departures from GR. In the 2PN expansion, 
the orbital phase evolution can be found by integrating the frequency evolution equation (see Eq. \eqref{dxdtGR} for the PN coefficients $b_i$)

\begin{equation}
\label{frequencyevolution}
 \frac{dx}{dt} = \frac{64\nu}{5} \frac{c^3}{GM} x^5 \left[ 1 + b_1 x + b_{3/2} x^{3/2} + b_2 x^2 \right].
\end{equation}
As thoroughly discussed by Yunes and Pretorius in the derivation of their ppE formalism \cite{yunespretorius2009}, in the adiabatic approximation the dimensionless frequency can be expressed as

\begin{equation}
\label{dxdt}
 \frac{dx}{dt} = \frac{\dot{E}}{dE/dx}.
\end{equation}
$E$ is the total binding energy or Hamiltonian (conservative part) of the system while $\dot{E}$ stands for the energy loss through gravitational 
waves or other physical degrees of freedom of energy loss (dissipative part). Considering the impact
of alternative theories on these two quantities leads to modifications of the gravitational wave phase. Certain theories such as Brans-Dicke theory introduce
scalar fields which lead to a difference in the self-gravitational binding energy $\mathpzc{G}$ per unit mass  \cite{will1993}, producing additional dipole radiation. 
The energy loss formula including dipole contributions can be expressed to leading quadrupole order as \cite{thorne1980,will1993,yunespretorius2009}:

\begin{equation}
\label{thorne}
 \dot{E} = - \frac{\mu^2 G^3 M^2}{c^5 r^4} \left[ \frac{8}{15} (\kappa_1 v^2 - \kappa_2 \dot{r}^2) + \frac{1}{3} \kappa_D \mathpzc{G}^2 \right] - \mathcal{L}_{\text{other}}.
\end{equation}
Here, $v$ and $r$ are the orbital velocity and separation of the system, respectively, while $\kappa_1$ and $\kappa_2$ are so-called Peter-Mathews parameters and 
$\kappa_D$ is a coefficient for the dipole contribution.
$\mathcal{L}_{\text{other}}$ stands for any other energy loss channel, either through other polarizations or as yet unknown physical
processes. Since we do not have any good parametrization for $\mathcal{L}_{\text{other}}$ so far, we do not consider it. In terms of dimensionless frequency, the dipole radiation term in Eq. \eqref{thorne} 
leads to an additional $x^{-1}$ term in the PN expansion \eqref{frequencyevolution}. 
\\
\\
We introduce a general parametrization 
where the effects on the phase are emphasized and no corrections to the wave amplitude are considered. 
The calculations are done for quasi-circular binaries with precession of 
both black hole spins described by the full 2PN waveform (2PN expansion of both the phase and the amplitude). 
We start by introducing corrections to the 2PN orbital frequency evolution $dx/dt$ which will lead to a corrected version of the 2PN orbital phase.
To do that, we introduce a correction term $a_i$ for every 2PN coefficient $b_i$ and an additional $x^{-1}$ and
$x^{1/2}$ term. 
Products of a correction term $a_{\text{-1}} x^{-1}$ with a PN expanded 
expression such as $1+b_1 x + b_{3/2} x^{3/2} + b_2 x^2$ result in $b_2$ featuring already at 1PN order. Hence for the final result to be consistent at 2PN order, 
we need to do all the calculations up to 3PN, truncating at 2PN only at the very end. The current 2.5PN expansion accounts for spin-orbit effects while
the 3PN expansion does not consider spin effects at all. Nevertheless, these higher order expansions can be used as approximations.
The 3PN evolution equations of the dimensionless orbital angular frequency are, motivated from \cite{blanchetbuonanno2006, blanchetfaye2002} 
(see appendix \ref{Appendix:PN:2.5and3PN})

\begin{eqnarray}
\label{dxdt3PN}
 \left( \frac{dx}{dt} \right)_{\text{3PN}}  & =  & \frac{64\nu}{5} \frac{c^3}{GM} x^5  \left[ 1 + b_1 x + b_{3/2} x^{3/2} + b_2 x^2 \right. \nonumber \\
                                            &    & \left. + b_{5/2} x^{5/2} + b_3 x^3 + b_{3,\text{log}} x^3 \log(x) \right] \text{ ,} \nonumber \\
\end{eqnarray}

with

\begin{eqnarray}
 b_1 & = & - \left( \frac{743}{336} + \frac{11\nu}{4} \right) \text{ ,}\nonumber \\
 b_{3/2} & = & \left( 4\pi - \frac{1}{12}  \beta(113,75)\right) \text{ ,} \nonumber \\ 
 b_2 & = & \left( \frac{34103}{18144} + \frac{13661\nu}{2016} + \frac{59\nu^2}{18} - \frac{1}{48} \sigma(247,721)\right) \text{ ,} \nonumber \\
 b_{5/2} & = & \pi \left( -\frac{4159}{672} - \frac{189\nu}{8}\right) +  \frac{1}{c} \left( -\frac{40127}{1008} + \frac{1465\nu}{28}\right)  \nonumber \\
         &   &\times \beta(1,0) + \frac{1}{c} \left( -\frac{583}{42} + \frac{3049\nu}{168}\right) \beta(-1,1) \text{ ,} \nonumber \\
 b_3 & = & \frac{16447322263}{139708800} - \frac{1712 \gamma_e}{105} + \frac{16\pi^2}{3} - \frac{56198689\nu}{217728} \nonumber \\
     &   & + \frac{451\pi^2 \nu}{48} +  \frac{541 \nu^2}{896} - \frac{5605\nu^3}{2592} - \frac{856}{105} \log(16), \nonumber \\
 b_{3,\text{log}} & = & - \frac{856}{105}, \nonumber \\ 
\end{eqnarray}
where $\beta$ and $\sigma$ are the spin-orbit and spin-spin couplings, respectively. To account for alternative theories, we generalize the frequency evolution to

\begin{eqnarray}
\label{dxdtCorrections}
 \left(\frac{dx}{dt}\right)_{\text{mod}} & = & \left(\frac{dx}{dt}\right)_{\text{3PN}} + \frac{64\nu}{5} \frac{c^3}{GM} x^5 \nonumber \\
  & & \times \left[ a_{\text{-1}} x^{-1} + a_0 +  a_{1/2} x^{1/2} + a_1 x \right. \nonumber \\
   & & \left. + a_{3/2} x^{3/2} + a_2 x^2 + a_{2\text{,log}} \text{ } x^2 \log(x) \right] \text{ ,} \nonumber \\
\end{eqnarray}
including corrections to every existing PN parameter and an additional $x^{-1}$ and $x^{1/2}$ term. The reason why $x^2 \log(x)$ appears is that a 
term proportional to $x^3 \log(x)$ enters the 3PN phase which has to be included in 2PN corrections because of couplings with $x^{-1}$ terms.

Note that we treat the $a_i$ as constants, i.e. we disregard any dependencies on binary parameters such as masses and spins, since we do not know
how they look like in general.

We now follow the steps for the derivation of the gravitational waveform presented in \cite{klein2009}, introducing these additional corrections, keeping them
at first order, and truncating at 3PN. 

By inverting and integrating Eq. (\ref{dxdtCorrections}) we find the time $t(x)$ as a function of the frequency to be of the form:

\begin{eqnarray}
\label{t-tc}
 t - t_c & \approx & t(x)\big|_{\text{3PN}} - t_c - \frac{5}{256\nu} \frac{GM}{c^3} \left[ T_{\text{-1}} x^{-1} + T_0  \right. \nonumber \\
         & & + \left. T_{1/2} x^{1/2} + T_1 x + T_{3/2} x^{3/2} + T_2 x^2 \right. \nonumber \\
         & & \left. + T_{2,\text{log}} x^2 \log(x) \right] .\nonumber \\
\end{eqnarray}
The coefficients $T_i$ are functions of $a_i$.
To find the orbital phase as function of frequency, we need to recast $t(x)$ into a series expansion for $x(t)$; we are then able to find the phase by integrating $\omega \propto x^{3/2}$
over time:

\begin{eqnarray}
  \left[ \varphi(x)\right]_{\text{mod}} & = & \left[ \varphi(x) \right]_{\text{2PN}} + \frac{1}{32 \nu} \frac{c^3}{GM} x^{-5/2} \left[ A_{\text{-1}} x^{-1}  \right. \nonumber \\
   & & + A_0 + A_{1/2} x^{1/2} + A_1 x + A_{3/2} x^{3/2}  \nonumber \\
   & & \left. + A_2 x^2 + A_{2,\text{log}} x^2 \log(x) \right] ,\nonumber \\
\end{eqnarray}
with the phase corrections $A_i(\{a_k\})$ as functions of the orbital frequency evolution corrections introduced in eq. \eqref{dxdtCorrections}.
% 
% \begin{eqnarray}
%  A_{-1} & = & \frac{5 a_{-1}}{7} \text{ ,} \nonumber \\
%  A_{0} & = & a_0 - 2 a_{-1} b_1 \text{ ,} \nonumber \\
%  A_{1/2} & = & \frac{5}{4} (a_{1/2} - 2 a_{-1} b_{3/2}) \text{ ,} \nonumber \\
%  A_1 & = &  \frac{5}{3} (a_1 - 2 a_0 b_1 + 3 a_{-1} b_1^2 - 2 a_{-1} b_2) \text{ ,} \nonumber \\
%  A_{3/2} & = & \frac{5}{2} \left(a_{3/2} - 2 (a_{1/2} b_1 + a_0 b_{3/2} - 3 a_{-1} b_1 b_{3/2} \right. \nonumber \\
%          &   & \left. + a_{-1} b_{5/2}) \right) \text{ ,} \nonumber \\
%  A_2 & = &  5 \left( a_2 + 2 a_{2,\text{log}} - 2 a_1 b_1 + 3 a_0 b_1^2 - 4 a_{-1} b_1^3 \right. \nonumber \\
%      &   & \left. - 2 a_0 b_2 + 6 a_{-1} b_1 b_2  b_{3/2} \right. \nonumber \\
%      &   & \left. - 2 a_{-1} b_3 - 2 a_{1/2} + 3 a_{-1} b_{3/2}^2 - 4 a_{-1} b_{3,\text{log}} \right) \text{ ,} \nonumber \\
%  A_{2,log} & = & 5 a_{2,\text{log}} - 10 a_{-1} b_{3,\text{log}} \text{ .} \nonumber \\
% \end{eqnarray}
At this point we choose not to consider the correction term $A_{2,log}$ in our implementation for simplicity and thus set $A_{2,log} = 0$ in the following.

\section{\label{sec:htildeModifications}Modifications to the 2PN Waveform}

Having found a 2PN expression for the orbital phase corrections, we are able to construct the gravitational waveform as a series of harmonics of the 
orbital frequency:

\begin{equation}
 h_{+,\times} = \frac{2GM\nu x}{D_L c^2} \left[ \sum_{n\geq 0} \left( A^{(n)}_{+,\times} \cos(n\phi) + B^{(n)}_{+,\times} \sin(n\phi) \right)\right] .
\end{equation}

\noindent Here, $\phi$ is the orbital phase of the binary with spin precession included: $\phi(t) = \left[\varphi(t)\right]_{mod} + \delta \varphi(t)$. 
The coefficients $A^{(n)}_{+,\times}$, $B^{(n)}_{+,\times}$ are both post-Newtonian series in $x$:

\begin{equation}
 A^{(n)}_{+,\times} = \sum_{i\geq 0} a^{(n,i/2)}_{+,\times} x^{i/2} \text{,\hspace{1cm} } B^{(n)}_{+,\times} = \sum_{i\geq 0} b^{(n,i/2)}_{+,\times} x^{i/2} .
\end{equation}
Explicit expressions for $A^{(n)}_{+,\times}$ and $B^{(n)}_{+,\times}$ can be found in \cite{klein2009}.
A three arm classic LISA will form two different detectors with uncorrelated noise:
for a detector $k$ with antenna pattern functions $F_k^+$ and $F_k^\times$, the response function can be written in the low frequency
approximation (LFA) as

\begin{eqnarray}
\label{waveformdetectork}
 h_k & = & \frac{\sqrt{3}}{2} \left( F^+_k h_+ + F^\times_k h_\times \right) \nonumber \\
     & = & \frac{\sqrt{3} GM\nu x}{D_L c^2} \sum_{n\geq 0} \left[  A_{k,n} \cos(n\psi) +  B_{k,n} \sin(n\psi)\right] ,\nonumber \\
\end{eqnarray}
with the antenna pattern functions

\begin{align}
 F_1^+(\theta_N, \phi_N, \psi_N) &= \frac{1}{2} \left( 1 + \cos^2 \theta_N
\right) \cos 2\phi_N \cos
2\psi_N \nonumber\\
&\phantom{=}- \cos \theta_N \sin 2\phi_N \sin 2\psi_N,\\
 F_1^\times(\theta_N, \phi_N, \psi_N) &= F_1^+(\theta_N,\phi_N,\psi_N-\pi/4),\\
 F_2^+(\theta_N, \phi_N, \psi_N) &= F_1^+(\theta_N,\phi_N-\pi/4,\psi_N),\\
 F_2^\times(\theta_N, \phi_N, \psi_N) &=
F_1^+(\theta_N,\phi_N-\pi/4,\psi_N-\pi/4).
\end{align}
$\theta_N$ and $\phi_N$ are the spherical angles of the position of the
binary
in the detector frame, and $\psi_N$ is defined through
\begin{equation}
 \tan \psi_N \equiv \frac{\uvec{L}\cdot\uvec{z} - (\uvec{L}\cdot\uvec{n})
(\uvec{z}\cdot\uvec{n})}{\uvec{n}\cdot( \uvec{L}\times\uvec{z} )},
\end{equation}
with $\psi = \left[\varphi\right]_{mod} + \delta \varphi + \phi_D$, including the LISA Doppler phase $\phi_D(t) = (\omega R / c) \sin\bar{\theta}_N \cos(\bar{\Phi}(t) - \bar{\phi}_N)$, 
where $R = 1$ AU and $\bar{\phi}(t) = 2\pi t / 1$ yr as explained in \cite{klein2009}. The harmonic coefficients are

\begin{eqnarray}
 A_{k,n} & = &  \sum_{i\geq 0} \left( F^+_k a^{(n,i/2)}_+ + F^\times_k a^{(n,i/2)}_\times \right) x^{i/2} \text{,} \nonumber \\
 B_{k,n} & = & \sum_{i\geq 0} \left( F^+_k b^{(n,i/2)}_+ + F^\times_k b^{(n,i/2)}_\times \right) x^{i/2} \text{ .} \nonumber \\
\end{eqnarray}
By changing the cosine+sine representation into a cosine+phase representation, we can write Eq. \eqref{waveformdetectork} as

\begin{eqnarray}
 h_k & = & \frac{\sqrt{3} GM\nu x}{D_L c^2} \left[ A^{(0)}_+ F^+_k + A^{(0)}_\times F^\times_k \right. \nonumber \\
     & & + \left. \sum_{n\geq 1} A^{\text{pol}}_{k,n} \cos\left(n\psi + \phi^{\text{pol}}_{k,n}\right)\right] \text{ ,} \nonumber \\
\end{eqnarray}
with 

\begin{equation}
 \tan{\phi^{\text{pol}}_{k,n}} = -\frac{B_{k,n}}{A_{k,n}} \text{,} \qquad A^{\text{pol}}_{k,n} = \text{sgn}(A_{k,n}) \sqrt{A^2_{k,n} + B^2_{k,n}} \text{ .}
\end{equation}
The Fourier transform of the response function is then, writing the cosine as an exponential and defining the new phase
$\psi_{k,n} \equiv n( \left[\varphi\right]_{mod} + \delta \varphi + \phi_D) + \phi^{\text{pol}}_{k,n}$:

\begin{eqnarray}
 \tilde{h}_k(f) & = & \frac{\sqrt{3} G M \nu}{2 D_L c^2}  \intAll  \bigg(   \sum_{n\geq 1} x A^{\text{pol}}_{k,n} \left[  e^{i(2\pi f t - \psi_{k,n})} \right. \nonumber \\
 & & \left. + \, e^{i(2\pi f t + \psi_{k,n})} \right]+ 2x  \left( A^{(0)}_+ F^+_k + A^{(0)}_\times F^\times_k \right) \nonumber \\
 & & \times \, e^{2\pi i f t} \bigg) dt \text{ .} \nonumber \\
\end{eqnarray}

\noindent The $n=0$ integral accumulates around frequencies different from the gravitational wave frequency and $e^{i(2\pi f t + \psi_{k,n})}$ around negative frequencies, so both can be neglected. Then the Fourier transform reduces to

\begin{equation}
 \tilde{h}_k(f) = \frac{\sqrt{3} G M \nu}{2 D_L c^2} \sum_{n\geq 1} \left[ \intAll x A^{\text{pol}}_{k,n} e^{i(2\pi f t - \psi_{k,n})} dt \right] \text{ .}
\end{equation}
In the stationary phase approximation (SPA, see e.g. \cite{yunesetal2009, droz1999}), $\tilde{h}_k(f)$ is approximated by

\begin{eqnarray}
 \tilde{h}_k(f) & \sim &  \frac{\sqrt{6\pi} G M \nu}{4 D_L c^2} \sum_{n\geq 1} x(t_n) A^{\text{pol}}_{k,n}(t_n) \, e^{i(2\pi f t_n - \psi_{k,n} - \frac{\pi}{4})} \nonumber \\
 & & \times \sqrt{\frac{1}{\left| \frac{d^2\psi_{k,n}}{dt^2} \right|}}, \nonumber \\
\end{eqnarray}
evaluated at the stationary points $t_n = t_{\text{2PN}}(f/n)$. The square root of the reciprocal of the second derivative of $\psi_{k,n}$ is found to be

\begin{equation}
 \sqrt{\frac{1}{\left| \frac{d^2\psi_{k,n}}{dt^2} \right|}} = \frac{\sqrt{5} \text{ } G M}{4 \sqrt{6\nu} \text{ } c^3 x^{11/4}} [S(f)]_{\text{mod}} \text{ ,}
\end{equation}
with $[S(f/n)]_{\text{mod}} = S_{\text{2PN}}(f/n) + \Delta S$ being a 2PN function with

\begin{eqnarray}
S_{\text{2PN}}(f) & = & \left[ 1+ \left( \frac{743}{336} + \frac{11\nu}{8}\right) x + \left( \frac{1}{24} \beta(113,75) \right. \right. \nonumber \\
& & \left. - 2\pi \right) x^{3/2} + \left( \frac{7266251}{8128512} + \frac{18913\nu}{16128} + \frac{1379\nu^2}{1152} \right. \nonumber \\
 &  & \left.\left. + \frac{1}{96} \sigma(247, 721) \right) x^2 \right],\nonumber \\
 & & \nonumber \\
 \Delta S & = & S_{\text{-1}} x^{-1} + S_0 + S_{1/2} x^{1/2} + S_1 x + S_{3/2} x^{3/2} \nonumber \\
 & & + S_2 x^2 + S_{2, \text{log}} x^2 \log(x).\nonumber \\
\end{eqnarray}
The $S_i$ are functions of the orbital phase corrections $A_i$. The waveform can then be written as

\begin{eqnarray}
 \label{htilde}
 \tilde{h}_k(f) & \sim & \frac{\sqrt{5\pi\nu} G^2 M^2}{8 D_L c^5} \sum_{n\geq 1}  A^{\text{pol}}_{k,n}(t(f/n)) x_n^{-7/4} [S(f/n)]_{\text{mod}} \nonumber \\
 & &  \times \text{ } \exp\left\{i[n([\Psi(f/n)]_{\text{mod}} - \delta \varphi(f/n) \right. \nonumber \\
 & & -\left. \phi_D[t(f/n)]) - \phi^{\text{pol}}_{k,n}[t(f/n)]]\right\}, \nonumber \\ 
\end{eqnarray}
where the modified phase is defined as $[\Psi(f/n)]_{\text{mod}} = [\Psi(f/n)]_{\text{2PN}} + \Delta \Psi$, with

\begin{eqnarray}
\label{phasecorrections}
 \Psi_{\text{2PN}} & = & \left( \frac{t_c c^3}{G M} \right) x^{3/2} - \phi_c - \frac{\pi}{4} \nonumber \\
                    & & + \frac{3x^{-5/2}}{256\nu} \left[ 1 + \left( \frac{3715}{756} + \frac{55\nu}{9}\right) x + \left( \frac{1}{3} \beta(113,75) \right. \right. \nonumber \\
                   & & \left. - 16\pi \right) x^{3/2} + \left( \frac{15293365}{508032} + \frac{27145\nu}{504} + \frac{3085\nu^2}{72} \right. \nonumber \\
                   & & + \left. \left. \frac{5}{24} \sigma(247,721) \right) x^2 \right], \nonumber \\
 & & \nonumber \\
 \Delta \Psi & = & \frac{3}{256 \nu} x^{-5/2} \left( \Psi_{\text{-1}} x^{-1} + \Psi_0 + \Psi_{1/2} x^{1/2} + \Psi_1 x \right. \nonumber \\
 & & + \left. \Psi_{3/2} x^{3/2} + \Psi_2 x^2 \right). \nonumber\\ 
\end{eqnarray}
The $\Psi_i$ are also functions of the orbital phase corrections $A_i$. It makes thus sense to work only with the phase correction parameters $\Psi_i$ from now on.
The coefficients of $\Delta S$ are then, given as functions of $\Psi_i$:

\begin{eqnarray}
\label{SofPsi}
 S_{\text{-1}} & = & -\frac{7}{24} \Psi_{\text{-1}} \text{ ,} \nonumber \\
 S_{0} & = & -\frac{35}{48} b_1 \Psi_{\text{-1}} - \frac{1}{2} \Psi_0 \text{ ,} \nonumber \\
 S_{1/2} & = & -\frac{49}{48} b_{3/2} \Psi_{\text{-1}} - \frac{7}{15} \Psi_{1/2} \text{ ,} \nonumber \\
 S_{1} & = & -\frac{21}{16} \left( \frac{b_1^2}{4} - b_2 \right) \Psi_{\text{-1}} +-\frac{7}{12} b_1 \Psi_{0} - \frac{3}{8} \Psi_1 \text{ ,} \nonumber \\
 S_{3/2} & = & \left( \frac{77}{96} b_1 b_{3/2} - \frac{49}{24} b_{5/2} \right) \Psi_{\text{-1}} - \frac{3}{4} b_{3/2} \Psi_{0}  \nonumber \\
         &   & - \frac{7}{15} b_1 \Psi_{1/2} - \frac{1}{4} \Psi_{3/2} \text{ ,} \nonumber \\
 S_2 & = & \left( -\frac{91}{384} b_1^3  + \frac{91}{96} b_1 b_2 - \frac{7}{3} b_3 + \frac{91}{192} b_{3/2}^2 \right. \nonumber \\ 
     &   & - \left. \frac{7}{12} b_{3, \text{log}} \right) \Psi_{\text{-1}} + \left( \frac{11}{48} b_1^2 - \frac{11}{12} b_2 \right) \nonumber \\
     &   & \Psi_0 - \frac{7}{12} b_{3/2} \Psi_{1/2} - \frac{27}{80} b_1 \Psi_1 - \frac{7}{60} \Psi_2 \text{ ,} \nonumber \\
 S_{2, \text{log}} & = & -\frac{7}{3} b_{3, \text{log}} \Psi_{\text{-1}} \text{ .} \nonumber \\
\end{eqnarray}

\noindent All the alternative theory parameters $\Psi_i$ are treated as constants. They will most probably depend on other binary parameters such as
masses and spins, but it is not possible at this point to find a general parametrization in terms of binary parameters. In practice this could lead to
further covariances between the alternative theory and binary parameters. Since in the PN expansion of the gravitational wave phase usually 
coefficients depending on the symmetric mass ratio of the form $\alpha_1+\alpha_2 \nu + \alpha_3 \nu^2 + \hdots$ appear, one could theoretically introduce a new set of parameters, as an attempt to disentangle binary and alternative theory parameters, but 
it would increase the number of parameters drastically, therefore reducing the accuracy of a single measurement. Since such a parametrization 
would not induce time varying couplings, and this study focuses on the measurement accuracy for individual systems, we chose not to take the mass ratio into account. 
However, the spins might lead to time varying modifications; we chose not to take them into account either, because of the lack of 
theoretical predictions for their form.

\section{\label{sec:ppE} Connection to the ppE formalism}

The idea of this work is based on the ppE formalism by Yunes and Pretorius \cite{yunespretorius2009}. To look for deviations from GR, they introduce
modifications to the amplitude and phase of the gravitational wave in the frequency domain \cite{cornishsampson2011}:

\begin{eqnarray}
 A(f) & = & \left( 1 + \sum_k \alpha_k \, u^{a_k} \right) A^{\text{GR}}(f), \nonumber \\
 \Psi(f) & = & \left( 1 + \sum_k \beta_k \, u^{b_k} \right) \Psi^{\text{GR}}(f). \nonumber \\
\end{eqnarray}
Here, $u = x^{3/2} \nu^{3/5}$ is the reduced frequency and $\alpha_k$, $\beta_k$ are alternative theory parameters which could depend on the binary parameters,
such as on the symmetric mass ratio or on some spin/angular-momentum quantities. These deviations results in a modification for the $n$-th harmonic of the gravitational
waveform (in the frequency domain) of the form

\begin{equation}
 \tilde{h}_n(f) = \tilde{h}_n^{\text{GR}}(f) \left[ 1 + \Delta A_n(f/n) \right] e^{i n \Delta \Psi(f/n)},
\end{equation}
where $\Delta A_n$ and $\Delta \Psi$ are power series in the frequency arising from the above modifications, and the overall waveform is 
the sum $\tilde{h}(f) = \displaystyle \sum_{n} \tilde{h}_n(f)$.

\noindent Previous studies \cite{yuneshughes2010, cornishsampson2011} used the restricted waveform ($n=2$) and investigated leading order deviations using
a waveform template of the form

\begin{equation}
 \tilde{h}(f) = \tilde{h}^{\text{GR}}(f) \left[ 1 +\alpha (4\nu)^A u^a \right] e^{i \beta (4\nu)^B u^b},
\end{equation}
where a dependency on the symmetric mass ratio $\nu$ is introduced. Let us relate this to our parametrization given in eq. \eqref{htilde}:

\begin{equation}
 \tilde{h}_n(f) = \tilde{h}^{\text{GR}}_{n}(f) \left( 1+ \frac{\Delta S(f/n)}{S(f/n)} \right) e^{i n \Delta \Psi(f/n)} \text{ .}
\end{equation}
Since in our implementation we start from the frequency evolution \eqref{dxdtCorrections}, the amplitude correction term $\Delta S / S$ entering through
the stationary phase approximation is only a pseudo
correction, as it can be expressed with phase correction parameters $\Psi_i$ \eqref{SofPsi}. Thus our implementation does not consider real amplitude
modifications, only the phase parameters $\Psi_i$ can be put into relation with the ppE formalism. The phase modifications $\Delta \Psi$ are, 
for the ppE formalism and our implementation respectively:

\begin{eqnarray}
 \Delta \Psi_{\text{ppE}}  & = & \sum_k \beta_k \, (4\nu)^{B_k} \, u^{b_k}, \nonumber \\
 \Delta \Psi_{\text{this work}} & = & \frac{3}{256 \nu} \sum_i \Psi_i \, x^{i-5/2}. \nonumber \\
\end{eqnarray}

Because of the special treatment of the symmetric mass ratio prefactor with a parameter $B_k$ and since the symmetric mass ratio enters the conversion 
between $u$ and $x$, there is no clear way how to put the parameter sets $\{\beta_k, B_k, b_k\}$ and $\{\Psi_i, i\}$ into relation. Only the frequency 
powers $b_k$ and $i$ where
the corrections enter can be compared: they relate as $b_k = \frac{2}{3} (i_k - \frac{5}{2})$, where the $i_k$ are our summation indices. Our implementation is thus
a subset of the ppE formalism with $b_k = \{-7/3, -5/3, -4/3, -1, -2/3, -1/3\}$.

This subset with fixed frequency does not cover the leading order contributions of every alternative to GR currently proposed. While it is able to catch 
leading order deviations originating from Brans-Dicke, massive graviton and quadratic curvature-type theories, it will not see the leading order imprints 
of Dynamical Chern-Simons gravity, Variable $G(t)$ theories and theories including extra dimensions (see \cite{cornishsampson2011} for an overview table 
of the leading order contributions of alternative theories). On the other hand, our implementation is able to investigate next-to-leading order effects 
and can quantify how the inclusion of alternative theory parameters with more than just one frequency power affects the measurement accuracy of a LISA-type
detector, including the effects of spin precession and higher harmonics.

\section{\label{sec:ParameterEstimation}Parameter Estimation}

To estimate how accurately LISA can measure deviations from the 2PN gravitational wave phase predicted by General Relativity, we use the standard Fisher information
formalism for gravitational wave experiments, as reviewed in \cite{cutlerflanagan1994, vallisneri2008}. The Fisher information formalism holds only in the limit of 
high SNR; this is true for a LISA-type mission, for which SNRs of a few thousands are expected. For low SNR, advanced Bayesian techniques exploring the whole parameter space such as Markov Chain Monte Carlo methods, (see e.g. \cite{littenbergcornish2009, cornishsampson2011}) 
are needed. Also, once data will become available, Bayesian statistics taking into account prior probability distributions will be the preferred framework 
\cite{lipozzo2012}.

We assume the gravitational wave signal to be buried in stationary Gaussian noise $n(t)$ such that the different Fourier components $\tilde{n}(f)$ are uncorrelated.
Moreover, we presume that the noise of the two detectors is totally uncorrelated.
Assuming flat priors, for a signal $h(t)$ described by a true parameter set $\vec{\theta}_t$, with noise with spectral density $S_n(f)$, the probability
for the measured data $d(t) = n(t) + h(t;\vec{\theta}_t)$ to take this specific form is proportional to

\begin{equation}
 p(d|\vec{\theta}_t) \propto e^{-(d-h(\vec{\theta}_t)|d-h(\vec{\theta}_t))},
\end{equation}

\noindent where the inner product $(g|h)$ is defined as

\begin{equation}
 (g|h) = 4 \text{ } Re \int_0^\infty \frac{\tilde{g}^*(f) \tilde{h}(f)}{S_n(f)} df \text{ .}
\end{equation}
The use of a waveform template with the parameter set $\vec{\theta}$ is inaccurate by $\Delta \theta^i = \theta_t^i - \theta^i$. 
The errors $\Delta \theta^i$ are then approximately given by maximizing the above likelihood distribution, expanding it in the errors assumed to be small and keeping
only first derivatives \cite{vallisneri2008}:

\begin{equation}
 \langle \Delta \theta_i \Delta \theta_j \rangle = \Sigma_{ij} = (\Gamma^{-1})_{ij} + \mathcal{O}\left(\frac{1}{\text{SNR}}\right) \text{ ,}
\end{equation}

\noindent where $\Sigma$ is the covariance matrix and

\begin{equation}
\label{fishermatrix}
 \Gamma_{ij} = \left( \pd{h}{\theta^i} \bigg| \pd{h}{\theta^j} \right)
\end{equation}

\noindent is the so-called \emph{Fisher matrix}. The expected measurement errors on the parameters $\theta^i$ can be expressed as

\begin{equation}
 \Delta \theta^i = \sqrt{(\Gamma^{-1})^{ii}} \text{. }
\end{equation}

We chose the same noise curve for classic LISA as in \cite{klein2009}, namely the piecewise fit used by the LISA parameter estimation community \cite{arunetal2009}
given by the instrumental noise

\begin{eqnarray}
 S_n(f) &= & \frac{1}{L^2} \Bigg\{ \left[ 1 + \frac{1}{2} \left( \frac{f}{f_*} \right)^2 \right] S_p\nonumber \\
        &  &  + \left[ 1 + \left( \frac{10^{-4}}{f} \right)^2 \right] \cdot \frac{4 S_a}{\left(2\pi f\right)^4} \Bigg\}, \\
\end{eqnarray}
and the confusion noise

\small
\begin{eqnarray}
S_{\mathrm{conf}}(f) &= \left\{
\begin{array}{l l}
 10^{-44.62} f^{-2.3} & (f \leqslant 10^{-3}), \\
 10^{-50.92} f^{-4.4} & (10^{-3} < f \leqslant 10^{-2.7}), \\
 10^{-62.8} f^{-8.8} & (10^{-2.7} < f \leqslant 10^{-2.4}), \\
 10^{-89.68} f^{-20} & (10^{-2.4} < f \leqslant 10^{-2}),\\
 0 & (10^{-2} < f),
 \end{array}
  \right.
\end{eqnarray}
\normalsize
where $L = 5 \times 10^9~\mathrm{m}$ is the arm length of classic LISA, $S_p = 4 \times
10^{-22}~\mathrm{m}^2~\mathrm{Hz}^{-1}$ is the white position noise level, $S_a
= 9 \times 10^{-30}~\mathrm{m}^2~\mathrm{s}^{-4}~\mathrm{Hz}^{-1}$ is the white
acceleration noise level, and $f_* = c/(2\pi L)$ is the arm transfer frequency.
The total noise curve is then $S_h(f) = S_n(f) + S_{\mathrm{conf}}(f)$.

\section{\label{sec:simulations}Simulations}

For our simulations, 21 parameters are needed: 15 GR parameters plus 6 alternative theory parameters. We use

\begin{enumerate}
 \item[(i)] $\log_{10} m_1/M_\odot$ and $\log_{10} m_2/M_\odot$, for the 
masses of the two black holes.
 \item[(ii)] $\mu_l = \cos \theta_l$ and $\phi_l$, for the spherical angles of
the
orbital angular momentum $\bm{L}$ at $\gamma=\frac{1}{6}$.
 \item[(iii)] $\mu_1 = \cos \theta_1$ and $\phi_1$ for the spherical angles of
the
spin of the first black hole
$\bm{S}_1$ at $\gamma=\frac{1}{6}$.
 \item[(iv)] $\chi_1 = \frac{c}{Gm_1^2}
\norm{\bm{S}_1}$ for the dimensionless strength of the spin of the first
black hole, which has to satisfy $0
\leqslant
\chi_1 < 1$.
 \item[(v)] $\mu_2 = \cos \theta_2$, $\phi_2$, and $\chi_2$ for the second
black hole, defined equivalently as for the first one.
 \item[(vi)] $\log t_c$, for the time of coalescence.
 \item[(vii)] $\varphi_c$, the phase at coalescence. As this phase is random and
its determination is not of any astrophysical interest, we can
safely neglect constants in the orbital phase, in particular $\delta\varphi_0$
from Eq.~\eqref{deltaphiprec}.
 \item[(viii)] $\mu_n=\cos\theta_n$ and $\phi_n$, the spherical angles of the
 position of the binary in the sky.
 \item[(ix)] $\log d_L$, for the luminosity distance between the source and the Solar
System.
 \item[(x)] $\Psi_i$ with $i \in \{-1,0,1/2,1,3/2,2\}$, the 6 alternative theory parameters defined in section \ref{sec:htildeModifications} 
\end{enumerate}

All angles are taken in the frame tied to the distant stars. Moreover, we set $t=0$ to be at the time when LISA will start operating.

We perform Monte Carlo simulations, keeping the masses $m_{1,2}$, the redshift $z$ and the alternative theory parameters $\Psi_i$ fixed, and randomizing all other
parameters using a flat probability distribution. The spin precession equations (\ref{spinprecession_omega}) are integrated using a fourth order adaptive Runge-Kutta algorithm to find the evolution
of $\vec{\hat{L}}(\omega)$ and $\vec{S}_{1,2}(\omega)$, going backwards in frequency. 

As generic starting point for $\omega$, we chose the frequency at the Schwarzschild ISCO (innermost stable circular orbit)
$r_{\text{\tiny ISCO}} = 6 \, GM/c^2$. Even though such a clear ISCO does not exist for black hole binaries with comparable mass and precessing spins, we find 
that this limit
is a good cut-off criterion, avoiding unphysical results. For more information about our considerations, the reader is referred to section \ref{sec:breakdown}
in the appendix.

We stop the evolution either at $t=0$ or when the frequency of the highest harmonic goes below the LISA band ($6\omega < 3 \times 10^{-5}$ Hz).
The upper and lower bounds on all the randomized parameters of the simulation are straightforward ($d_L$ is just a function of the redshift $z$, defined in (\ref{dL})), except for 
$t_c$ for which we set a lower bound of $t_c = t_{\text{2PN}}( \omega(r = r_{\text{ISCO}} ) )$ using Eq. (\ref{t-tc}) and an upper bound of $t_c = 2$yr, which is the minimum 
science requirement for the LISA mission running time.

Using the angular momentum, spin and orbital time evolution we are able to compute the Fisher matrix elements (\ref{fishermatrix}), taking the analytical derivatives
with respect to $\log t_c$, $\log d_L$, $\phi_c$, $\mu_n$, $\phi_n$ and all the GR correction parameters $\Psi_i$. The first three derivatives are easy to compute:

\begin{align}
 \frac{\partial\tilde{h}_k(\theta^j,f)}{\partial \log t_c} &= 2 \pi i f \, t_c \,
\tilde{h}_k(\theta^j,f), \\
 \frac{\partial\tilde{h}_k(\theta^j,f)}{\partial \log d_L} &= -
\tilde{h}_k(\theta^j,f), \\
 \frac{\partial\tilde{h}_k(\theta^j,f)}{\partial \varphi_c} &= -i \sum_n n
\tilde{h}_{k,n}(\theta^j,f),
\end{align}
where $\tilde{h}_{k,n}$ is the $n$th harmonic of $\tilde{h}_k$. The derivatives with respect to the corrections $\Psi_i$ are of the form

\begin{eqnarray}
\label{dhtildedPsii}
 \pd{\tilde{h}_k}{\Psi_i}(f) & = & \frac{\sqrt{5\pi\nu} G^2 M^2}{8 D_L c^5} \sum_{n\geq 1}  A^{\text{pol}}_{k,n} x_n^{-7/4} \nonumber \\
  & & \times \, e^{i\left[n(\Psi_{\text{GR}} + \Delta \Psi - \delta \varphi - \phi_D) - \phi^{\text{pol}}_{k,n}\right]} \nonumber \\
  & & \times \left( i \text{ } n \text{ } \left(S_{\text{2PN}} + \Delta S \right)\pd{\Delta\Psi}{\Psi_i} + \pd{\Delta S}{\Psi_i}\right), \nonumber \\
\end{eqnarray}
and can be calculated in a straightforward way. The derivatives which we could not compute analytically are approximated by

\begin{equation}
 \frac{\partial \tilde{h}_k (\theta^j,f)}{\partial \theta^i} \approx
\frac{\tilde{h}_k(\theta^j + \epsilon \delta^{ij} /2, f) - \tilde{h}_k(\theta^j
-\epsilon \delta^{ij}/2, f)}{\epsilon},
\end{equation}
where $\epsilon$ is a small displacement of the parameter $\theta_i$ which we chose to be of the constant value $\epsilon = 10^{-7}$ for every parameter, except
for $\phi_l$ for which $\epsilon$ was divided by $2 - 2|\mu_l|$, $\mu_i$
($i \in \{1,2\}$)  for which $\epsilon$ was divided by $5 \chi_i$,
and $\phi_i$ for which $\epsilon$ was divided by $10 \chi_i(1 - |\mu_i|)$. The
formula is accurate up to $O(\epsilon^2)$.

For each set of parameters we then compute the Fisher matrix using \emph{Clenshaw-Curtis quadrature} and then invert it in order to find the corresponding errors on 
the parameters which we analyze in section \ref{sec:results}. In order to avoid 
matrix inversion problems, we use a normalization of the Fisher-Matrix so that all diagonal elements 
are $A_{ii} = 1$ and all off-diagonal elements are in the range $A_{ij} \in [-1;1]$:

\begin{equation}
 A_{ij} \equiv \frac{1}{\sqrt{\Gamma_{ii} \Gamma_{jj}}} \text{ } \Gamma_{ij} \text{ .}
\end{equation}

\noindent After inversion, the covariance matrix can then be recovered with

\begin{equation}
 \Sigma_{ij} = \frac{1}{\sqrt{\Gamma_{ii} \Gamma_{jj}}} \text{ } A^{-1}_{ij} \text{ .}
\end{equation}
In situations where $\vec{\hat{L}} \cdot \vec{\hat{n}}$ is close to 1, the Runge Kutta method fails to converge because

\begin{equation}
 \frac{d\delta \varphi}{d\omega} = \frac{\uvec{L} \cdot \uvec{n}}{1 -
\left(
\uvec{L} \cdot \uvec{n} \right)^2} \left( \uvec{L} \times \uvec{n} \right) \cdot
\frac{d\uvec{L}}{d\omega} \underset{\vec{\hat{L}} \cdot \vec{\hat{n}} \rightarrow 1}{\longrightarrow} \infty \text{.}
\end{equation}
Whenever this happens, we take the approximate value

\begin{multline}
 \delta\varphi(\omega + \delta\omega) \approx \\
 \delta\varphi(\omega) +
\mbox{angle}
\left[ \left( \uvec{L}(\omega+\delta\omega) \times \uvec{n} \right) ,  \left(
\uvec{L}(\omega) \times \uvec{n} \right) \right],
\end{multline}
as explained in \cite{klein2009}.

\section{\label{sec:results}Results}

\begin{figure}[!hbp]
 \includegraphics[width=\columnwidth]{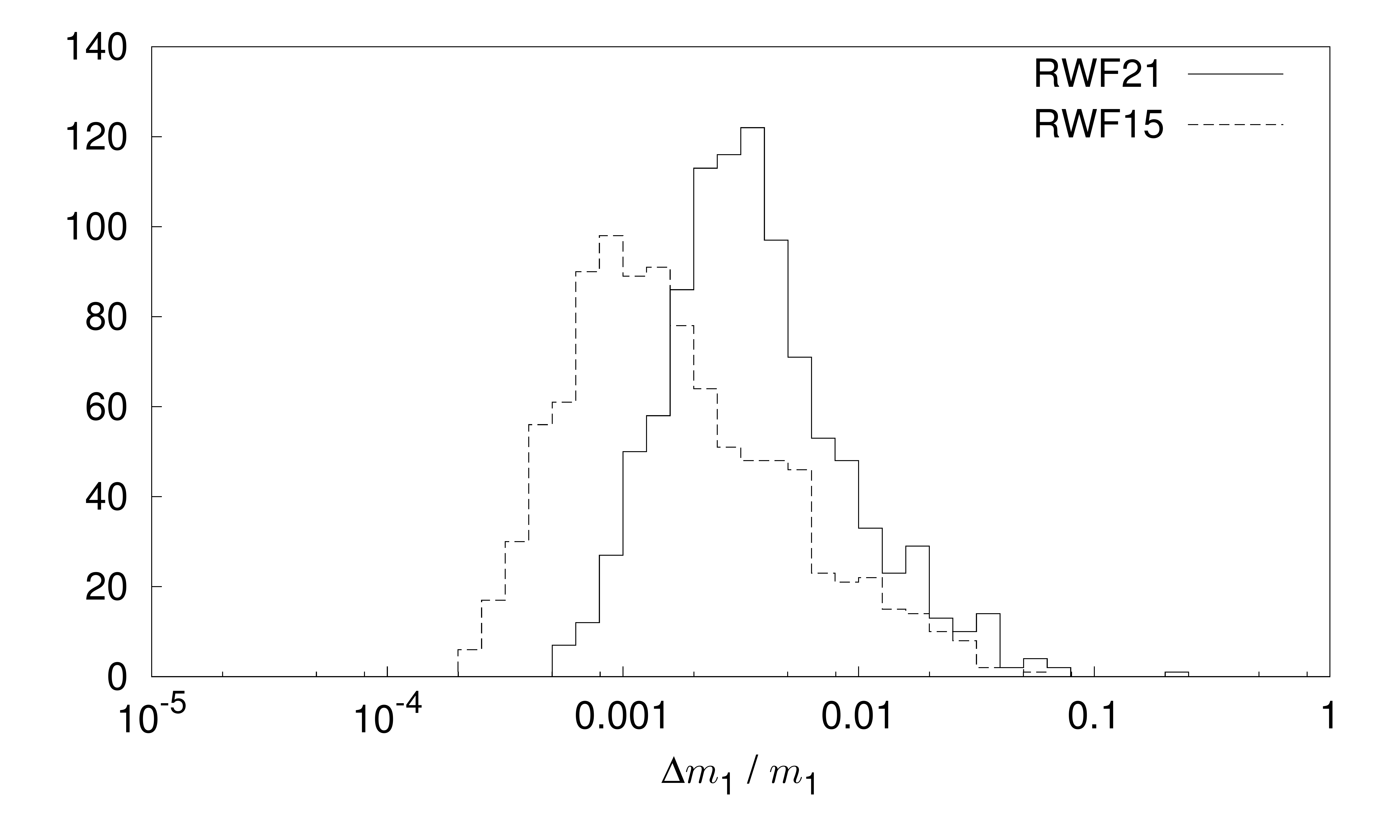}
  \caption{Comparison between the estimated distributions of the measurement error on
 $m_1$ for a low-mass binary system $m_1 = 1 \times 10^6 \Msun$ and $m_2 = 3 \times
10^5 \Msun$ with (RWF21) and without (RWF15) including alternative theory parameters, using only the restricted waveform. \label{fig:m11635_RWF}}
\end{figure}

\begin{figure}[!hbp]
 \includegraphics[width=\columnwidth]{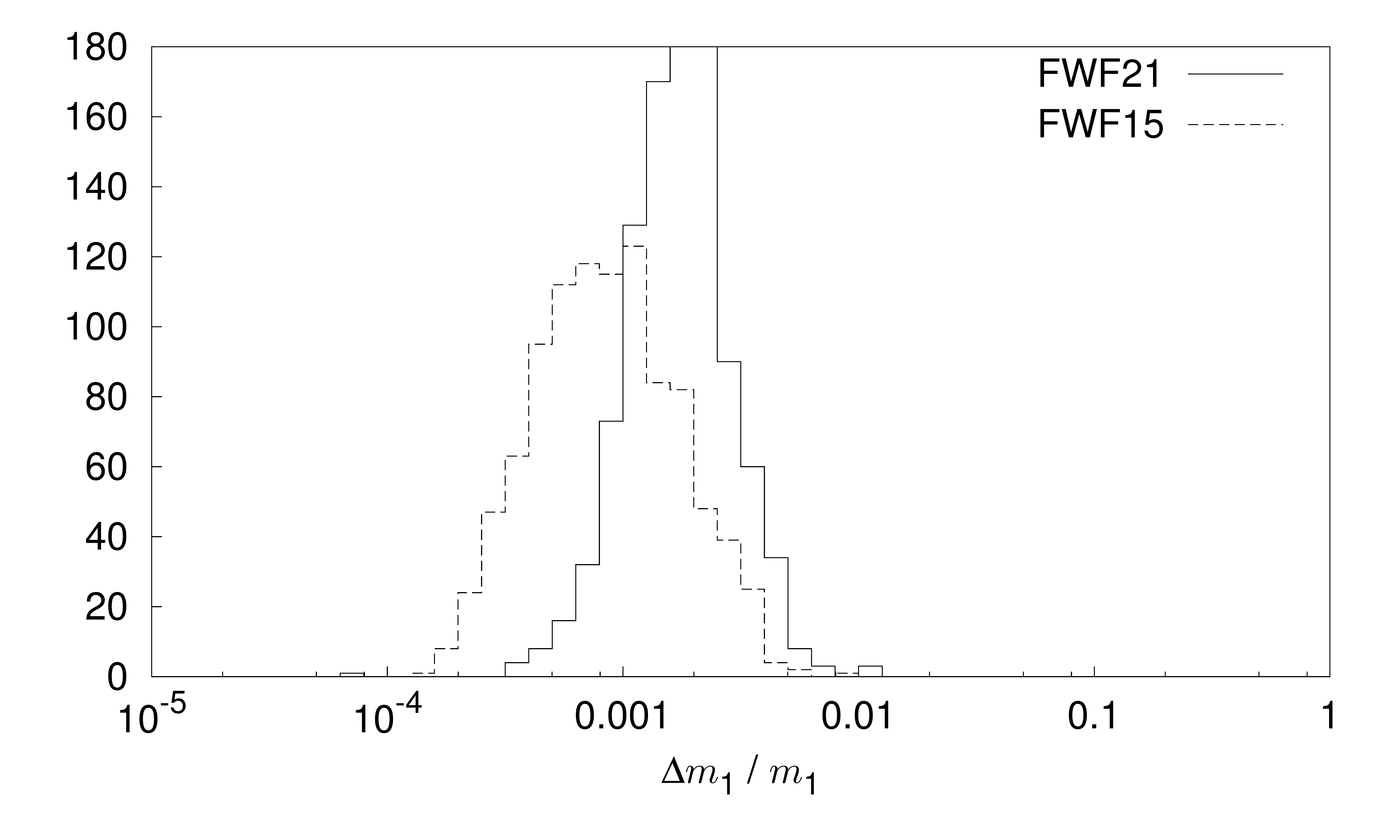}
  \caption{Comparison between the estimated distributions of the measurement error on
 $m_1$ for a low-mass binary system $m_1 = 1 \times 10^6 \Msun$ and $m_2 = 3 \times
10^5 \Msun$ with (FWF21) and without (FWF15) including alternative theory parameters and using the full waveform. \label{fig:m11635_FWF}}
\end{figure}

We performed simulations for 17 different mass configurations, with total masses between $10^5 \Msun$ and $10^8 \Msun$, mass ratios varying between 1:1 and 1:10, 
and using $10^3$ points in the parameter space for each configuration. 

The redshift has been kept fixed to $z=1$ since it is not possible to disentangle redshift, mass and distance. 
The signal coming from a binary with masses $m_{1,2}$ at redshift $z$ and luminosity distance $d_L(z)$ can be expressed with one from an apparent binary with 
$\tilde{m}_{1,2} = \frac{1+z}{1+z_0} m_{1,2}$ at redshift $z_0$ and luminosity distance $d_L(z_0)$ multiplied by an overall factor of $d_L(z_0)/d_L(z)$. Thus every BBH inspiral producing a signal at redshift $z$
can be described with a waveform template at redshift $z_0$. The Fisher matrix scales as

\begin{eqnarray}
 \Gamma_{ij}^{(z)} & = & \left( \pd{h}{\theta^i}(m_1,m_2,z) \bigg| \pd{h}{\theta^j}(m_1,m_2,z) \right) \nonumber \\
                   & = & \left( \frac{d_L(z_0)}{d_L(z)}\right)^2 \left( \pd{h}{\theta^i}(\tilde{m}_1,\tilde{m}_2,z_0) \bigg| \pd{h}{\theta^j}(\tilde{m}_1,\tilde{m}_2,z_0) \right) \nonumber \\
                   & = & \left( \frac{d_L(z_0)}{d_L(z)}\right)^2 \Gamma_{ij}^{(z_0)}. \nonumber \\
\end{eqnarray}
The errors on the parameters scale then with

\begin{equation}
\label{errorredshift}
 \Delta \theta^i(z) = \frac{d_L(z)}{d_L(z_0)} \Delta \theta^i(z_0).
\end{equation}

%\todo{and not with $(1+z) d_L(z) / ((1+z_0) d_L(z_0))$ as incorrectly stated in \cite{klein2009}.}

Since we choose to work in a picture where General Relativity is the theory assumed to be true and we are keen to know how well LISA will be able to measure
deviations from its post-Newtonian expansion terms $\psi_i$, we fixed the alternative theory parameters to the fiducial values $\Psi_i = 0$.

For each of the 17 binaries we computed the  best-case measurement error (5\% quantile), the typical error (median) and the worst-case error (95\% quantile) for the full (FWF)
and restricted waveforms (RWF) and present them in tables \ref{table:m1_21}-\ref{table:dl_15}. For each BBH parameter we are interested in, we give an error table with 
(21 parameters in total) and without (15 parameters in total) including the alternative theory parameters $\Psi_i$. We do this to show how much accuracy is lost by introducing 
alternative theory corrections into a GR waveform template. For binaries where no signal can be extracted from the dataset,
we fix the error to infinity.

We give the errors on the sky localization not in terms of errors on $\mu_n$ and $\phi_n$ but instead in terms of an error ellipse with principal axes $2a$ and
$2b$, enclosing the region outside of which there is an $1/e$ probability of finding the binary, following \cite{langhughes2006}.

Moreover, in tables \ref{table:Psim1_21}-\ref{table:Psi2_21} we give measurement errors on the alternative theory parameters, using both the RWF and FWF.

We roughly divide the binaries into two classes: low-mass binaries ($M \lesssim 10^7 \Msun$) and high-mass binaries ($M \gtrsim 10^7 \Msun$). 
Below we discuss these two cases, using BBHs with $m_1 = 10^6 \Msun$, $m_2 = 3 \times 10^5 \Msun$
and $m_1 = 3 \times 10^7 \Msun$, $m_2 = 10^7 \Msun$ as representative examples for low-mass and high-mass binaries respectively. We find when using both the
RWF and the FWF, the error distributions of the mass and spin parameters behave similarly, losing a factor $1.2-5$ of accuracy when alternative theory parameters are included.
The error on the sky location of the binary $2a$ and $2b$ is at maximum an order of magnitude worse. For high-mass binaries, factors of $\sim10$ and $\sim100$ are lost 
in the determination of the luminosity distance $d_L$, using the FWF and RWF respectively. While the RWF/FWF errors on the alternative theory parameters are almost equal for low-mass binaries, the RWF errors
are about $100$ times higher for high-mass binaries

\subsection{\label{sec:results_lowmass}Low-mass binaries}

For low-mass binaries with total masses below $10^7 \Msun$ we find that in general, using the FWF instead of the RWF improves the measurement errors $\Delta \Psi_i$ on the alternative
theory parameters by a factor of $\sim1.5-3$. The correlation with the new parameters causes a decrease in the accuracy
of the 15 binary parameters. For both the FWF and the RWF, the errors on the mass and spin parameters are typically worse by 
a factor of $2-5$ while the luminosity distance is approximately half as accurate. The sky location errors increase only by $\sim10\%$;
this is reasonable, since we do not expect alternative theories to correlate strongly with rotations on a large scale. Therefore it is not necessary to use the FWF instead
of the RWF for the sole purpose of measuring alternative gravity parameters in the low-mass regime.

\begin{figure}[!htp]
 \includegraphics[width=\columnwidth]{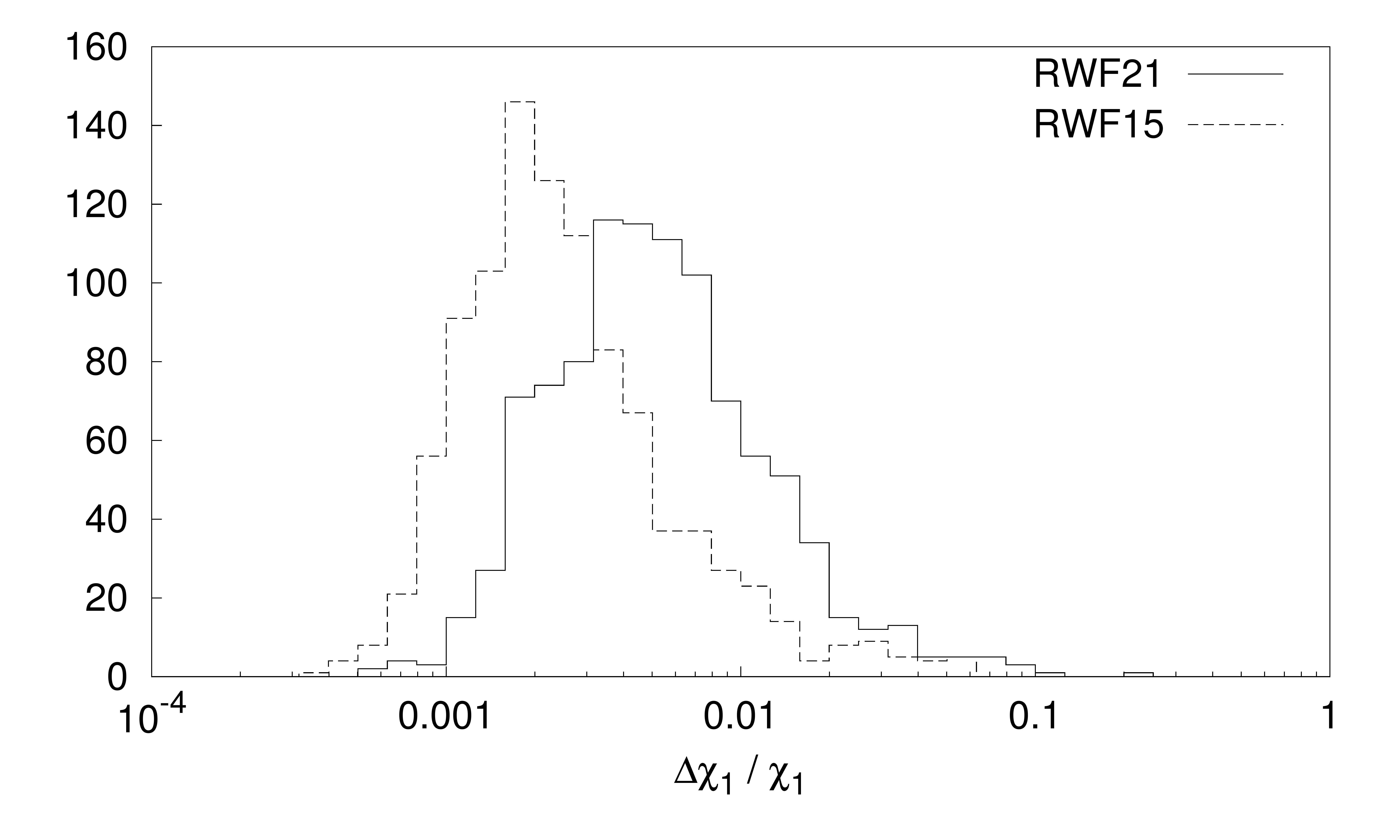}
  \caption{Comparison between the estimated distributions of the measurement error on
 $\chi_1$ for a low-mass binary system $m_1 = 1 \times 10^6 \Msun$ and $m_2 = 3 \times
10^5 \Msun$ with (RWF21) and without (RWF15) including alternative theory parameters, using only the restricted waveform. \label{fig:chi1635_RWF}}
\end{figure}

\begin{figure}[!htp]
 \includegraphics[width=\columnwidth]{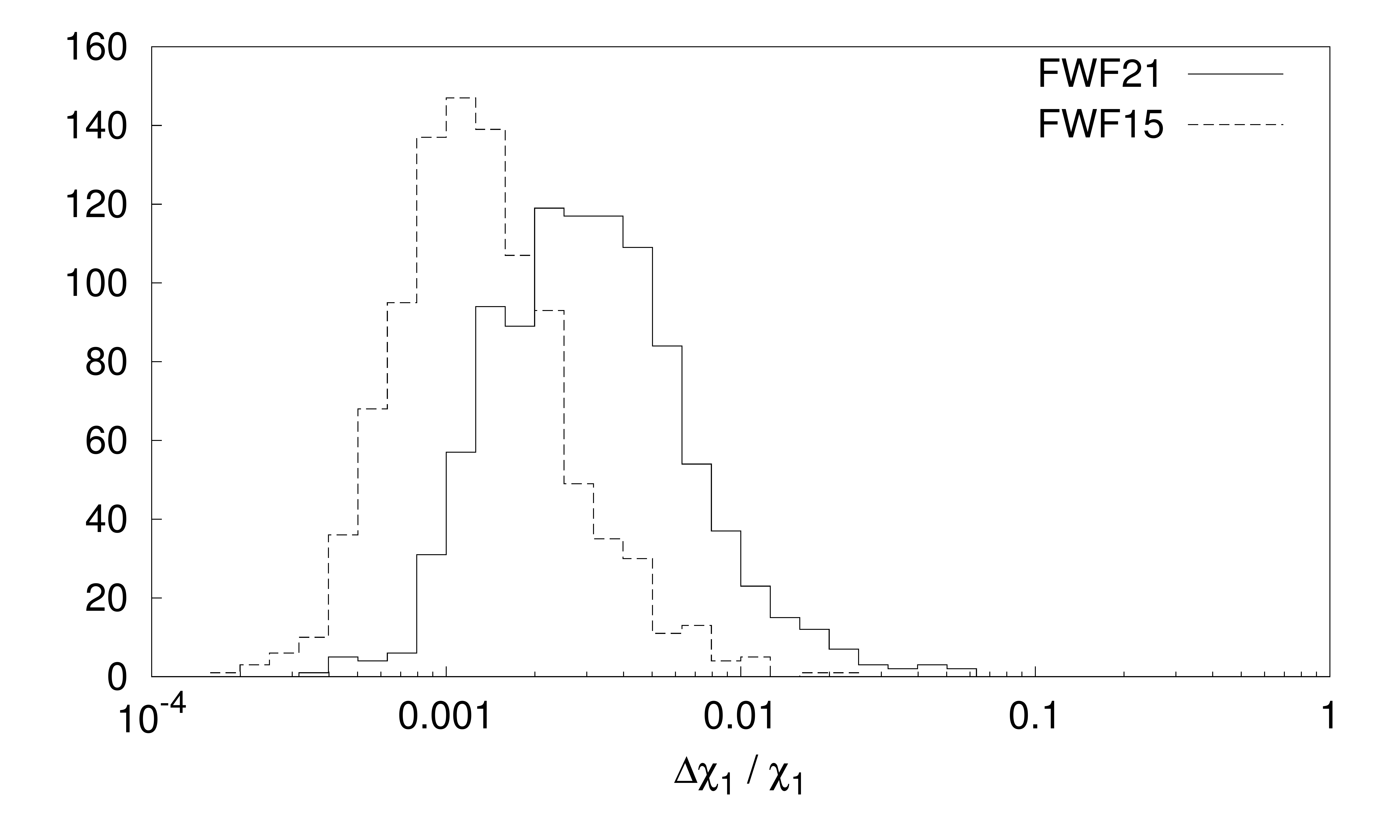}
  \caption{Comparison between the estimated distributions of the measurement error on
 $\chi_1$ for a low-mass binary system $m_1 = 1 \times 10^6 \Msun$ and $m_2 = 3 \times
10^5 \Msun$ with (FWF21) and without (FWF15) including alternative theory parameters and using the full waveform. \label{fig:chi1635_FWF}}
\end{figure}

\begin{figure}[!htp]
 \includegraphics[width=\columnwidth]{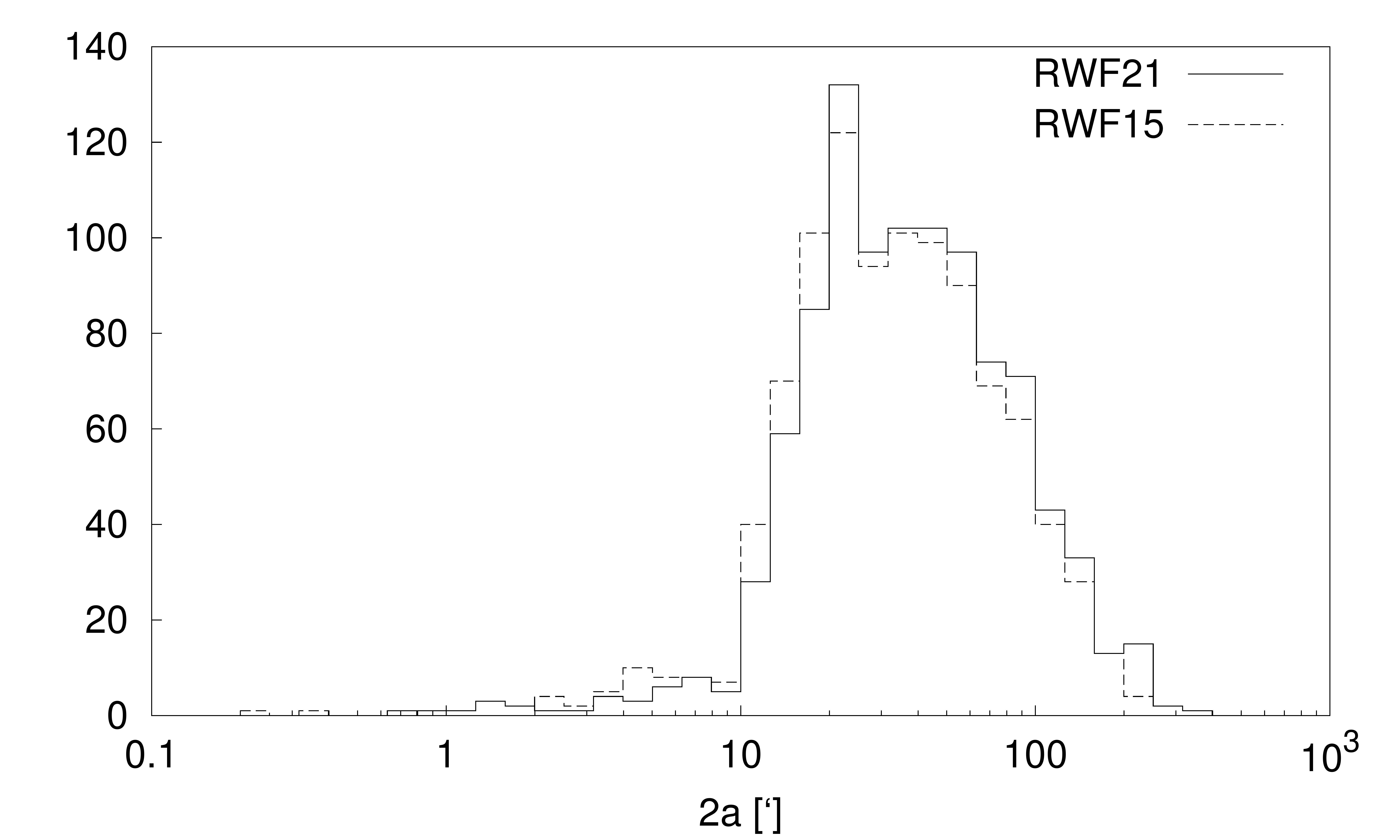}
  \caption{Comparison between the estimated distributions of the major axis of the positioning error ellipse for a low-mass binary system $m_1 = 1 \times 10^6 \Msun$ and $m_2 = 3 \times
10^5 \Msun$ with (RWF21) and without (RWF15) including alternative theory parameters, using only the restricted waveform. \label{fig:2a1635_RWF}}
\end{figure}

\begin{figure}[!htp]
 \includegraphics[width=\columnwidth]{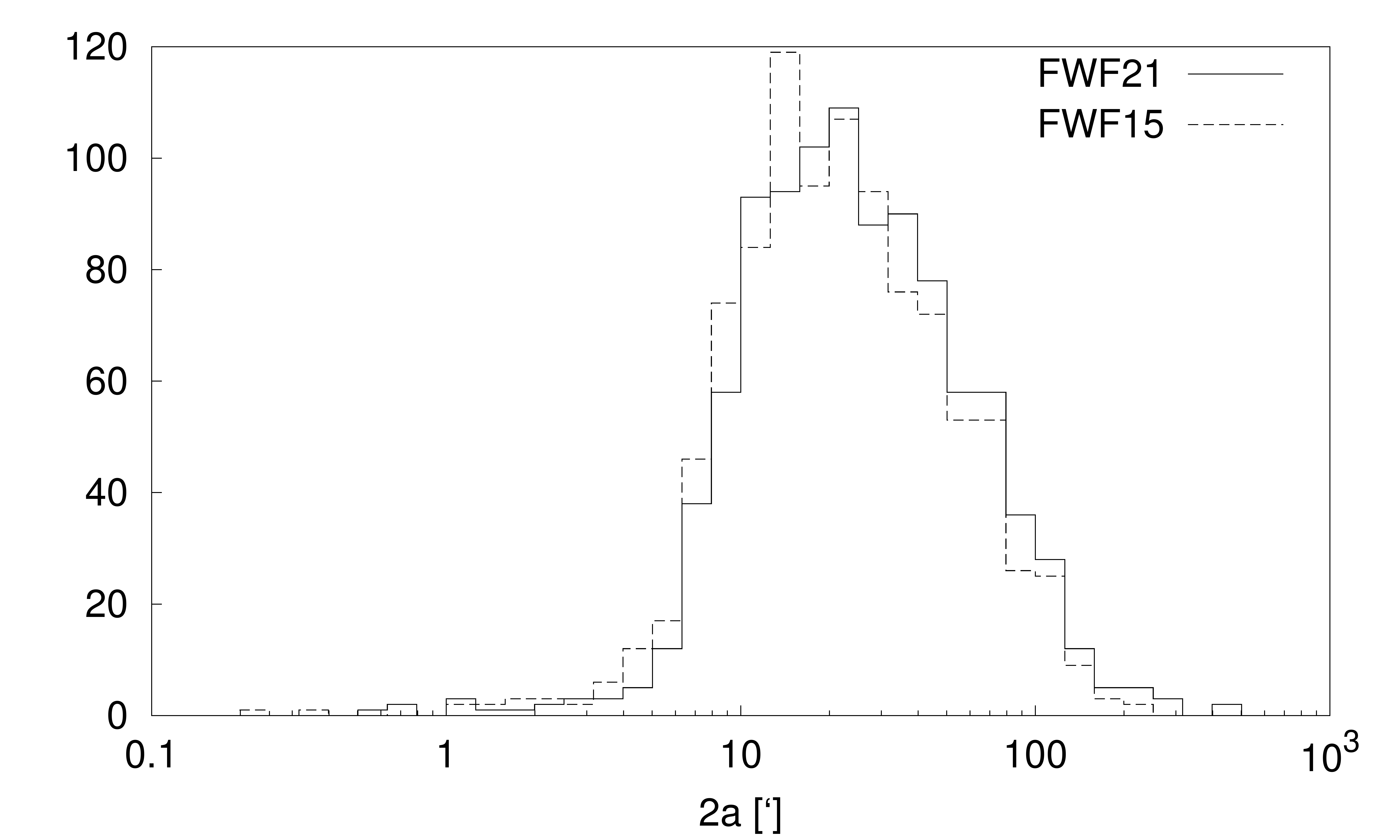}
  \caption{Comparison between the estimated distributions of the major axis of the positioning error ellipse for a low-mass binary system $m_1 = 1 \times 10^6 \Msun$ and $m_2 = 3 \times
10^5 \Msun$ with (FWF21) and without (FWF15) including alternative theory parameters and using the full waveform. \label{fig:2a1635_FWF}}
\end{figure}

\begin{figure}[!htp]
 \includegraphics[width=\columnwidth]{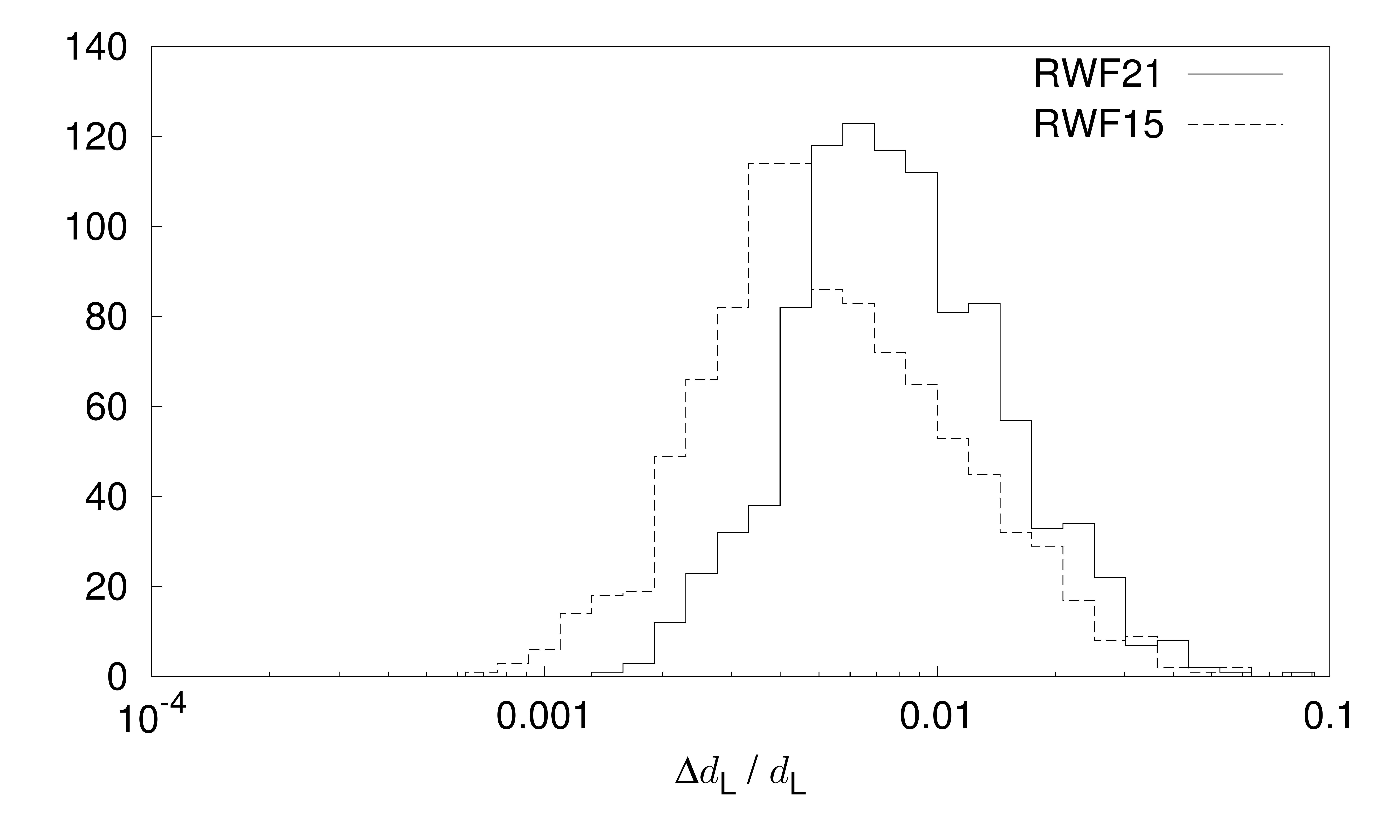}
  \caption{Comparison between the estimated distributions of the measurement error on
 $d_L$ for a low-mass binary system $m_1 = 1 \times 10^6 \Msun$ and $m_2 = 3 \times
10^5 \Msun$ with (RWF21) and without (RWF15) including alternative theory parameters, using only the restricted waveform. \label{fig:dl1635_RWF}}
\end{figure}

\begin{figure}[!htp]
 \includegraphics[width=\columnwidth]{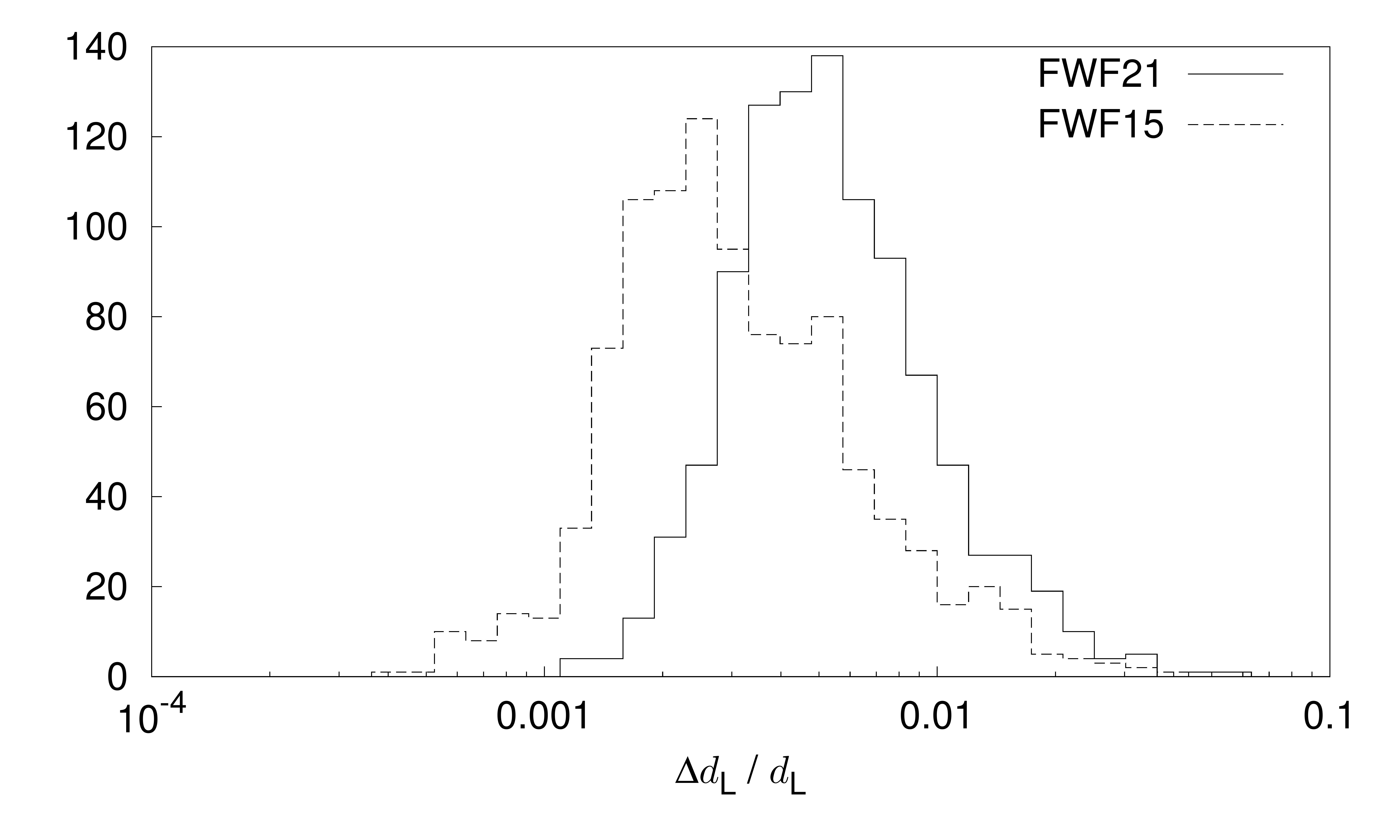}
  \caption{Comparison between the estimated distributions of the measurement error on
 $d_L$ for a low-mass binary system $m_1 = 1 \times 10^6 \Msun$ and $m_2 = 3 \times
10^5 \Msun$ with (FWF21) and without (FWF15) including alternative theory parameters, using only the restricted waveform. \label{fig:dl1635_FWF}}
\end{figure}

\begin{figure}[!htp]
 \includegraphics[width=\columnwidth]{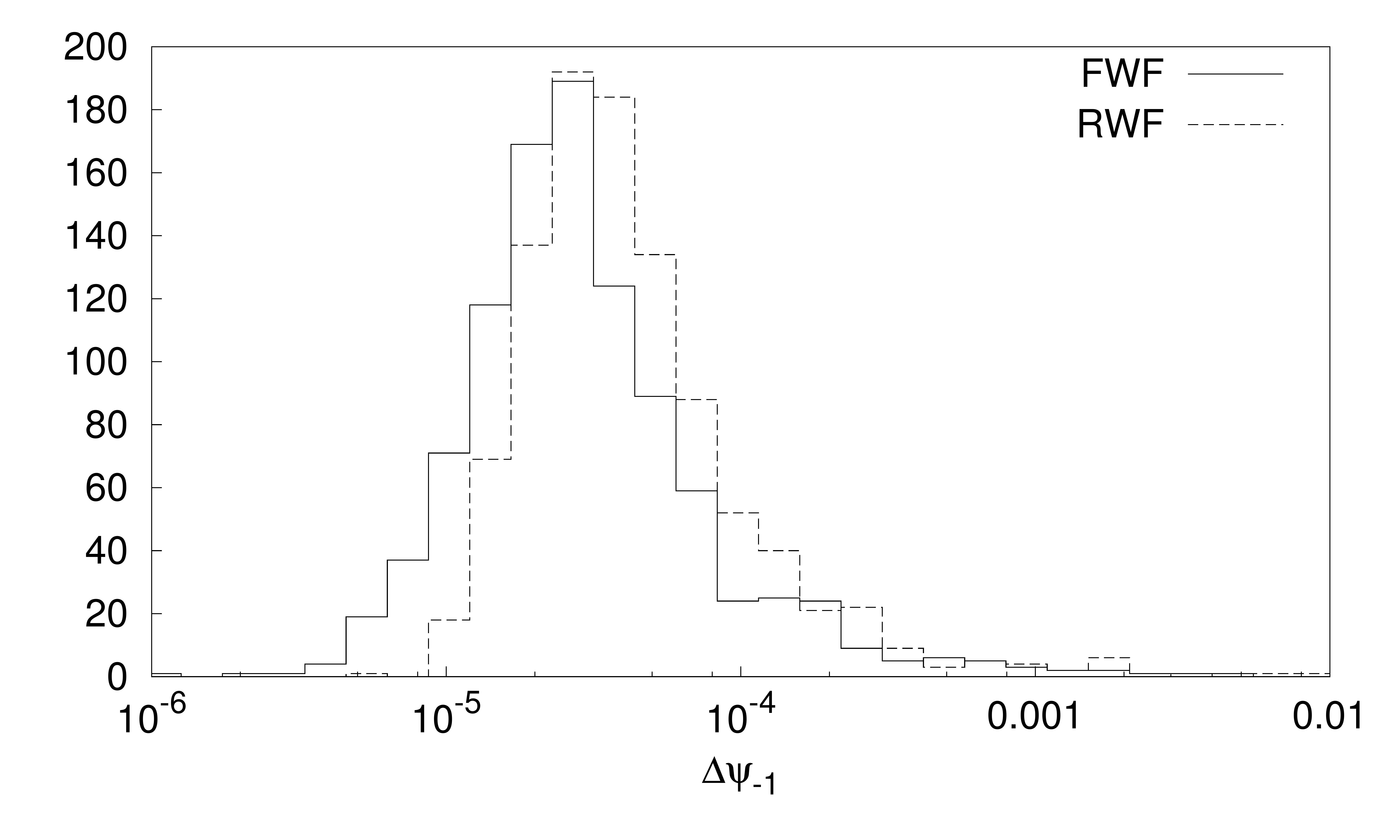}
  \caption{Comparison between the estimated distributions of the measurement error on the alternative theory parameter
 $\Psi_{\text{-1}}$ for a low-mass binary system $m_1 = 1 \times 10^6 \Msun$ and $m_2 = 3 \times
10^5 \Msun$, using the restricted waveform (RWF) and the full waveform (FWF). \label{fig:psim11635}}
\end{figure}

\begin{figure}[!htp]
 \includegraphics[width=\columnwidth]{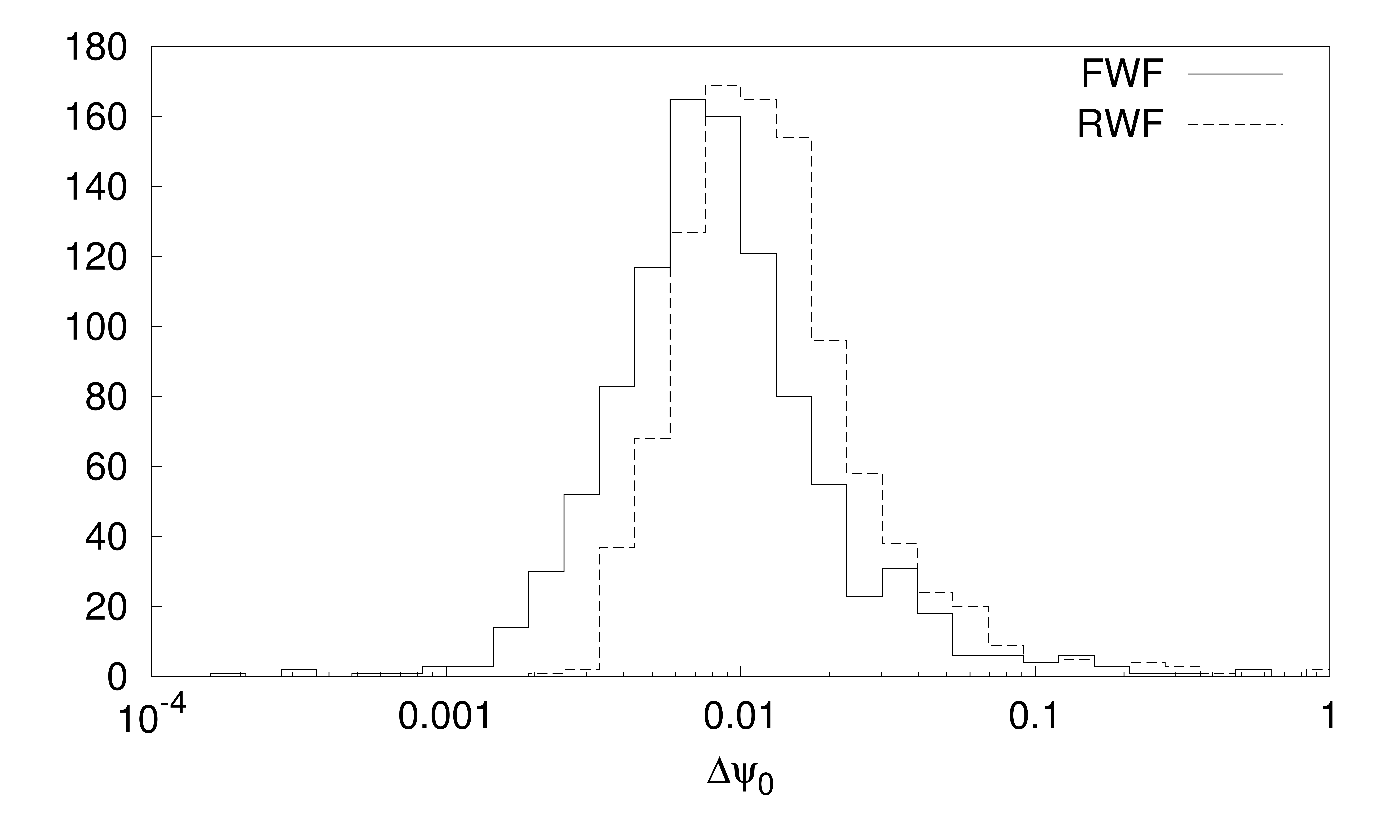}
  \caption{Comparison between the estimated distributions of the measurement error on the alternative theory parameter
 $\Psi_{0}$ for a low-mass binary system $m_1 = 1 \times 10^6 \Msun$ and $m_2 = 3 \times
10^5 \Msun$, using the RWF and the FWF. \label{fig:psi01635}}
\end{figure}

\begin{figure}[!htp]
 \includegraphics[width=\columnwidth]{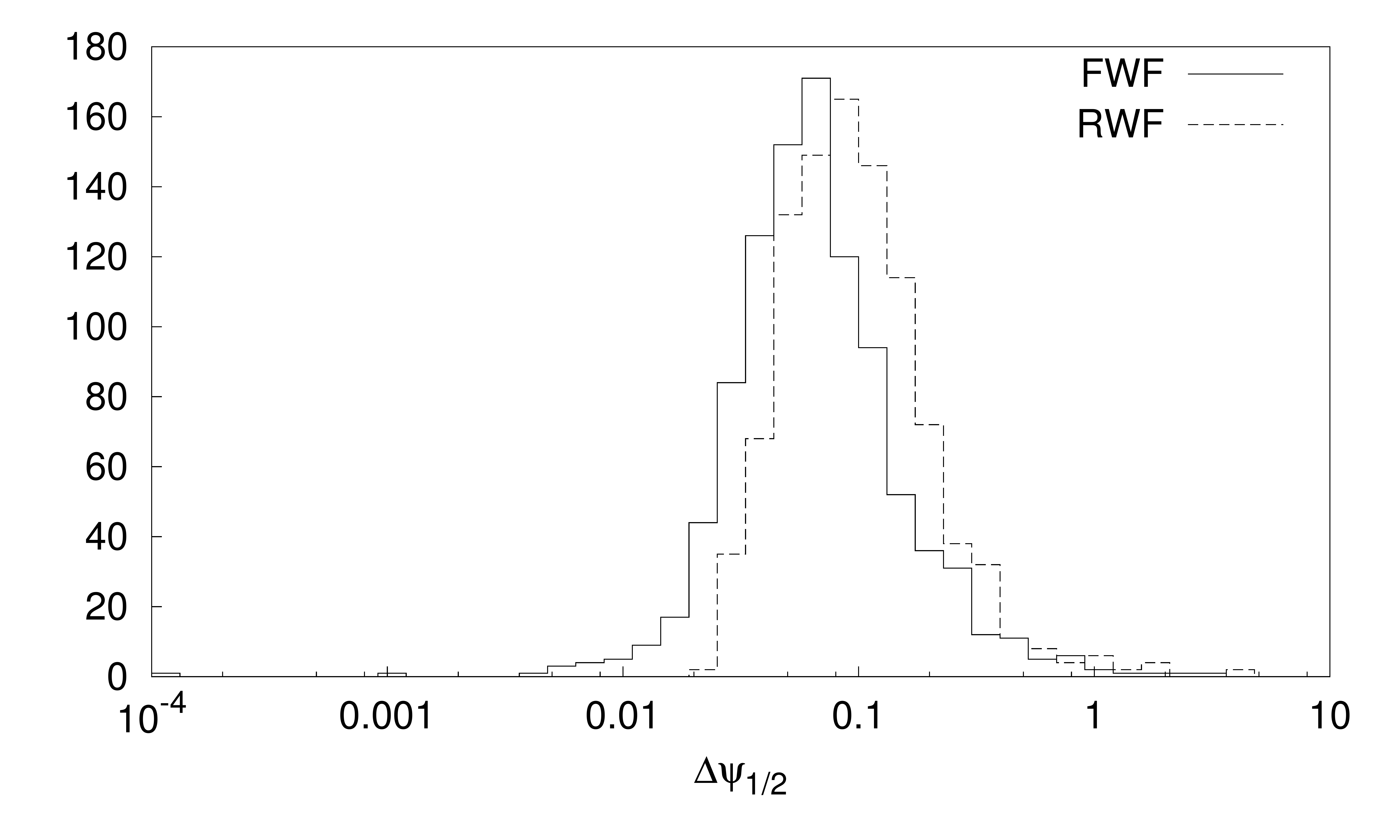}
  \caption{Comparison between the estimated distributions of the measurement error on the alternative theory parameter
 $\Psi_{1/2}$ for a low-mass binary system $m_1 = 1 \times 10^6 \Msun$ and $m_2 = 3 \times
10^5 \Msun$, using the RWF and the FWF. \label{fig:psi121635}}
\end{figure}

\begin{figure}[!htp]
 \includegraphics[width=\columnwidth]{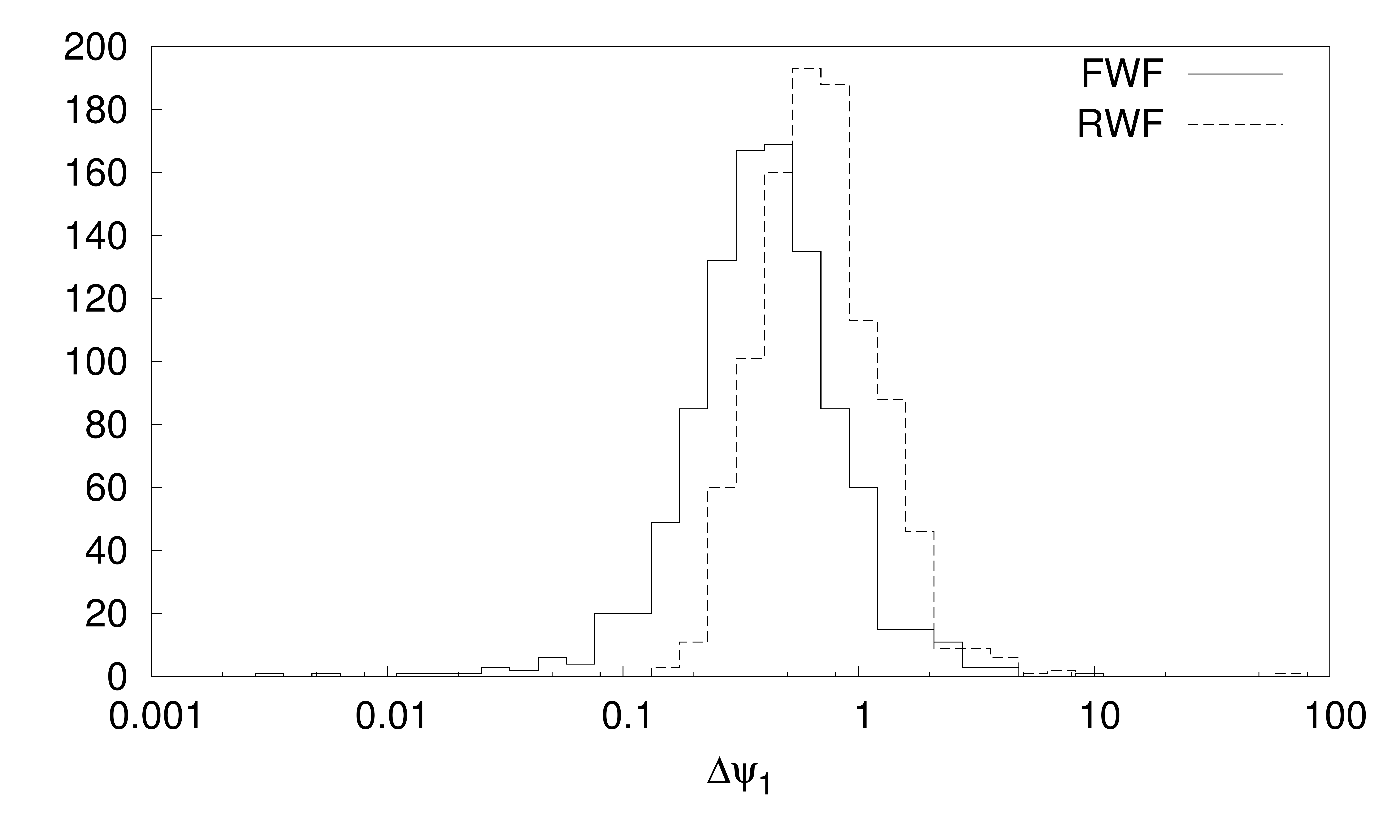}
  \caption{Comparison between the estimated distributions of the measurement error on the alternative theory parameter
 $\Psi_{1}$ for a low-mass binary system $m_1 = 1 \times 10^6 \Msun$ and $m_2 = 3 \times
10^5 \Msun$, using the RWF and the FWF. \label{fig:psi11635}}
\end{figure}

\begin{figure}[!htp]
 \includegraphics[width=\columnwidth]{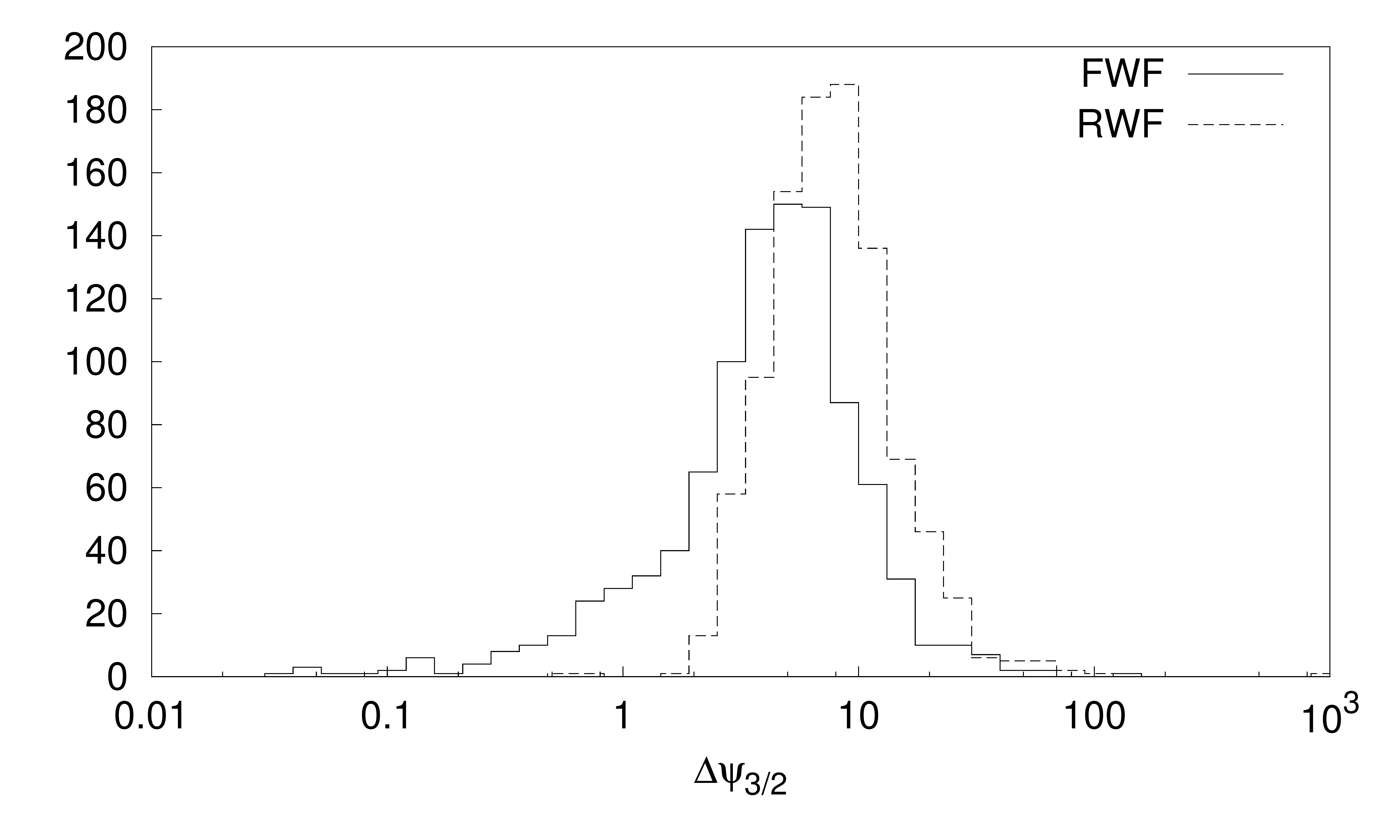}
  \caption{Comparison between the estimated distributions of the measurement error on the alternative theory parameter
 $\Psi_{3/2}$ for a low-mass binary system $m_1 = 1 \times 10^6 \Msun$ and $m_2 = 3 \times
10^5 \Msun$, using the RWF and the FWF. \label{fig:psi321635}}
\end{figure}

\begin{figure}[!htp]
 \includegraphics[width=\columnwidth]{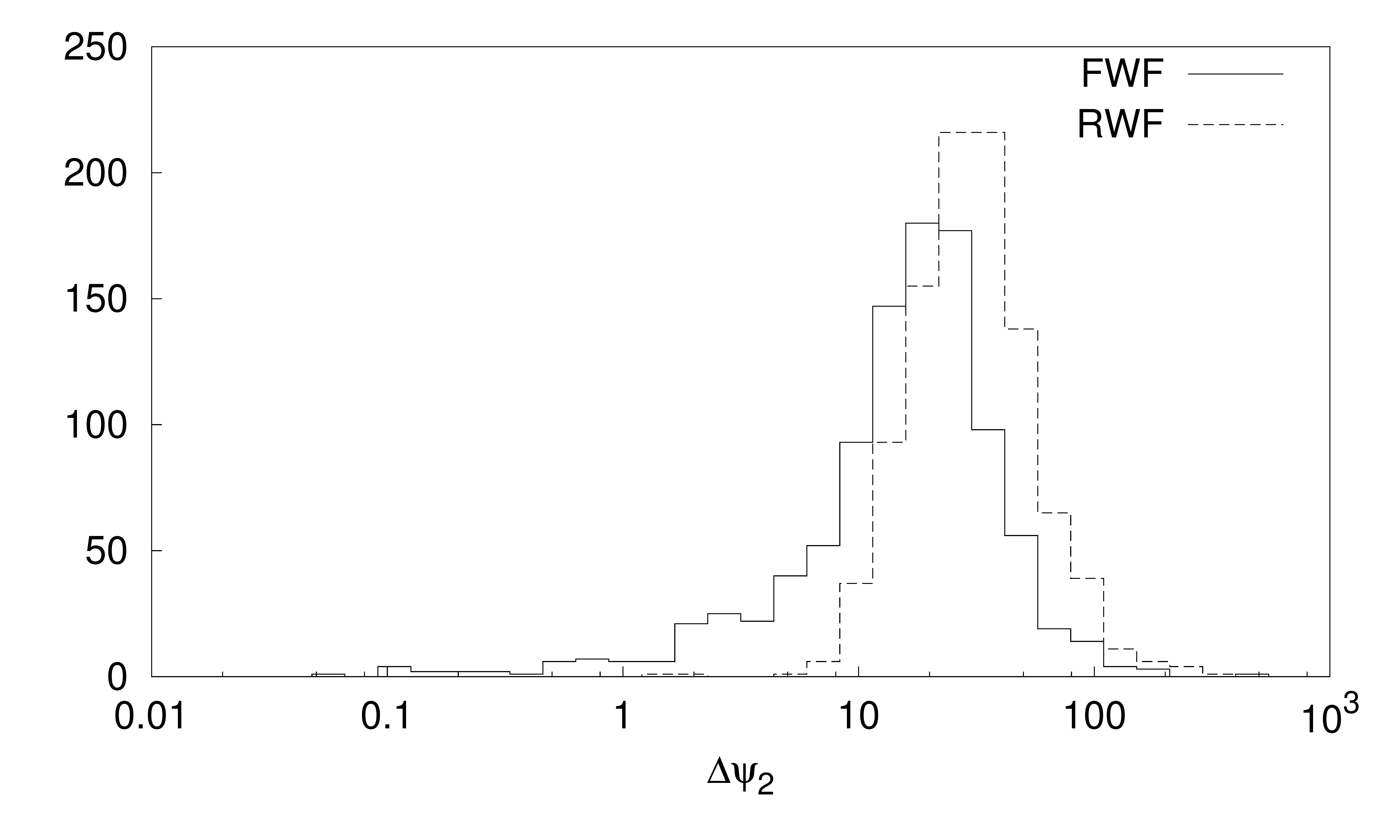}
  \caption{Comparison between the estimated distributions of the measurement error on the alternative theory parameter
 $\Psi_{2}$ for a low-mass binary system $m_1 = 1 \times 10^6 \Msun$ and $m_2 = 3 \times
10^5 \Msun$, using the RWF and the FWF. \label{fig:psi21635}}
\end{figure}

We present selected distributions of the measurement errors $\Delta m_1 / m_1$, $\Delta \chi_1/\chi_1$, $2a$, $\Delta d_L / d_L$ and all the six
$\Delta \Psi_i$ in figures \ref{fig:m11635_RWF}-\ref{fig:psi21635}. The error distributions of $\Delta m_2 / m_2$, $\Delta \chi_2/\chi_2$ and $2b$ are similar to the ones
of $\Delta m_1 / m_1$, $\Delta \chi_1/\chi_1$ and $2a$.

It is important to recall that we used the low frequency approximation (LFA) \cite{cutler1998, vecchiowickham2004, seto2002, rubbocornish2004} to generate the LISA detector response. This approximation
holds as long as the wavelength of the gravitational wave is much larger than the arm length $L$ of the LISA-type detector, in other words:
as long as $f_{GW} \ll f_* = \frac{c}{2 \pi L}$, where $c$ is the speed of light and $f_*$ is the so called \textit{transfer frequency}. As soon as the
wavelength is comparable to the arm length, the detector response function begins to depend strongly on the sky location and orientation of the source.
Effects neglected before, such as the cartwheel motion of LISA, become important, resulting in a modulation of the
the waveforms: more information about orientation and sky location is encoded in the signal. Consequently, the errors
on extrinsic parameters such as the angles $\mu_n$, $\phi_n$, the luminosity distance $d_L$ and the angular momentum orientation $\mu_l$,
$\phi_l$ effectively decrease compared to the LFA, while the intrinsic parameter errors differ only slighly. Usually, the problems with the approximation
start around $3$ mHz \cite{seto2002, rubbocornish2004, langhughescornish2011}: in our case the first three mass configurations with total masses of $3.3\times10^5$, $4\times10^5$ and 
$6\times10^5 \Msun$ are above this limit, with frequencies (at $f_{\text{ISCO}} = 6 \, GM/c^2$ and redshift z=1) of $6.6$, $5$ and $3.6$ mHz, 
respectively. Following fig. 2 in \cite{seto2002}, this means that our results for these three configurations should be too pessimistic, the relative
errors on the luminosity distance would in general be smaller by $\sim 10\%$, $20\%$ and $50\%$ for the respective configurations. Also the errors
on sky location and angular momentum orientation will be better by up to $\sim 50\%$ for the $3.3\times10^5$ binary.

%\cleardoublepage

\subsection{\label{sec:results_highmass}High-mass binaries}

\begin{figure}[!htp]
 \includegraphics[width=\columnwidth]{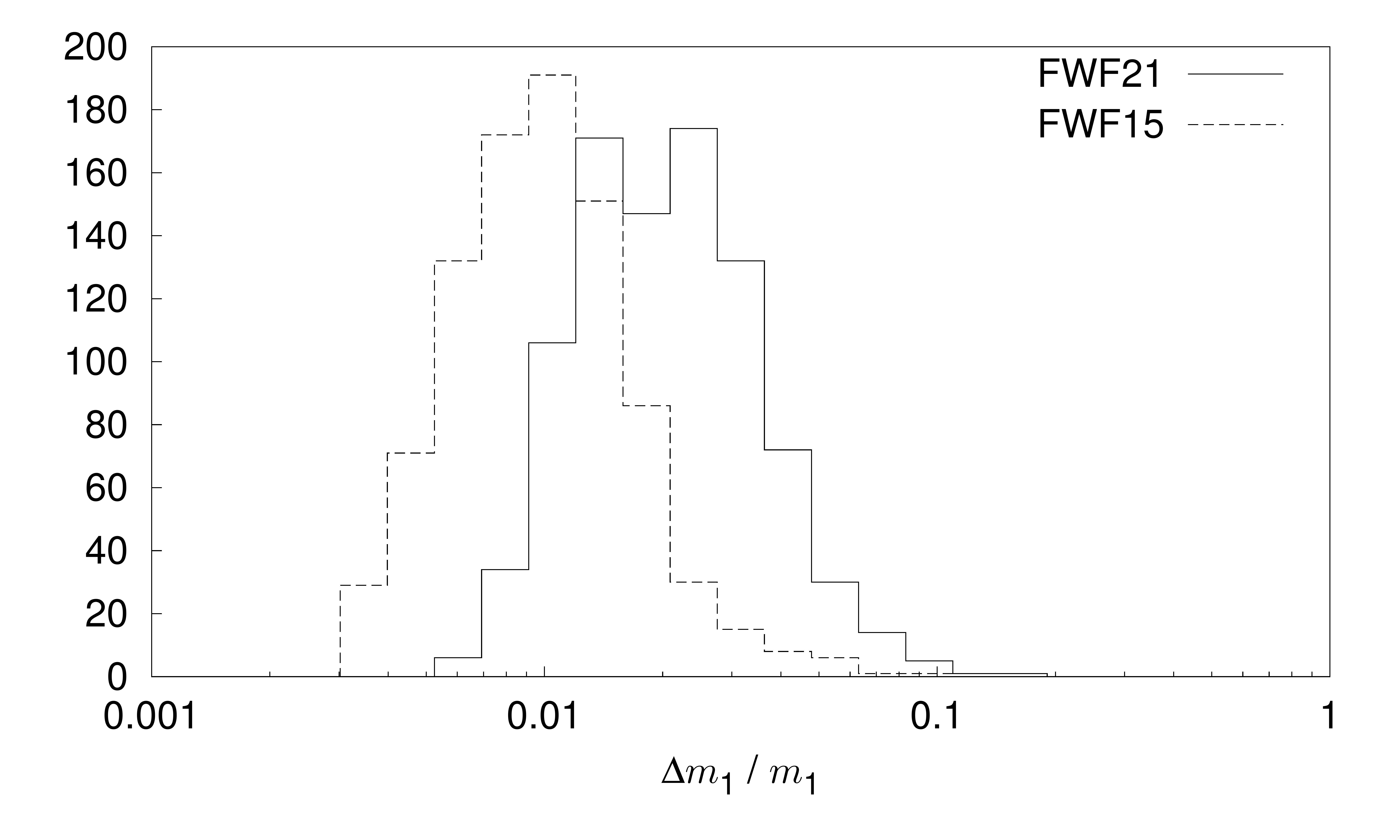}
  \caption{Comparison between the estimated distributions of the measurement error on
 $m_1$ for a high-mass binary system $m_1 = 3 \times 10^7 \Msun$ and $m_2 = 1 \times
10^7 \Msun$ with (FWF21) and without (FWF15) including alternative theory parameters and using the full waveform. \label{fig:m13717_FWF}}
\end{figure}

\begin{figure}[!htp]
 \includegraphics[width=\columnwidth]{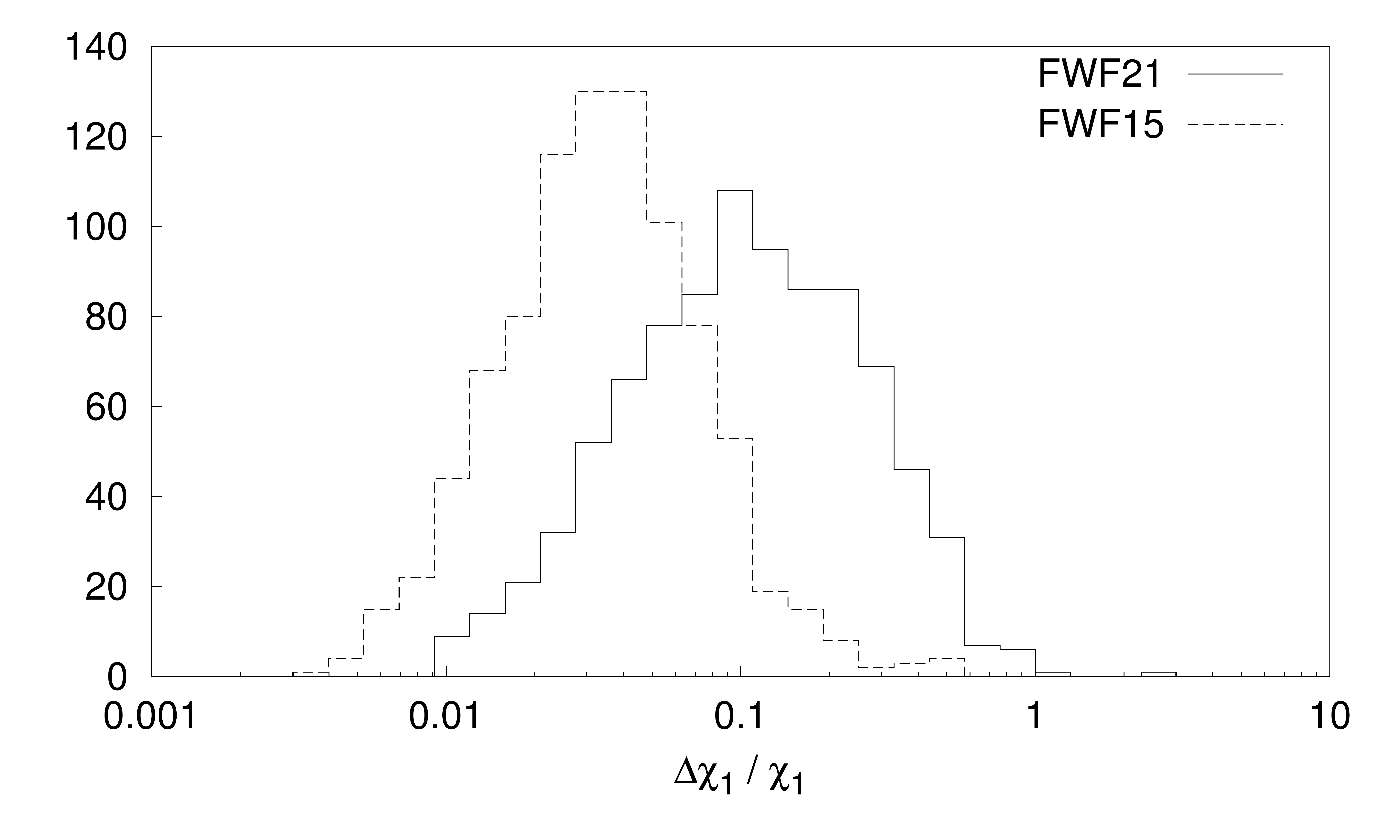}
  \caption{Comparison between the estimated distributions of the measurement error on
 $\chi_1$ for a high-mass binary system $m_1 = 3 \times 10^7 \Msun$ and $m_2 = 1 \times
10^7 \Msun$ with (FWF21) and without (FWF15) including alternative theory parameters and using the full waveform. \label{fig:chi3717_FWF}}
\end{figure}

\begin{figure}[!htp]
 \includegraphics[width=\columnwidth]{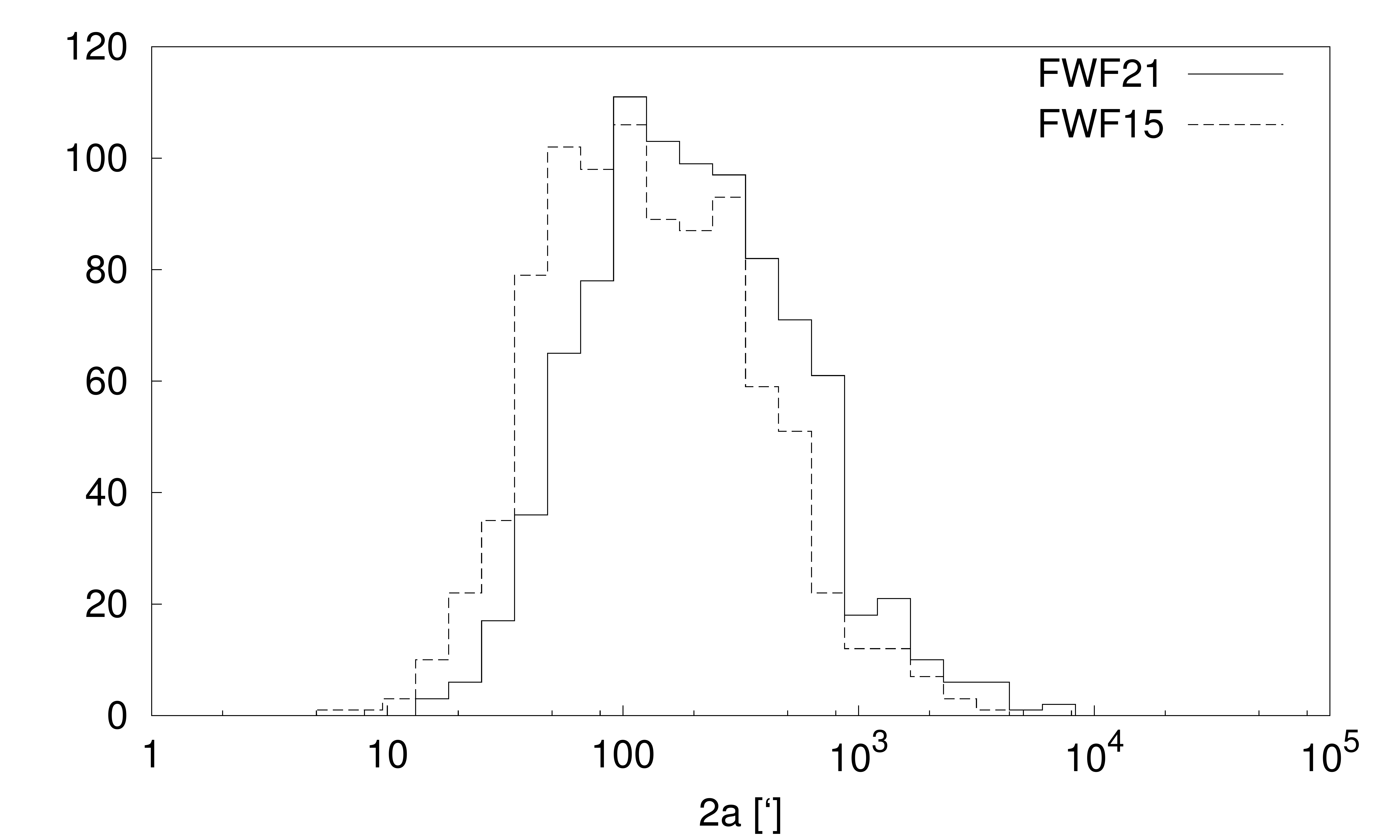}
  \caption{Comparison between the estimated distributions of the major axis of the positioning error ellipse for a high-mass binary system $m_1 = 3 \times 10^7 \Msun$ and $m_2 = 1 \times
10^7 \Msun$ with (FWF21) and without (FWF15) including alternative theory parameters and using the full waveform. \label{fig:2a3717_RWF}}
\end{figure}

By using the FWF instead of the RWF for high-mass binaries with total masses $\gtrsim 10^7 \Msun$, we find significant improvements for the measurement
errors of the alternative theory parameters by factors of $\sim100-1000$ for $\Delta \Psi_{\text{-1}}$, $\sim30-60$ for $\Delta \Psi_{0}$ and $\Delta \Psi_{1/2}$, and 
$\sim10-100$ for $\Delta \Psi_{1}$, $\Delta \Psi_{3/2}$ and $\Delta \Psi_{2}$. This makes it clear that it is inevitable to use the FWF in the high-mass regime to perform precision tests of GR. In any case, since the second
harmonic spends only a few orbits in the LISA band, the use of the RWF is not trustworthy. Moreover, for BBHs with total masses higher than $10^8 \Msun$, 
LISA will not be able to see the second harmonic at all and so the RWF cannot be used. For both the FWF and the RWF, the errors on the mass and spin parameters
are typically worse by a factor of $\sim1.2-4$ when accounting for alternative gravity parameters. The luminosity distance is about $50-1000$ times less accurate for the RWF while for the FWF
it is only $\sim 10-100$ times worse. For the FWF, the sky location error is at maximum $5$ times worse while the RWF loses up to a factor of $\sim10$
in accuracy.

\begin{figure}[!htp]
 \includegraphics[width=\columnwidth]{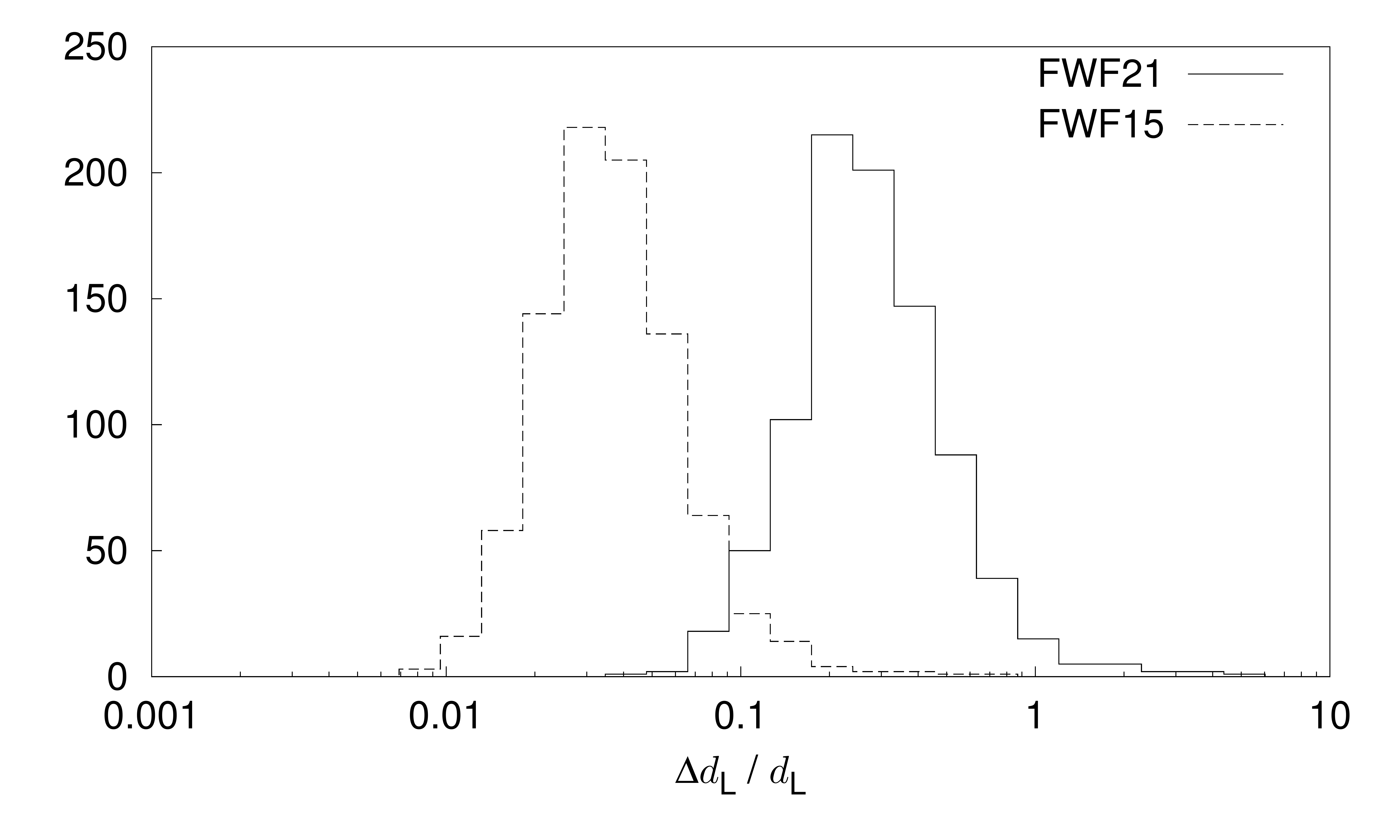}
  \caption{Comparison between the estimated distributions of the measurement error on
 $d_L$ for a high-mass binary system $m_1 = 3 \times 10^7 \Msun$ and $m_2 = 1 \times
10^7 \Msun$ with (FWF21) and without (FWF15) including alternative theory parameters, using only the restricted waveform. \label{fig:dl3717_FWF}}
\end{figure}

\begin{figure}[!htp]
 \includegraphics[width=\columnwidth]{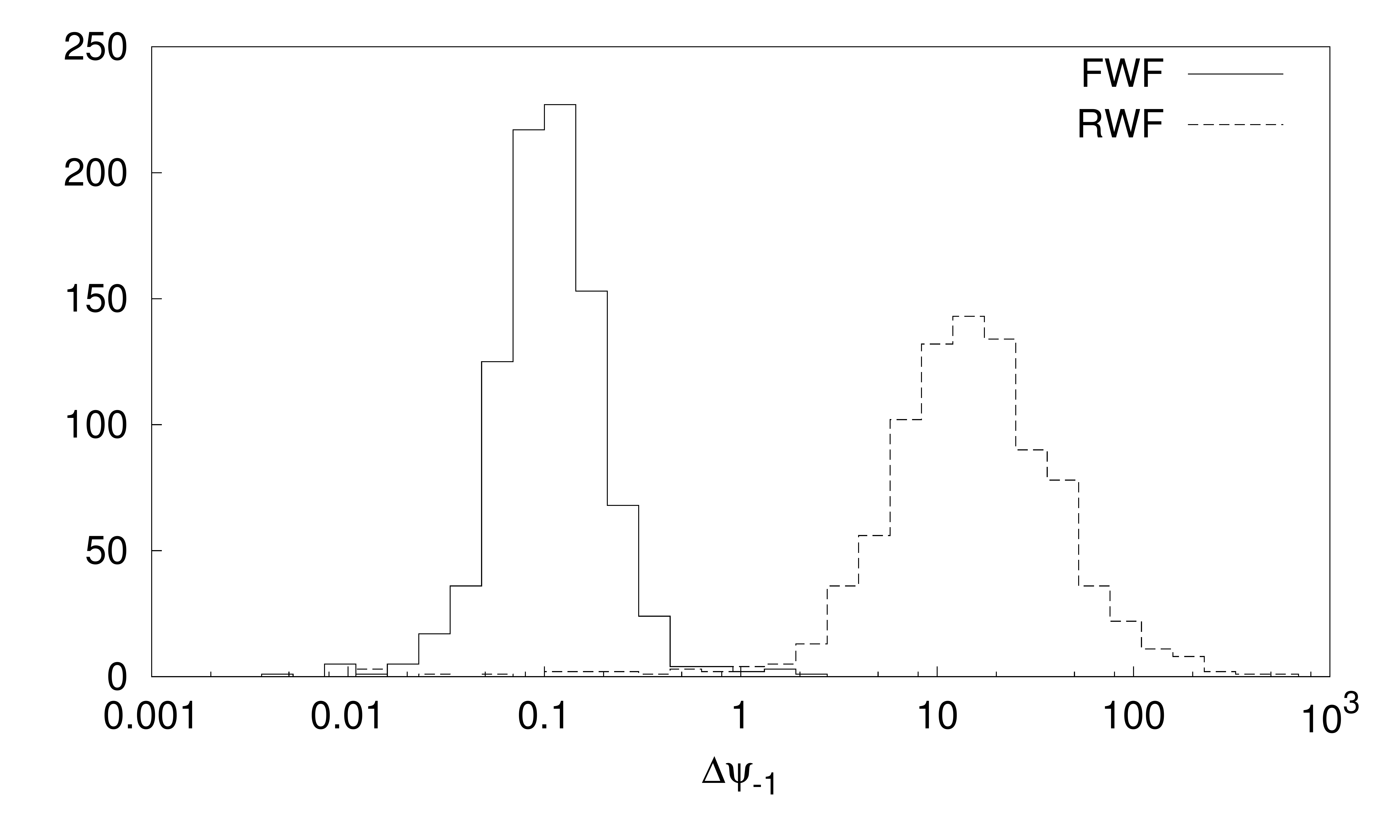}
  \caption{Comparison between the estimated distributions of the measurement error on the alternative theory parameter
 $\Psi_{\text{-1}}$ for a high-mass binary system $m_1 = 3 \times 10^7 \Msun$ and $m_2 = 1 \times
10^7 \Msun$, using the RWF and the FWF. \label{fig:psim13717}}
\end{figure}

\begin{figure}[!htp]
 \includegraphics[width=\columnwidth]{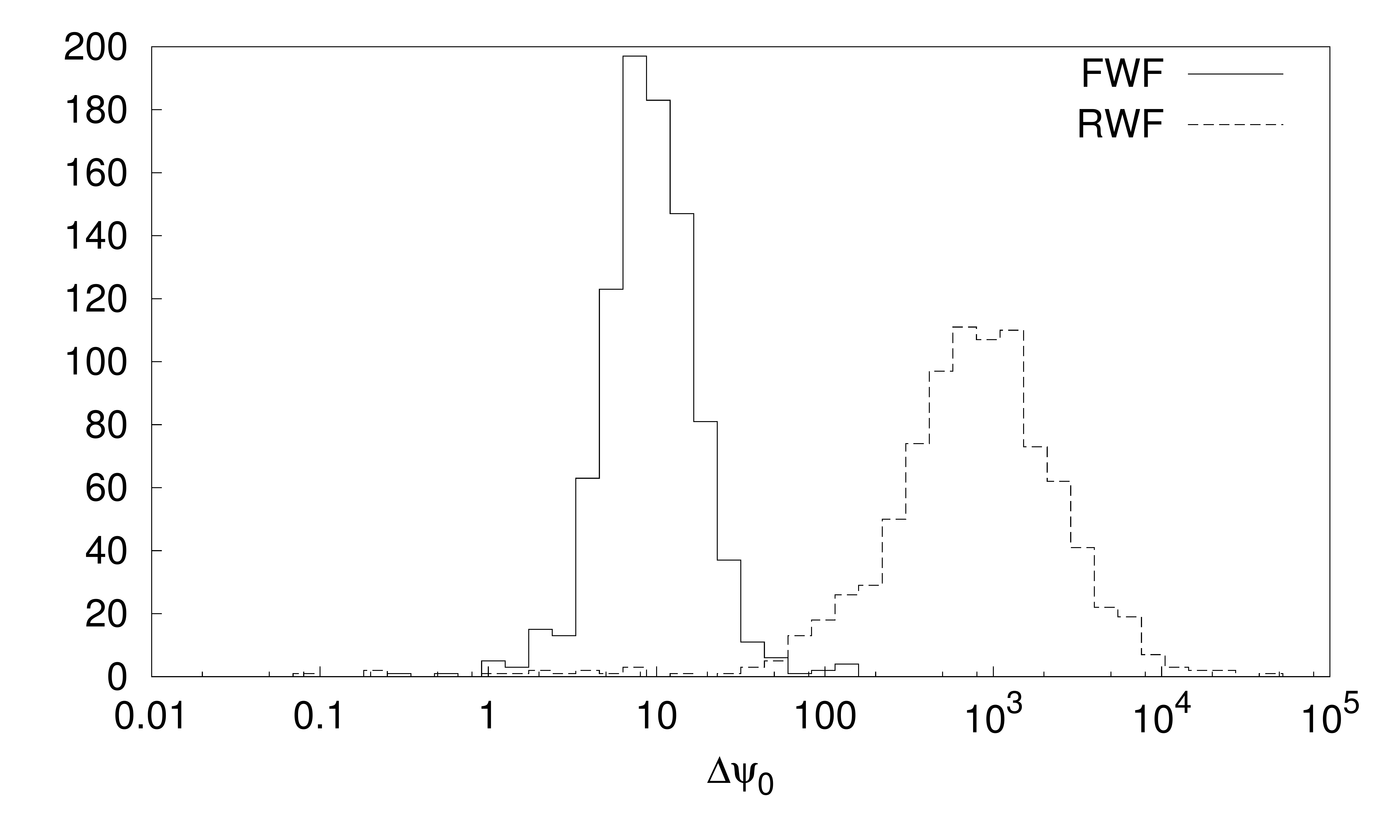}
  \caption{Comparison between the estimated distributions of the measurement error on the alternative theory parameter
 $\Psi_{0}$ for a high-mass binary system $m_1 = 3 \times 10^7 \Msun$ and $m_2 = 1 \times
10^7 \Msun$, using the RWF and the FWF. \label{fig:psi03717}}
\end{figure}

\begin{figure}[!htp]
 \includegraphics[width=\columnwidth]{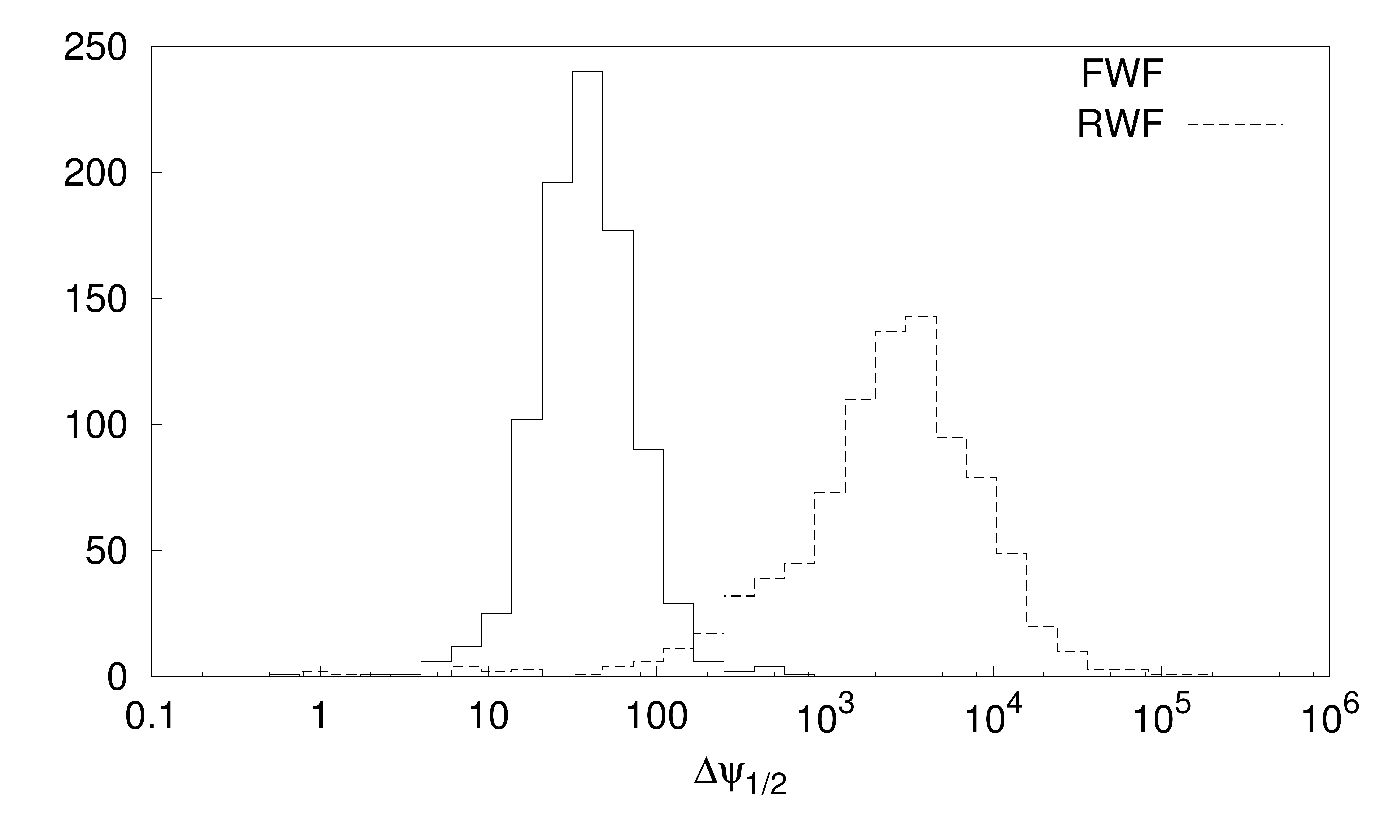}
  \caption{Comparison between the estimated distributions of the measurement error on the alternative theory parameter
 $\Psi_{1/2}$ for a high-mass binary system $m_1 = 3 \times 10^7 \Msun$ and $m_2 = 1 \times
10^7 \Msun$, using the RWF and the FWF. \label{fig:psi123717}}
\end{figure}

\begin{figure}[!htp]
 \includegraphics[width=\columnwidth]{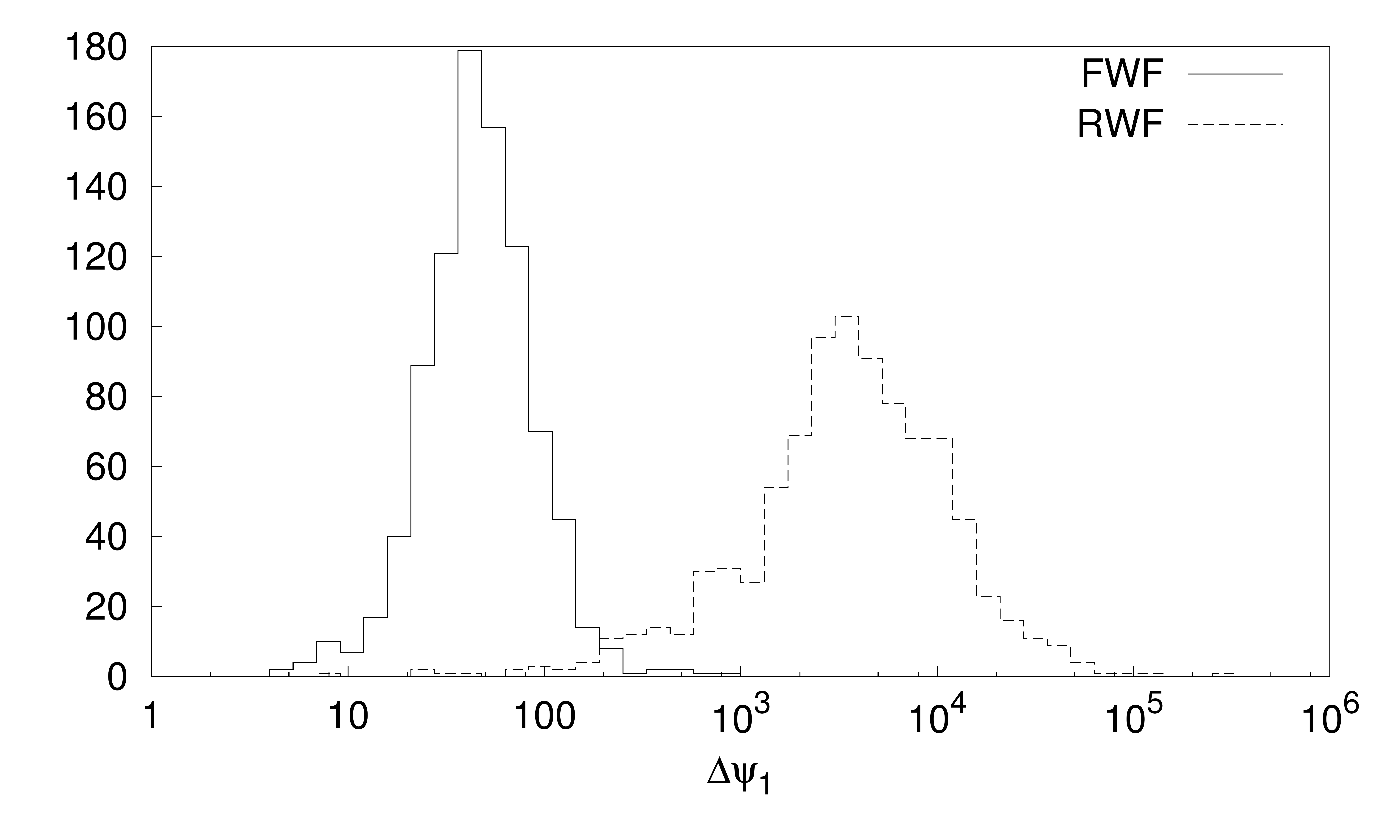}
  \caption{Comparison between the estimated distributions of the measurement error on the alternative theory parameter
 $\Psi_{1}$ for a high-mass binary system $m_1 = 3 \times 10^7 \Msun$ and $m_2 = 1 \times
10^7 \Msun$, using the RWF and the FWF. \label{fig:psi13717}}
\end{figure}

\begin{figure}[!htp]
 \includegraphics[width=\columnwidth]{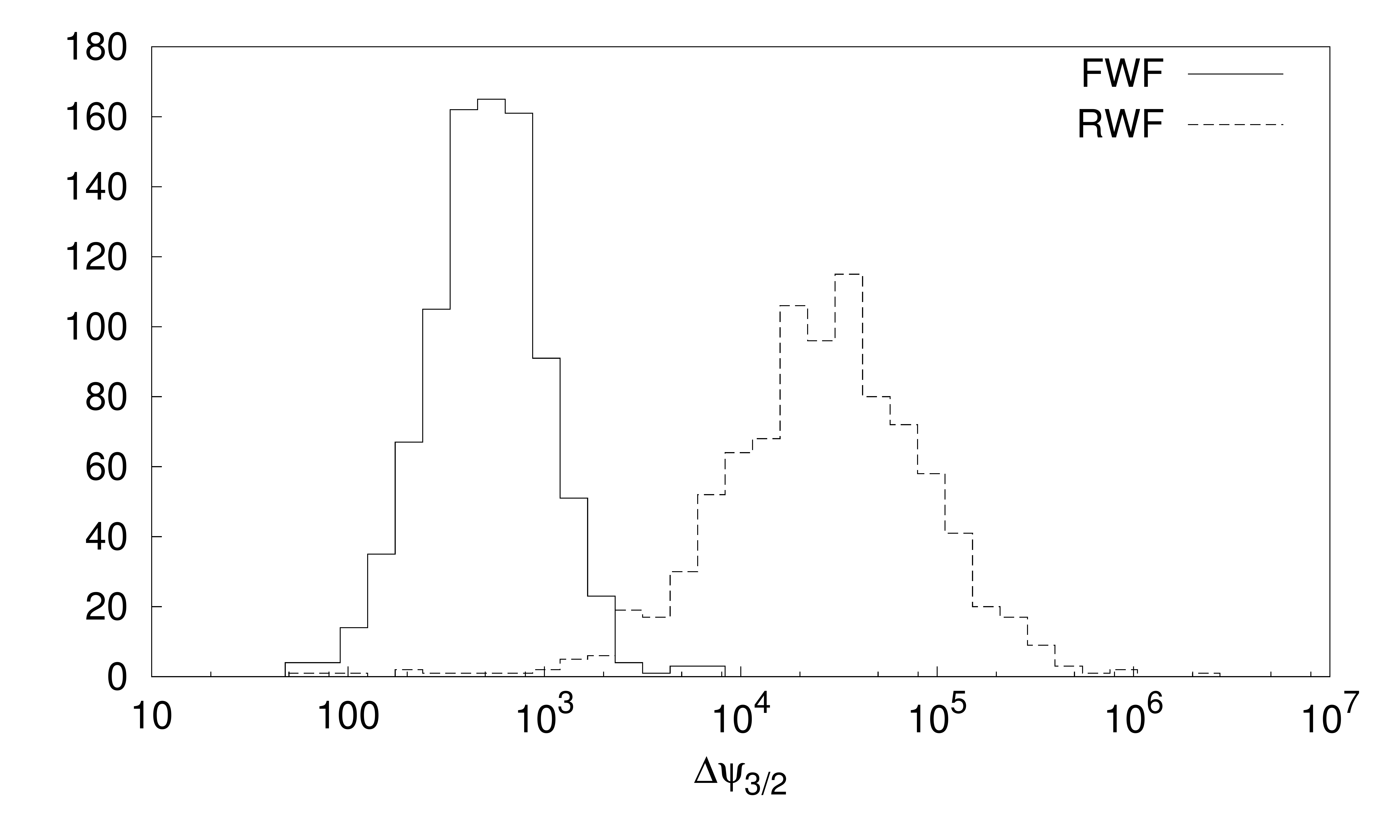}
  \caption{Comparison between the estimated distributions of the measurement error on the alternative theory parameter
 $\Psi_{3/2}$ for a high-mass binary system $m_1 = 3 \times 10^7 \Msun$ and $m_2 = 1 \times
10^7 \Msun$, using the RWF and the FWF. \label{fig:psi323717}}
\end{figure}

\begin{figure}[!htp]
 \includegraphics[width=\columnwidth]{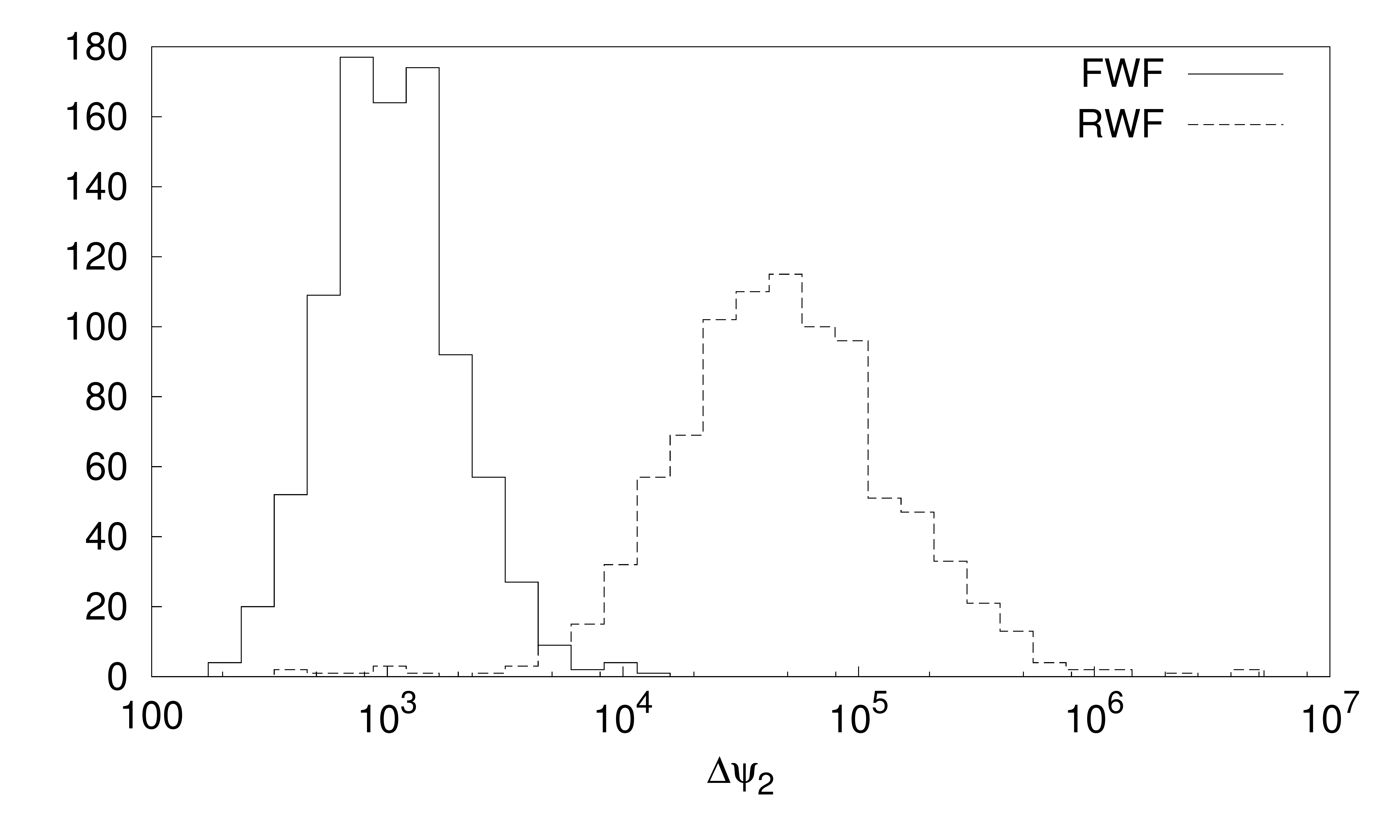}
  \caption{Comparison between the estimated distributions of the measurement error on the alternative theory parameter
 $\Psi_{2}$ for a high-mass binary system $m_1 = 3 \times 10^7 \Msun$ and $m_2 = 1 \times
10^7 \Msun$, using the RWF and the FWF. \label{fig:psi23717}}
\end{figure}

We present selected distributions of the measurement errors $\Delta m_1 / m_1$, $\Delta \chi_1/\chi_1$, $2a$, $\Delta d_L / d_L$ and all the six
$\Delta \Psi_i$ in figures \ref{fig:m13717_FWF}-\ref{fig:psi23717}.

%\cleardoublepage

\subsection{\label{sec:correlations}Correlations between alternative theory parameters}

The correlation coefficients for two parameters $\theta_i$ and $\theta_j$ are given by the normalized covariance matrix as

\begin{equation}
 C_{ij} = \frac{\Sigma_{ij}}{\sqrt{\Sigma_{ii} \Sigma_{jj}}},
\end{equation}

and are in a range between $-1$ (perfectly anti-correlated) and $1$ (perfectly correlated). Since we are only interested in the mere
presence of correlations, we will focus on the absolute value $|C_{ij}|$ varying in the range between $0$ (no correlation) and $1$.

Because of their simple form in the gravitational wave phase, the alternative theory parameters are expected to correlate highly among each other
and with the rest of the phase parameters, especially with the ones which have a similar simple dependency on frequency (and are already highly correlated) like 
the phase or time at coalescence, $\phi_c$ and $t_c$. Often, the use of higher harmonics makes the resulting errors and correlations more complicated and 
unpredictable: the mostly narrow and symmetric RWF distribution is smeared out over the whole range of possible correlations, usually with a long tail. Also, higher harmonics
can in principle introduce new correlations among certain parameters that have not been there before. 
Below we shortly investigate correlations among the alternative theory parameters and between alternative theory and binary parameters.

\subsubsection{Correlations between alternative theory parameters}

\begin{figure}[!htp]
 \includegraphics[width=\columnwidth]{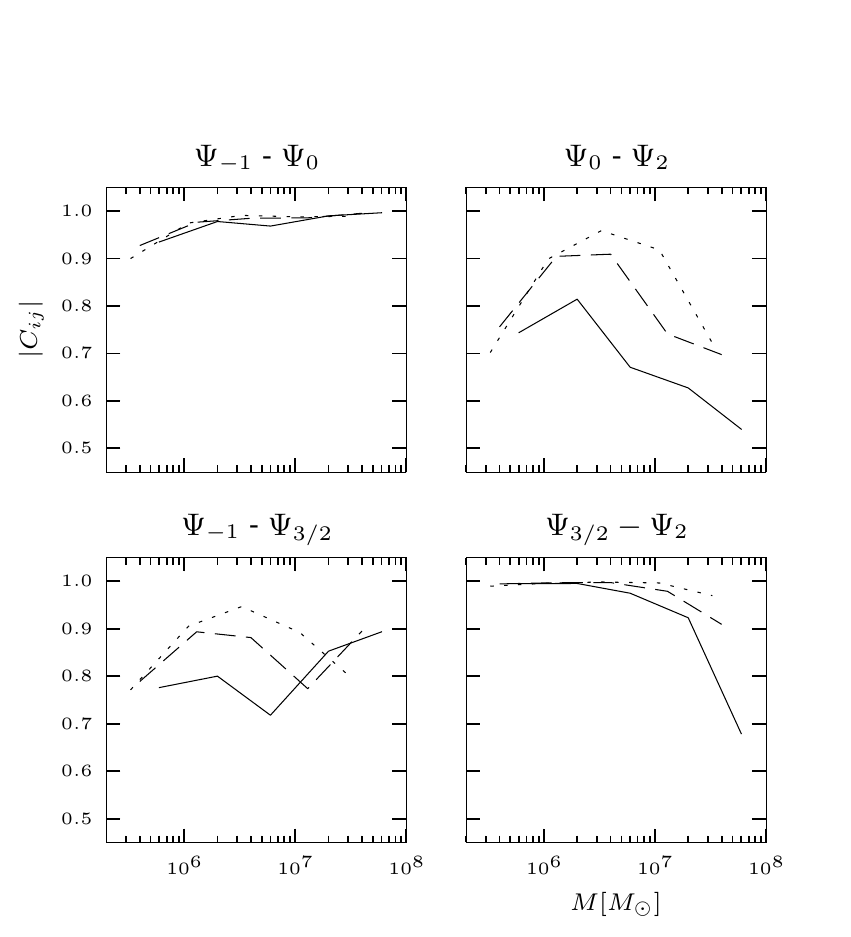}
  \caption{Median correlations (FWF) between selected alternative theory parameters varying with total mass and shown for each mass ratio independently 
  (1:1 - solid line, 1:3 - dashed line, 1:10 - dotted line). The sets $\{\Psi_{\text{-1}}, \Psi_0, \Psi_{1/2}\}$ and $\{\Psi_{1}, \Psi_{3/2}, \Psi_{2}\}$
  show very high correlations among themselves (top-left, bottom-right) while correlations between theory parameters belonging to different sets
  are lower (top-right, bottom-left).
   \label{fig:masses_medianplot_theory}}
\end{figure}

We find that the alternative theory parameters can be subdivided into two sets: $\Psi_{\text{low}} \equiv \{\Psi_{\text{-1}}, \Psi_0, \Psi_{1/2}\}$ 
and $\Psi_{\text{high}} \equiv \{\Psi_{1}, \Psi_{3/2}, \Psi_{2}\}$. The parameters in every set show very high correlations among each other, but less
correlation with the parameters of the other set. 
The parameters in $\Psi_{\text{low}}$ have either no fiducial GR phase equivalent with the same frequency power ($\Psi_{\text{-1}}$ and $\Psi_{1/2}$) or one which 
is fixed to 1 ($\Psi_0$). In contrast, every parameter in $\Psi_{\text{high}}$ can correlate to intrinsic binary parameters with the same 
frequency dependency, such as masses and spins. Since one integrates over the frequency to compute the Fisher matrix, two parameters have higher correlation if
the frequency powers proportional to which they appear in the phase or amplitude are close. So we expect parameters from $\Psi_{\text{high}}$ to have higher correlation
with the intrinsic binary parameters appearing in the GR phase with the same frequency power than with the $\Psi_{\text{low}}$ parameters appearing with lower 
frequency powers. Consequently, we expect high correlations among the parameters within both sets and also high, but slightly lower correlations between 
parameters belonging to a different set each.
In fig. \ref{fig:masses_medianplot_theory}, we plotted the median FWF correlations for selected
parameters of both sets against the total mass to illustrate this finding. For two parameters drawn from different sets, the mass ratio also plays an important role
for the resulting correlations, while for parameters from the same set, the correlations mainly depend on the total mass.

\begin{figure}[!htp]
 \includegraphics[width=\columnwidth]{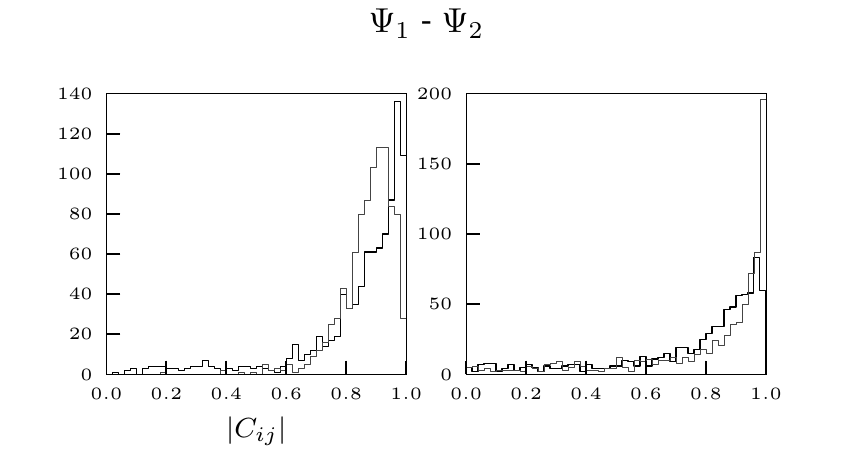}
  \caption{Correlation breaking for a low-mass $m_1 = 10^6 \Msun$, $m_2 = 3\times10^5 \Msun$ binary (left) and a $m_1 = 3\times10^7 \Msun$, $m_2 = 10^7 \Msun$ binary (right). 
           The results of the RWF are indicated with the light-thin line and the results for the FWF with the dark-bold line. For this selected combination
           of theory parameters, there is modest breaking for high masses.\label{fig:histogram_psi1_psi2}}
\end{figure}

\begin{figure}[!htp]
 \includegraphics[width=\columnwidth]{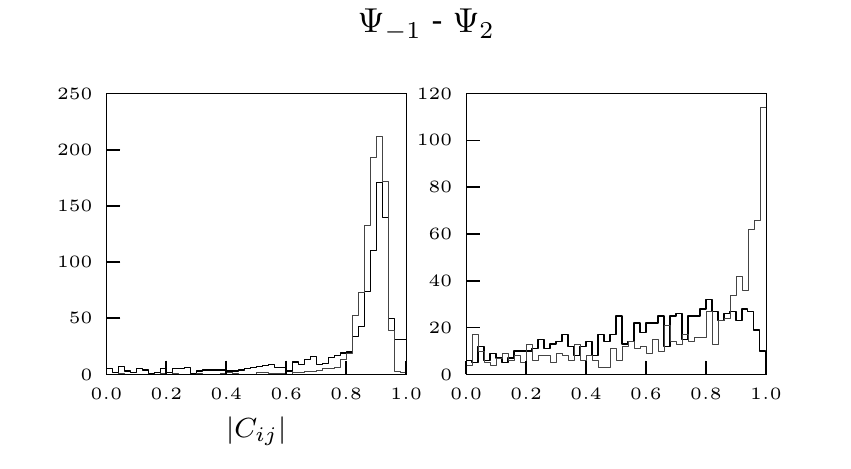}
  \caption{Correlation breaking for a low-mass $m_1 = 10^6 \Msun$, $m_2 = 3\times10^5 \Msun$ binary (left) and a $m_1 = 3\times10^7 \Msun$, $m_2 = 10^7 \Msun$ binary (right). 
           The results of the RWF are indicated with the light-thin line and the results for the FWF with the dark-bold line. For this selected combination
           of theory parameters, there is stronger correlation breaking for high masses. \label{fig:histogram_psi0_psi2}}
\end{figure}

Within the set $\Psi_{\text{low}}$, the FWF is not very effective in breaking the correlations that are present using the RWF model, in some cases it
even introduces further correlation. Among theory parameters from the set $\Psi_{\text{high}}$, there is a modest correlation breaking for high total masses 
while for low masses the FWF model stretches out the nearly symmetric RWF correlation distributions by providing them with a long tail on the left-hand side 
and slightly shifting the peak to the right-hand side (fig. \ref{fig:histogram_psi1_psi2}). 
For correlations between two parameters coming out from different sets, there is the same stretching effect and modest correlation breaking for high-masses as
for parameters in $\Psi_{\text{high}}$, but only for parameters from $\Psi_{\text{low}}$ in combination with $\Psi_2$, a stronger breaking of correlations is 
achieved by the FWF (fig. \ref{fig:histogram_psi0_psi2}).

\subsubsection{Correlations between binary and theory parameters}

Although there are mass and spin-dependent terms that are proportional to the same frequency power as the alternative theory parameters, mass, spin and angular momentum parameters
show only absolute correlations of $\lesssim0.5$ with the theory parameters, because they enter non-linearly and in several different frequency powers.

\begin{figure}[!htp]
 \includegraphics[width=\columnwidth]{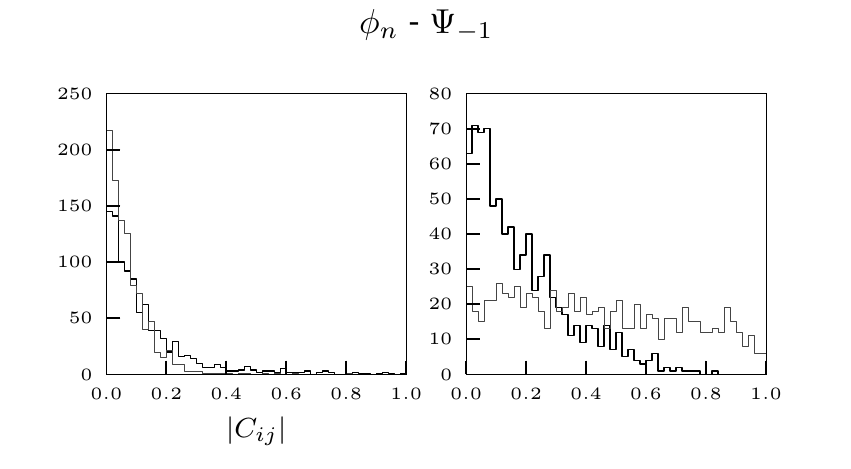}
  \caption{Correlation breaking for a low-mass $m_1 = 10^6 \Msun$, $m_2 = 3\times10^5 \Msun$ binary (left) and a $m_1 = 3\times10^7 \Msun$, $m_2 = 10^7 \Msun$ binary (right). 
           The results of the RWF are indicated with the light-thin line and the results for the FWF with the dark-bold line. When accounting for higher
           harmonics, correlations of $\phi_n$ with alternative theory parameters are strongly broken for high masses. \label{fig:histogram_psim1_phn}}
\end{figure}

\begin{figure}[!htp]
 \includegraphics[width=\columnwidth]{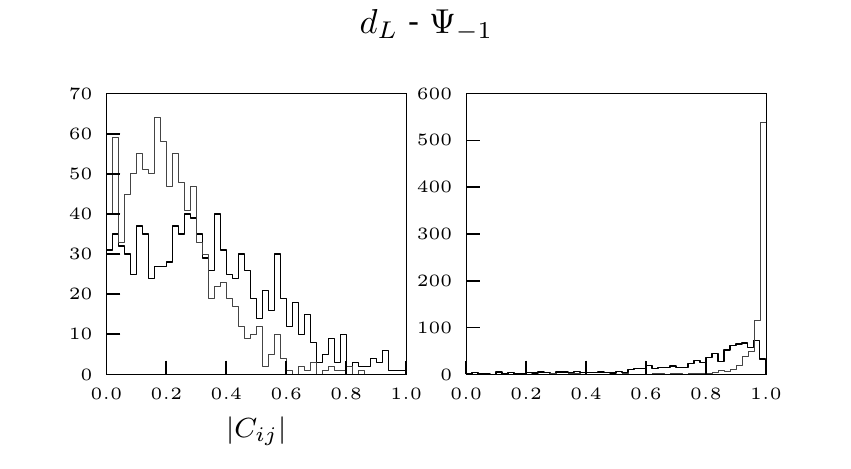}
  \caption{Correlation breaking for a low-mass $m_1 = 10^6 \Msun$, $m_2 = 3\times10^5 \Msun$ binary (left) and a $m_1 = 3\times10^7 \Msun$, $m_2 = 10^7 \Msun$ binary (right). 
           The results of the RWF are indicated with the light-thin line and the results for the FWF with the dark-bold line. For low masses, the correlation with
           the luminosity distance parameter increases while there is modest breaking for high masses when introducing higher harmonics. \label{fig:histogram_psim1_dl}}
\end{figure}

The phase and time at coalescence $\phi_c$ and $t_c$ are formally equivalent to $\Psi_{2.5}$ and $\Psi_4$, respectively, and are therefore 
highly correlated with the theory parameters.
Especially for tightly correlated parameters, correlations can be
broken easily through the introduction of extra structure with higher harmonics.
Also the correlations with the 
sky position parameters $\mu_n$ and $\phi_n$ can be strongly broken for high masses (fig. \ref{fig:histogram_psim1_phn}) when using higher harmonics.
Interestingly, correlations with the luminosity distance parameter $d_L$ increase for low masses (extra structure can in principle also
introduce additional correlations), while there is a modest breaking for high masses (fig. \ref{fig:histogram_psim1_dl}).

\subsection{\label{sec:results_redshifts}Upper limits for redshifted masses}

\begin{figure}[!hbp]
 \includegraphics[width=\columnwidth]{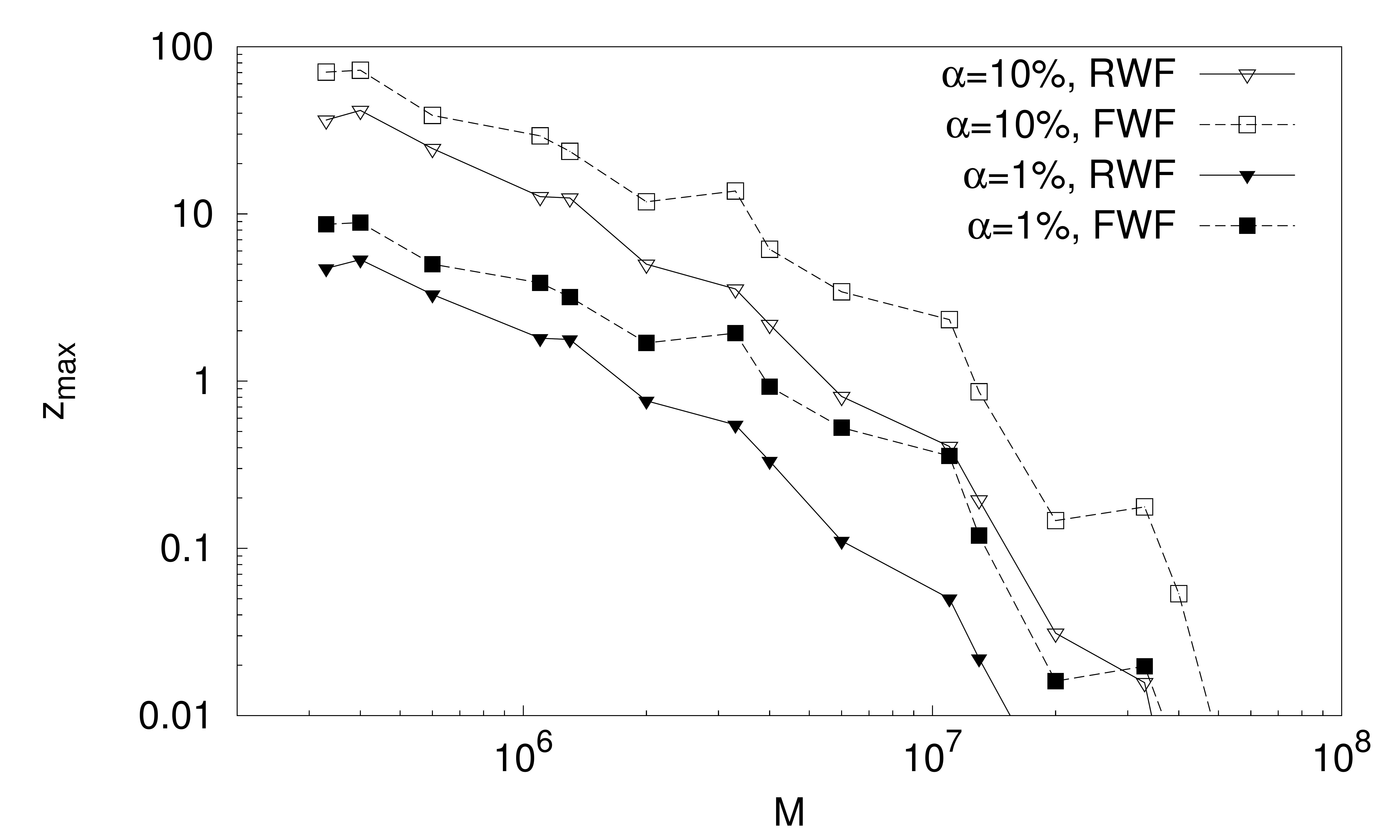}
  \caption{Maximal redshifts for the alternative theory parameter $\Psi_0$ such that the relative error $\Delta \Psi_0 / \psi_0$ 
is smaller than $\alpha$. $\psi_0$ is the corresponding fiducial 2PN phase coefficient. 
For a relative error of $1\%$, low-mass binaries are suitable up to
redshifts $z\sim 1-10$ while high-mass binaries can be observed up to $z\sim0.01-0.1$. \label{fig:zmaxPsi0}}
\end{figure}

\begin{figure}[!hbp]
 \includegraphics[width=\columnwidth]{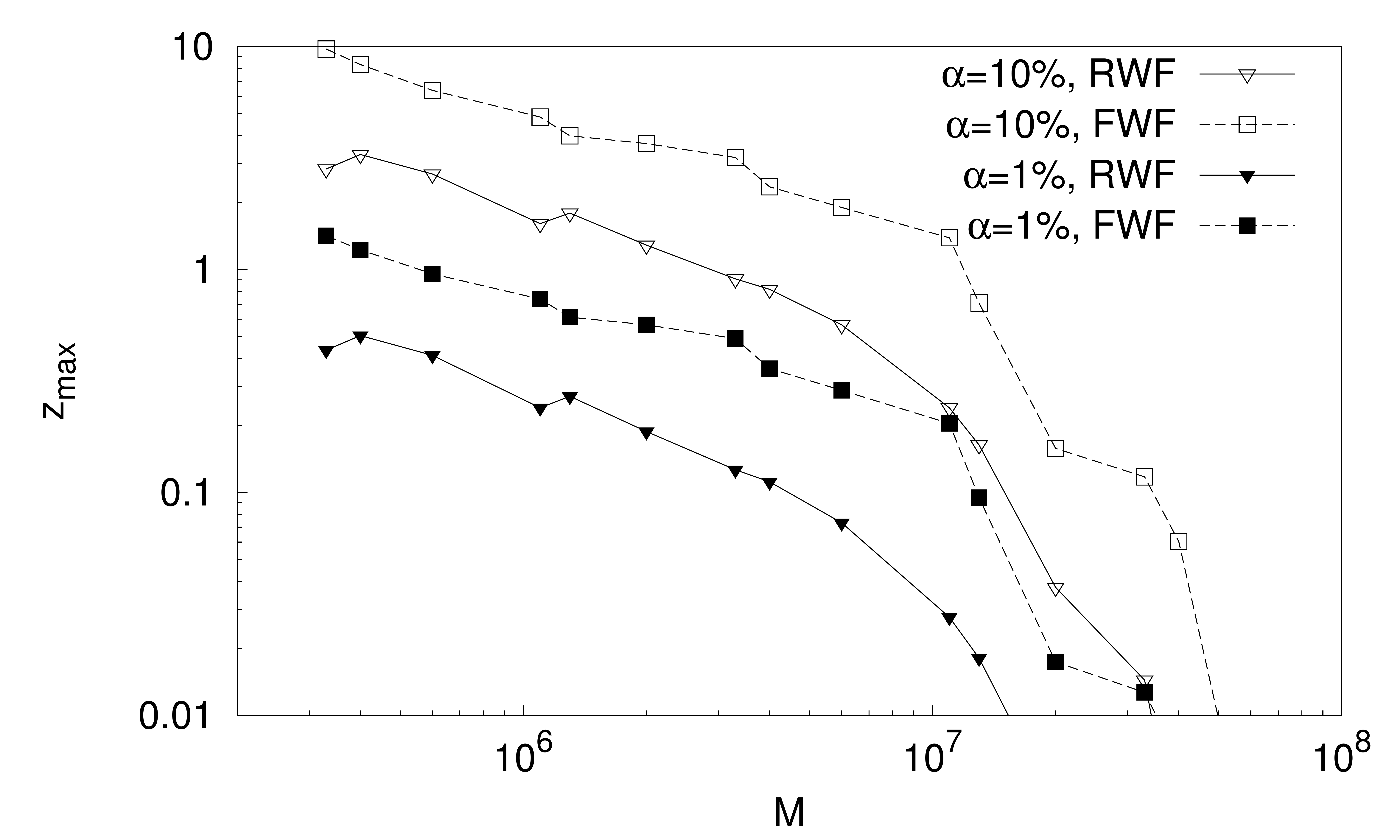}
  \caption{Maximal redshifts for the alternative theory parameter $\Psi_1$ such that the relative error $\Delta \Psi_1 / \psi_1$ 
is smaller than $\alpha$. $\psi_1$ is the corresponding fiducial 2PN phase coefficient. For a relative error of $1\%$, low-mass binaries are suitable up to
redshifts $z\sim 0.1-1$ while high-mass binaries can be observed up to $z\sim0.01-0.1$.\label{fig:zmaxPsi1}}
\end{figure}

\begin{figure}[!hbp]
 \includegraphics[width=\columnwidth]{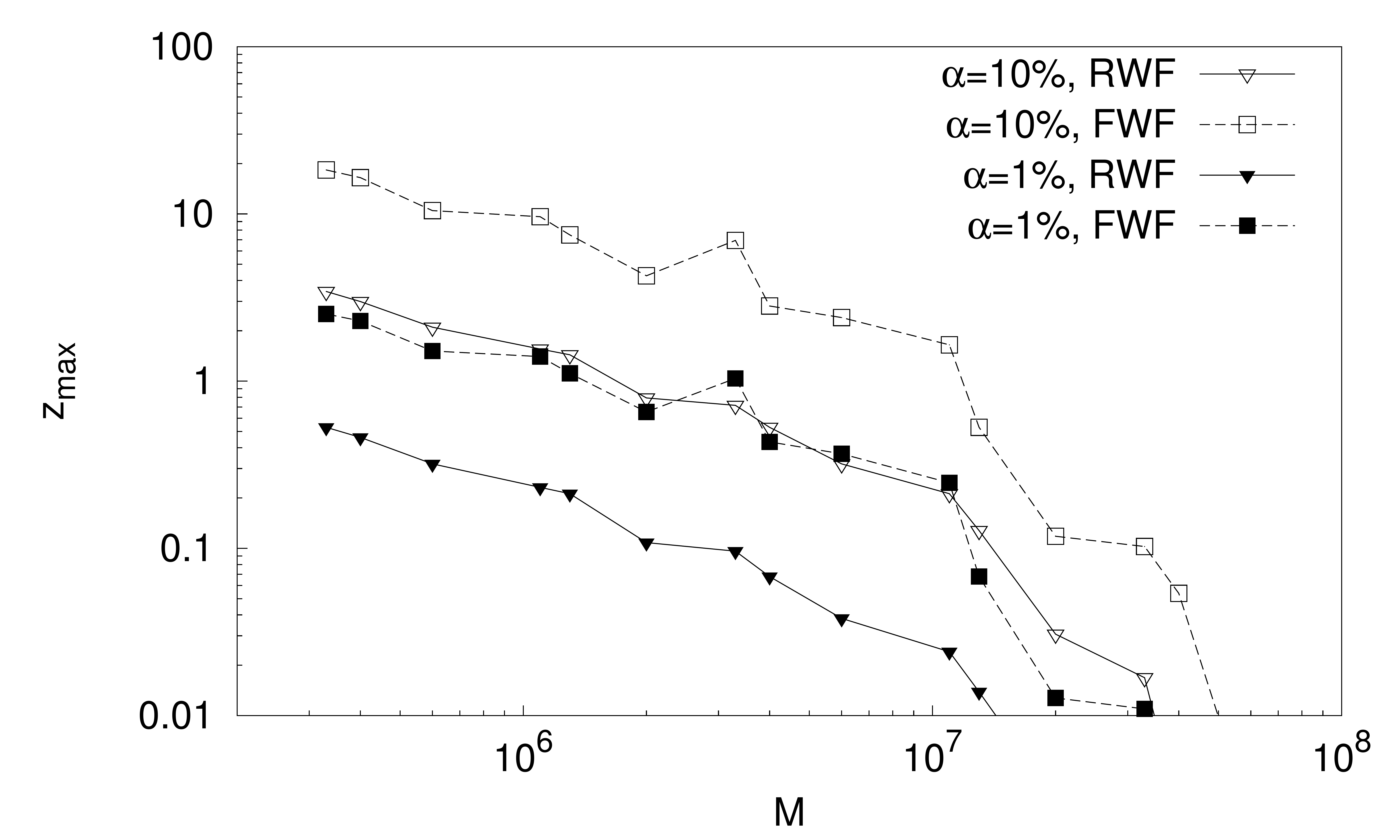}
  \caption{Maximal redshifts for the alternative theory parameter $\Psi_{3/2}$ such that the relative error $\Delta \Psi_{3/2} / \psi_{3/2}$ 
is smaller than $\alpha$. $\psi_1$ is the corresponding fiducial 2PN phase coefficient. For a relative error of $1\%$, low-mass binaries are suitable up to
redshifts $z\sim 0.1-1$ while high-mass binaries can be observed up to $z\sim0.01-0.1$.\label{fig:zmaxPsi32}}
\end{figure}

\begin{figure}[!hbp]
 \includegraphics[width=\columnwidth]{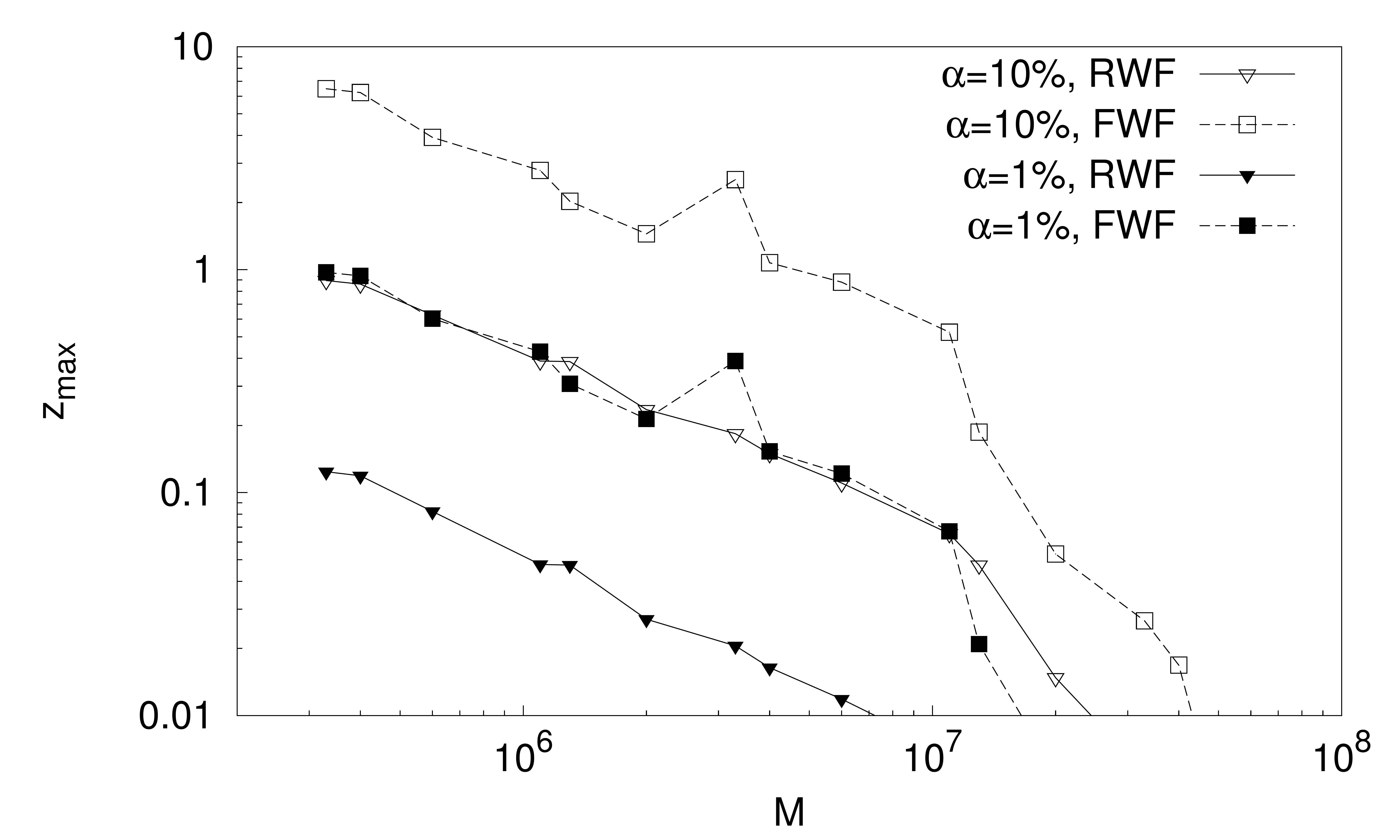}
  \caption{Maximal redshifts for the alternative theory parameter $\Psi_2$ such that the relative error $\Delta \Psi_2 / \psi_2$ 
is smaller than $\alpha$. $\psi_2$ is the corresponding fiducial 2PN phase coefficient. For a relative error of $1\%$, low-mass binaries are suitable up to
redshifts $z\sim 0.1-1$ while high-mass binaries can be observed up to $z\sim0.01-0.1$.\label{fig:zmaxPsi2}}
\end{figure}

All the errors tabularized in appendix \ref{sec:tables} are given for the fixed redshift $z=1$. Some of them in the high-mass regime are apparently too high 
at $z=1$. Nevertheless, since the measurement accuracy of the parameters is correlated with the redshift as given in Eq. \eqref{errorredshift},
for an equivalent mass configuration at a lower redshift the errors should reduce to reasonable values. 
Since the actual values of the alternative theory parameters are not known, we cannot fix the 
accuracy with which we want to measure $\Psi_i(z)$. For this reason, we introduce the relative accuracy parameter $\alpha$ such that $\Delta \Psi_i(z) / \psi_i < \alpha$
where $\psi_i$ is the fiducial 2PN phase coefficient from $\Psi_{\text{2PN}}$ in Eq. \eqref{phasecorrections}. The maximal redshift is then given as

\begin{equation}
 z_{\text{max}} = z\left( \alpha \, d_L(z_0)  \, \left|\Delta \Psi_i(z_0) / \psi_i\right|^{-1} \right),
\end{equation}
where $z(d_L)$ is the inverse of \eqref{dL} and can be computed numerically. 
We use here the $5\%$-quantile for $\Delta \Psi_i(z_0 = 1)$ as given in the tables in appendix \ref{sec:tables}, i.e. we define the (optimistic) maximal redshift as the redshift where
$5\%$ of the binaries in the sample can still be seen with relative accuracy less than $\alpha$. Since we expect corrections to the 2PN
phase parameters of GR to be small (at least for the lower PN orders), we focus here on a relative accuracy below $10\%$. At redshift $z=1$ this accuracy is already
difficult to reach for binaries with masses above $10^6 \Msun$ (see also \cite{arunetal2006}). It is important to emphasize that we concentrate here on actually
measuring the alternative theory parameters instead of just setting bounds upon them. In figures \ref{fig:zmaxPsi0}-\ref{fig:zmaxPsi2} we present the maximal 
redshifts at which LISA can still measure the alternative theory parameters $\Psi_0$, $\Psi_1$, $\Psi_{3/2}$ and $\Psi_2$ for certain mass configurations with 
relative accuracies of $\alpha = 10\%$ and $\alpha = 1\%$. Since for $\Psi_{\text{-1}}$ and $\Psi_{1/2}$ the fiducial 2PN phase coefficients are zero, we do not consider them. 
We checked that the error roughly scales with the redshift. For a relative accuracy of $1\%$, $\Psi_0$ is measurable up to redshifts of $z\sim1-10$
for low-mass binaries and up to redshifts of $z\sim0.01-0.1$ for high-mass binaries. $\Psi_1$, $\Psi_{3/2}$ and $\Psi_2$ can all be detected with a relative accuracy of
$1\%$ up to redshifts of $z\sim0.1-1$ for low masses and $z\sim0.01-0.1$ for high masses. For $\Psi_0$, the use of the FWF improves the maximal redshifts by about a 
factor of 2 for low masses and up to a factor of $10$ for high masses, while the maximal redshifts are improved by almost an order of magnitude for the rest of the 
alternative theory parameters. If we were lucky and LISA could find a low-mass black hole binary at very low redshift $z=0.1$, we would be able to recover the alternative theory parameters
with $\sim10$ times smaller errors than given in tables \ref{table:Psim1_21}-\ref{table:Psi2_21}.

\subsection{\label{sec:massivegraviton}Example: Lower bound on graviton Compton wavelength}

In order to compare our results with previous work in the field, we present here a lower bound 
on a possible graviton Compton wavelength from our results at redshift $z=1$. The term 'massive graviton' is commonly used to state that the speed of
gravitational waves depends on frequency rather than being constant. According to \cite{will1998}, the effect of a 'massive graviton' can 
be accounted for by introducing a gravitational wave phase correction

\begin{equation}
 \Delta \Psi_{\text{MG}}(z) = -\beta(z) \, \nu^{-3/5} \, x^{-3/2},
\end{equation}
where $x$ is the dimensionless frequency, $\nu$ is the symmetric mass ratio and the parameter $\beta(z)$ is defined as

\begin{equation}
 \beta(z) = \frac{G}{c^2} \frac{\pi^2 D(z) \mathcal{M}}{\lambda_g^2 (1+z)}.
\end{equation}
Here $\lambda_g$ is the Compton wavelength of the graviton, $z$ is the redshift, $\mathcal{M} = (1+z) M \nu^{3/5}$ is the \emph{measured} chirp mass
affected by redshift, and $D(z)$ is the distance given as

\begin{equation}
 D(z) = (1+z) \frac{c}{H_0} \int_0^z \frac{dz'}{(1+z')^2 \sqrt{\Omega_M (1+z')^3 + \Omega_\Lambda}},
\end{equation}
where $H_0$, $\Omega_M$ and $\Omega_\Lambda$ are defined as in section \ref{sec:evolution}. In our implementation, this is similar to the correction in Eq. \eqref{phasecorrections}:

\begin{equation}
 \Delta \Psi_{\text{MG}}(z) = \frac{3}{256 \nu} x^{-3/2} \Psi_1(z).
\end{equation}
Hence the errors on $\beta$ and $\Psi_1$ can be related with

\begin{equation}
 \Delta \beta(z) = \frac{3}{256} \nu^{-2/5} \Delta \Psi_1(z).
\end{equation}

We take the fiducial value $\beta = 0$, thus the error $\Delta \beta$ sets an upper bound on possible values for $\beta$. A lower bound on the Compton wavelength
of the graviton can then be calculated at redshift $z$ as

\begin{equation}
 \lambda_g(z) > \sqrt{\frac{256}{3} \frac{G}{c^2} \frac{\pi^2 D(z) M \nu}{(1+z) \Delta \Psi_1(z)}},
\end{equation}
where $M$ is the redshifted total mass of the binary. At redshift $z=1$ we find that optimal lower bounds on $\lambda_g$ originate from a $(3 \times 10^6 + 1 \times 10^7) \Msun$
binary for the FWF and from a $(1 \times 10^6 + 1 \times 10^6) \Msun$ binary for the RWF. Including all six alternative theory parameters $\Psi_i$, the resulting
average bounds are $\lambda_g > 1.2 \times 10^{21}$ cm (FWF) and $\lambda_g > 7.8 \times 10^{20}$ cm (RWF). These bounds are both lower than the one Yagi and Tanaka 
\cite{yagitanaka2010} found ($\lambda_g > 4.9 \times 10^{21}$ cm) using the RWF and simple precession at a distance of 3 Gpc; this is because the presence of the other five alternative 
theory parameters increases correlations among the parameters. If we consider only one correction parameter $\Psi_1$ which among other things accounts for massive gravity,
the bounds increase to $\lambda_g > 7.6 \times 10^{21}$ cm (FWF) and $\lambda_g > 4.9 \times 10^{21}$ cm (RWF). The RWF bound is slightly higher than the one by Yagi and 
Tanaka for a $(10^6+10^7) \Msun$ binary; for this mass configuration we found a lower RWF bound of $\lambda_g > 2.8 \times 10^{21}$ cm. Cornish et al. 
\cite{cornishsampson2011} found a similar optimal RWF bound of $\lambda_g > 3.8 \times 10^{21}$ cm. 
The use of the FWF improves the bound on the graviton Compton wavelength by a factor of $\sim1.6$ with respect to the RWF, regardless whether only one or all the alternative
theory parameters are included into the simulations. Approximately this factor of accuracy will be lost when going from classic LISA to eLISA/NGO \cite{bertietal2011}.

\section{\label{sec:conclusion}Conclusion and Outlook}

We analyzed the expected measurement error distributions of 17 different mass configurations of supermassive black hole binaries with masses between $10^5-10^8\Msun$.
We found that the black hole binaries can roughly be divided into two groups: low-mass binaries with $M \lesssim 10^7 \Msun$ and high-mass binaries with $M \gtrsim 10^7 \Msun$.
Comparing the results of the simulations using the FWF and the RWF, we found that the RWF errors on the alternative theory parameters $\Psi_i$ are a factor of $\sim100$
times higher than the FWF errors for high-mass binaries, while they are almost comparable for low-mass binaries. Due to the dilution of the available information
through the introduction of six extra parameters, the original parameters lose accuracy. For masses and spins this is only a factor of 1.2-5 for both low- and high-mass binaries
regardless of whether the FWF or RWF is used. The loss of accuracy on the position of the black hole binary on the sky is at maximum $10\%$ for low-mass binaries and up to
a factor of $5$ for high-mass binaries. However, the accuracy of the luminosity distance is affected more seriously for high-mass binaries, using the RWF results in a loss
of a factor of $\sim50-1000$ while using the FWF reduces it to factors of $\sim10-100$. For low-mass binaries it is only about a factor of 2 worse. 
The use of the FWF is therefore mandatory for high-mass binaries, while the parameter estimation is more efficient for low-mass binaries and only up to a factor of $5$ 
times worse when the RWF is used instead of the FWF.

Since the error distributions were all calculated at fixed redshift $z=1$ but the errors increase with redshift, we computed typical maximal redshifts
up to which the alternative theory parameters are detectable with a relative accuracy smaller than $1\%$ for the best $5\%$ of the binaries in the sample. 
We found that for a deviation of $1\%$ from the fiducial value, $\Psi_0$ is detectable up to redshifts of $z \sim 1-10$ for low total masses and up to $z \sim 0.01-0.1$ for 
high total masses. The rest of the alternative theory parameters $\Psi_1$, $\Psi_{3/2}$ and $\Psi_2$ with a fiducial 2PN phase coefficient unequal zero are detectable 
up to redshifts of $z \sim 0.1-1$ for low-mass binaries and $z \sim 0.01-0.1$ for high-mass binaries with the same relative accuracy. The use of the FWF improves the maximal redshifts up to a factor 
of $10$ for high total masses.

The FWF enables us to increase the optimal lower bound on the Compton wavelength of the graviton by about a factor of $1.6$ compared to the one reached by the RWF. 
We achieve an optimal lower bound of $\lambda_g > 7.6 \times 10^{21}$ cm for the classic LISA detector design if only the alternative theory parameter $\Psi_1$ is considered.

Since the proposed eLISA/NGO mission will most certainly fly as a reduced variant of classic LISA, it is important to investigate
the reassessment of certain aspects of the mission. A broad range of LISA variants are currently reviewed by the community.
To account for the technical 'shortcomings' it is thus of great importance to use as accurate waveform templates as possible to restore the lost accuracy with computational power on Earth. The use of the FWF improves the accuracy of the alternative theory 
parameters by at least an order of magnitude compared to the RWF. As shown by \cite{arunwill2009}, the use of
hybrid inspiral-merger-ringdown templates instead of inspiral-only templates improves the accuracy by an order of magnitude for the RWF; it would be interesting
to find out how much such templates are improved when the FWF is used. The accuracy can further be enhanced by about an order of magnitude when considering
combined observations instead of just extracting alternative theory parameters from individual black hole binaries \cite{bertietal2011}. Also effects of eccentric orbits should be accounted for to make the model more realistic.

Future work could include the introduction of amplitude corrections such as in \cite{yunespretorius2009}, since certain alternative theories have dominant contributions
in the gravitational wave amplitude (e.g. Chern-Simons-modified gravity \cite{alexanderyunes2009}). Also, the underlying mechanism of spin precession should be analyzed
for effects originating from possible alternative theories. In this paper we neglected the energy loss of black hole binaries through unexpected physical effects such as 
further degrees of freedom in the propagation of gravitational waves arising from additional polarizations (e.g. longitudinal modes). 
It would be interesting to introduce a parametrized model for these effects \cite{chatziioannou2012} into our simulations.
Also, since we studied a search for modifications at different PN orders at the same time, one could use the results of this work to investigate how the use
of next-to-leading order modifications of GR could affect the determination of alternative theory parameters. The impact of turning off and on correction 
parameters also needs further studies (following e.g. \cite{lipozzo2012}).

\begin{acknowledgments}
C.~H. would like to thank Sylvain Marsat for interesting discussions about the post-Newtonian expansion and Ed Porter for his suggestions for improvements
on our code. The authors appreciate the valuable comments by the anonymous referee. C.~H. and A.~K. are supported by the Swiss National Science Foundation. 
\end{acknowledgments}

% Use appendix* for single appendix
\appendix

\section{\label{sec:breakdown}Breakdown conditions}

Since in previous work different viewpoints are taken on the choice of a critical orbit at which the integrations need to be stopped for binary black holes with
precessing spins, we give here a quick summary of the approximations we used for the gravitational wave signal generation and indicate at which point we consider them to have failed.
The three major assumptions are that orbits can be considered to be quasi-circular (adiabatic approximation), the spins can be treated as
constants over one orbit (orbit-averaged spin precession) and the weak field or post-Newtonian approximation,
which assumes typical velocities to be smaller than the speed of light, which enables us to perform a PN expansion in terms of powers of $v/c$. 
We shall discuss below how to estimate at which point the breakdown of these assumptions occurs; in
particular, the breakdown of the PN approximation can be estimated
using different methods, among which the use of the minimum energy circular
orbit (MECO) or the PN energy flux is common.

We decided to stop our
integrations always at the ISCO of $6 \, GM/c^2$, since orbit-averaged spin precession can already start
to be inaccurate at this point and the authors do not trust the PN expansion below this limit. Also we did not find any binary system with a 
minimum energy circular orbit, flux or adiabatic breakdown higher than this radius. In the following subsection we list four different approximations
criteria and discuss the limits of their validity.

\subsection{Adiabatic approximation}

The adiabatic approximation assumes that the time needed for one orbit is much smaller than the timescale for orbit shrinkage. In other words, 
the orbit shrinkage velocity $\dot{r} = \frac{dr}{dt}$ is required to be much smaller than the orbital velocity $\omega r$, then the orbits can be considered to
be quasi-circular. The orbital separation is given (expanded in terms of the dimensionless frequency $x$ up to 2PN order) by

\begin{eqnarray}
 r(x) & = & \frac{GM}{c^2 x} \left[ 1 + \frac{1}{3} (-3+\nu) x - \frac{1}{3} \beta(2,3) x^{3/2} \right. \nonumber \\
      & + & \left. \left(\frac{19}{4}\nu + \frac{1}{9} \nu^2 - \frac{1}{2} \sigma(1,3)\right) x^2 \right], \nonumber \\
\end{eqnarray}
where $\beta$ and $\sigma$ (expected to vary only slowly on one orbit) have been treated as constants. As an indicator for the faithfulness of the adiabatic approximation, 
we choose the expression

\begin{equation}
 \frac{|\dot{r}|}{\omega r} < \kappa_{\text{adiab}}.
\end{equation}

\begin{figure}[!htp]
 \includegraphics[width=\columnwidth]{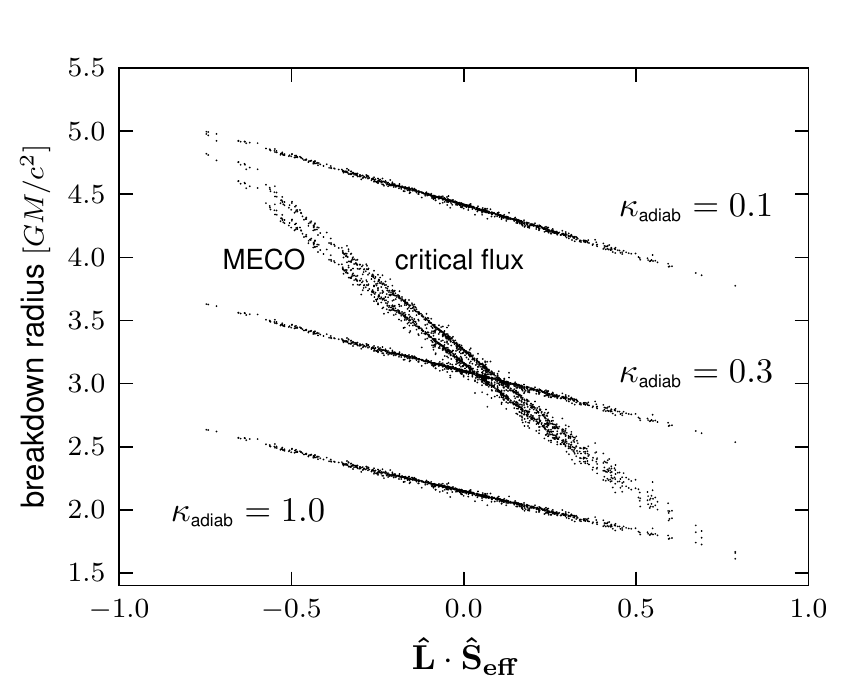}
  \caption{Plot of MECO radius, critical flux radius and adiabatic breakdown radii (for different limits $\kappa_{\text{\tiny adiab}}$) against the initial effective spin orientation for 1000 simulated systems with two equal mass $2 \times 10^6 \Msun$ black holes (binary of two $10^6 \Msun$ black holes seen at redshift $z=1$).\label{fig:mecoradius_equal}}
\end{figure}

\begin{figure}[!htp]
 \includegraphics[width=\columnwidth]{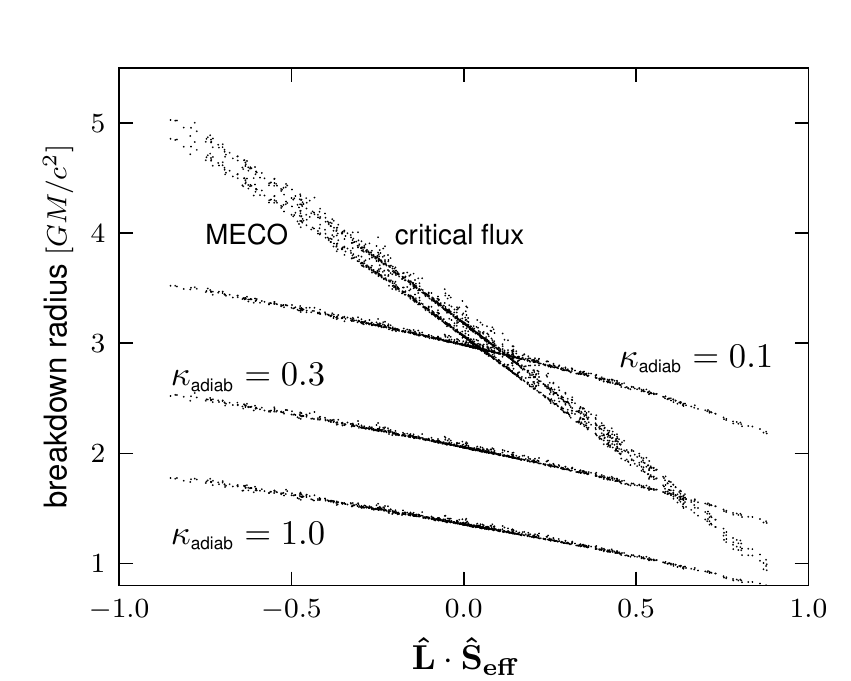}
  \caption{Plot of MECO radius, critical flux radius and adiabatic breakdown radii (for different limits $\kappa_{\text{\tiny adiab}}$) against the initial effective spin orientation for 1000 simulated systems with black hole binaries of mass ratio 1:10 ($m_1 = 2 \times 10^7 \Msun$, $m_2 = 2 \times 10^6 \Msun$).\label{fig:mecoradius_1_10}}
\end{figure}

The quantities $\omega r$ and $\dot{r} = \frac{dr}{dx} \frac{dx}{dt}$ can be computed to stop the integration when a certain adiabatic breakdown 
limit $\kappa_{\text{adiab}}$ of our choice is reached. The breakdown radius for constant $\kappa_{\text{\tiny adiab}}$ shows almost linear dependency on
the initial value of $\vec{\hat{L}} \cdot \vec{S}_{\text{eff}}$ (when the binary enters the LISA band). In figs. \ref{fig:mecoradius_equal} and \ref{fig:mecoradius_1_10}, the adiabatic breakdown limits for 
$\kappa_{\text{\tiny adiab}} = 0.1,0.3$ and $1.0$ are plotted for $10^3$ randomly distributed systems in the parameter space with equal masses and a mass ratio of 1:10
respectively. The figures indicate that the adiabatic approximation is still quite reasonable ($\kappa_{\text{\tiny adiab}} < 0.1$) for orbital separations larger than
$r = 5 \, GM/c^2$, so we do not have to consider it since we already stop before this limit.

\subsection{MECO}

The last stable circular orbit (ISCO) for test masses orbiting a non-spinning, Schwarzschild black hole takes place at the minimum of the effective gravitational potential 
$\frac{dV_{\text{eff}}}{dr} = 0$, corresponding to an orbital separation of $6 \text{ } GM/c^2$. This is of course different for black hole binaries with
comparable masses and non-zero spins; there, the total energy is only known in terms of a PN expansion \cite{buonannochenvallisneri2003, kidderwillwiseman1993, kidder1995}

\begin{eqnarray}
 E & = & - \frac{\mu c^2}{2} x \left( 1 - \frac{1}{12} (9+\nu) x + \frac{c}{G} \frac{4}{3M^2} \,\vec{\hat{L}} \cdot \vec{S}_{\text{eff}} \; x^{3/2} \right. \nonumber \\
    & + & \left. \left[ \frac{1}{24} (-81+57\nu-\nu^2) + \frac{c^2}{G^2} \frac{1}{\nu M^4} (\vec{S}_1 \cdot \vec{S}_2 \right. \right. \nonumber \\
    &  & - \left. \left. 3(\hat{\vec{L}}\cdot\vec{S}_1)(\hat{\vec{L}}\cdot\vec{S}_2)) \right] x^2 \right), \nonumber \\
\end{eqnarray}
including leading order spin-spin and spin-orbit couplings. The effective spin $\vec{S}_{\text{eff}}$ is defined as the combination

\begin{equation}
\label{effectiveSpin}
 \vec{S}_{\text{eff}} = \left( 2 + \frac{3 m_2}{2 m_1} \right) \vec{S}_1 +  \left( 2 + \frac{3 m_1}{2 m_2} \right) \vec{S}_2.
\end{equation}
The last stable circular orbit is then thought to take place at the point where

\begin{equation*}
 \frac{dE}{dx} = 0,
\end{equation*}
the \textit{minimum energy circular orbit} (MECO). Afterwards the binaries are thought to plunge and quasi-circular orbit approximations will certainly fail. In figures 
\ref{fig:mecoradius_equal} and \ref{fig:mecoradius_1_10}, the MECO radii for $10^3$ randomly distributed systems in the parameter space are plotted
for mass ratios of 1:1 and 1:10 respectively. The MECO radius is always below the radius where the gravitational wave energy flux reaches a critical limit (defined
in the next subsection), so we do not consider the MECO as a breakdown criterion for our simulations but instead use the flux condition worked out in the next
subsection.

\subsection{Flux}

The energy flux of a gravitational wave can be expressed as \cite{blanchet2006}

\begin{eqnarray}
\label{Flux}
\mathcal{L} & = & -\frac{dE}{dt} = -\frac{dx}{dt} \frac{dE}{dx} \nonumber \\
  & =  &\frac{32 c^5}{5G} \nu^2 x^5 \left[ 1 - \left( \frac{1247}{336} + \frac{35}{12} \nu \right) x + \alpha_{3/2} x^{3/2} + \alpha_{2} x^2 \right], \nonumber \\
\end{eqnarray}
where $\frac{dx}{dt}$ and $E$ are the 2PN expressions used in this paper. For the expressions $\alpha_{3/2}$ and $\alpha_2$ containing spin-orbit and spin-spin
couplings, the reader is referred to \cite{blanchet2006}. As long as $x$ is small, this flux will stay close to its leading order contribution. As soon as $x$ 
gets close to $1$, the 1PN term will grow stronger, decrease the flux and eventually make it negative \cite{buonannochenvallisneri2003_2}. One can thus infer that the PN series tends to breakdown if $\mathcal{L}$ deviates
significantly from its leading order contribution and has for sure broken down if the flux is negative. 

We decided to stop the integrations if the flux is smaller than 10\% of its leading order contribution (with spin-angular momentum and spin-spin terms included).
The plots in figures \ref{fig:mecoradius_equal} and \ref{fig:mecoradius_1_10} show that the critical flux is never reached above $r=5 \, GM/c^2$, which means
that there are no black hole binaries with a MECO higher than $r=6 \, GM/c^2$ in our mass range which could potentially lead to unphysical results. Nevertheless, we use a catch in our
code to stop the integration if the flux gets by an unforeseen chance smaller than 10\% of its leading order contribution. Especially for parallel spins, 
one could theoretically try to go even down to $2-4 \, GM/c^2$. In these regions a lot more SNR could be accumulated, resulting in a $\sim10$ times higher overall SNR 
and sometimes several orders of magnitude smaller errors. This is very dangerous, since we do not expect post-Newtonian theory to be physically 
accurate enough in these regions and one should be suspicious of such small errors.

\subsection{Orbit-averaged spin precession}

Since we use orbit-averaged spin precession equations \cite{apostolatos1994}, we need to assure that the underlying assumption of the timescale for 
precession always being smaller than the orbital time still holds. Like other recent studies (see e.g. \cite{langhughescornish2011}), 
we do not consider the breakdown of this approximation in our integrations, since both timescales are comparable only around $2-3 \; GM/c^2$.
We are however not sure, how strongly errors in the spin precession affect the matched filtering process. Since large spin precession occurs 
only in the late inspiral (where the largest part of the SNR is accumulated), an improper treatment of orbit-averaged spin precession creates a
theoretical error in the waveform template and thus could result in a significant loss of SNR, despite the fact that the Fisher matrix gave
an optimistic error estimate. We plan to quantify this theoretical error in a future publication.

In this subsection, we present the breakdown radii corresponding to certain limits on the angular momentum precession timescale, i.e. the critical orbits
where the integration should be stopped. 

The timescale for one full orbit is

\begin{equation}
\label{Torb}
 T_{\text{orb}} = 2\pi \sqrt{ \frac{r^3}{G M} }.
\end{equation}
Ignoring spin-spin terms, the precession of the angular momentum unit vector can then be written as (see e.g. \cite{cornishkey2010})

\begin{equation}
 \dot{\vec{\hat{L}}} = \frac{G}{c^2 r^3} \vec{S}_{\text{eff}} \times \vec{\hat{L}},
\end{equation}
with the effective spin vector $\vec{S}_{\text{eff}}$ defined in \eqref{effectiveSpin}. Thus $\vec{\hat{L}}$ precesses with an angular frequency of approximately $\omega_{\text{prec}} = \frac{G}{c^2 r^3} |\vec{S}_{\text{eff}}|$ which corresponds to a time of

\begin{equation}
 T_{\text{prec}} = 2 \pi \frac{c^2 r^3}{G  |\vec{S}_{\text{eff}}|}
\end{equation}
for one precession. A good indicator for the breakdown of orbit-averaged spin precession is thus the fraction

\begin{equation}
\label{TorbTprec}
\frac{T_{\text{orb}}}{T_{\text{prec}}} < \kappa_{\text{prec}},
\end{equation}
where $\kappa_{\text{prec}}$ is the critical limit of our choice.
In the case where the two timescales are equal ($\kappa_{\text{prec}} = 1$), this corresponds to a full precession in one orbit. At this point one certainly cannot speak
of 'orbit-averaged' spin precession anymore. 

\begin{figure}[!hbt]
 \includegraphics[width=\columnwidth]{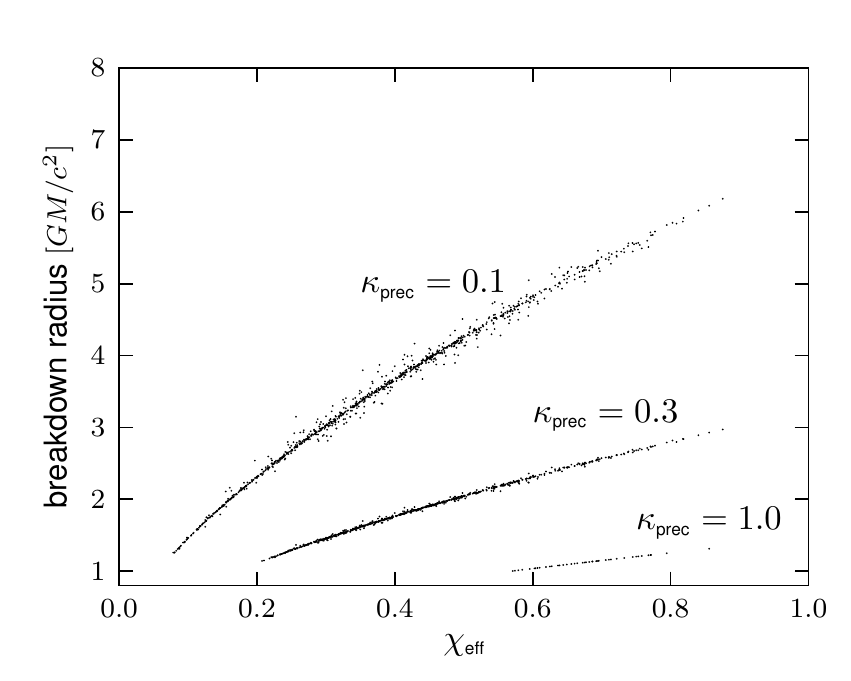}
  \caption{Plot of orbit-averaged precession breakdown radii (for different limits $\kappa_{\text{\tiny prec}}$) against the initial effective spin strength for 1000 simulated systems with two equal mass $2 \times 10^6 \Msun$ black holes.\label{fig:precession_equal}}
\end{figure}

\begin{figure}[!hbt]
 \includegraphics[width=\columnwidth]{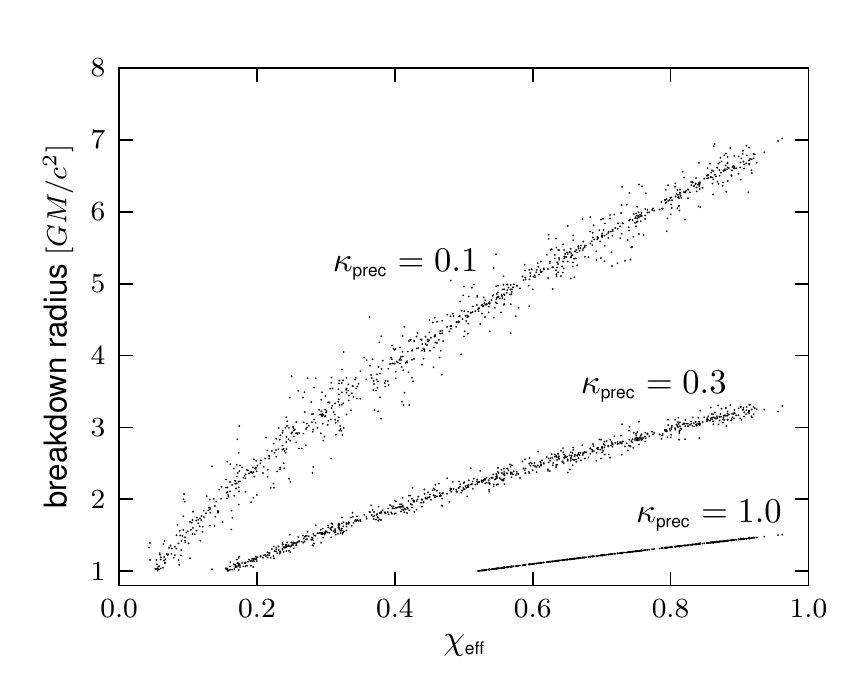}
  \caption{Plot of orbit-averaged precession breakdown radii (for different limits $\kappa_{\text{\tiny prec}}$) against the initial effective spin strength for 1000 simulated systems with black hole binaries of mass ratio 1:10 ($m_1 = 2 \times 10^7 \Msun$, $m_2 = 2 \times 10^6 \Msun$).\label{fig:precession_1_10}}
\end{figure}

The maximum absolute value which the effective spin is able to reach can be found to be $|\vec{S}_{\text{eff}}| = \frac{GM^2}{c} (2-\nu)$, for two aligned, maximally spinning 
black holes. Hence we can write the effective spin introducing a dimensionless strength $0 \leq \chi_{\text{eff}} < 1$ as

\begin{equation}
 |\vec{S}_{\text{eff}}| = \chi_{\text{eff}} \frac{GM^2}{c} (2-\nu).
\end{equation}
From eqs. \eqref{Torb} - \eqref{TorbTprec} we can then infer the critical radius where the orbit-averaged precession equations break down (slightly perturbed by fluctuations coming from neglected spin-spin terms):

\begin{equation}
 \label{radiusprec}
 r = \left(\frac{(2-\nu)\chi_{\text{eff}}}{\kappa_{\text{prec}}}\right)^{2/3} \frac{GM}{c^2}.
\end{equation}
In figures \ref{fig:precession_equal} and \ref{fig:precession_1_10}, numerical simulations (including spin-spin terms) are shown, where $10^3$ binary systems with mass ratios
1:1 and 1:10 (and uniformly distributed parameters) are used, respectively. The simulations match with the predictions by eq. \eqref{radiusprec}. For high effective spins, 
the integrations should be stopped already around $r=6 \, GM/c^2$ in the conservative limit ($\kappa_{\text{prec}} = 0.1$) and $r=2 \, GM/c^2$ in a very optimistic limit ($\kappa_{\text{prec}} = 1$). Since we stop at $r=6 \, GM/c^2$, we chose to ignore
the breakdown of orbit-averaged spin precession in the current work, but emphasize that theoretical errors arising from this assumption should be investigated in the future.

\section{\label{sec:PN}The 2.5PN and 3PN orbital frequency evolution equations}
\label{Appendix:PN:2.5and3PN}

The inclusion of dipole radiation corrections proportional to $x^{-1}$ requires the knowledge of higher PN orders to be consistent to 2PN order, namely
2.5PN and 3PN contributions. Since the current 2.5PN expansion just considers spin-orbit contributions and no spin-spin effects, and the 3PN
expansion does not account for any spin coupling effects at all, these are of course only approximations.

\subsection{2.5PN from Blanchet et al. 2006}

Blanchet et al. 2006 \cite{blanchetbuonanno2006} compute the angular frequency evolution for a binary with symmetric mass ratio $\nu$ as

\begin{eqnarray}
\label{blanchetbuonanno6.14}
 \frac{\dot{\omega}}{\omega^2} & = & \frac{96}{5} \nu x^{5/2} \left\{ 1+ x \left( -\frac{743}{336} - \frac{11}{4} \nu \right) + 4\pi x^{3/2} \right. \nonumber \\
 &  & + x^2 \left( \frac{34103}{18144} + \frac{13661}{2016} \nu + \frac{59}{18} \nu^2 \right) + \nonumber \\
 &  & \pi x^{5/2} \left( -\frac{4159}{672} - \frac{189}{8} \nu \right) + \frac{x^{3/2}}{Gm^2} \left[ -\frac{47}{3} S_l \right. \nonumber \\
 &  & \left. - \frac{25}{4} \frac{\delta m}{m} \Sigma_l \right] + \frac{x^{5/2}}{Gm^2} \left[ \left( -\frac{40127}{1008} + \frac{1465}{28} \nu \right) S_l \right. \nonumber \\
 &  & + \left. \left( - \frac{583}{42} + \frac{3049}{168} \nu \right) \frac{\delta m}{m} \Sigma_l\right] + O\left(\frac{1}{c^6}\right) \text{ , }\nonumber \\
\end{eqnarray}

\noindent where $\omega = c^3/(GM) \text{ } x^{3/2}$, $\delta m = m_1 - m_2$ is the mass difference and $m = m_1 + m_2$ is the total mass. The spin interaction terms are expressed with 

\begin{equation}
 \vec{\Sigma} = m \left[ \frac{\vec{S}_2}{m_2} - \frac{\vec{S}_1}{m_1}\right] \text{, } \hspace{1cm} S_l = \vec{S} \cdot \vec{l} \text{, } \hspace{1cm} \Sigma_l = \vec{\Sigma} \cdot \vec{l},
\end{equation}

\noindent where $\vec{S} = \vec{S}_1 + \vec{S}_2$ is the total spin and $\vec{l} = \frac{\vec{L}}{|\vec{L}|}$ is the angular momentum unit vector. 
This enables us to write Eq. (\ref{blanchetbuonanno6.14}) in the same form as Eq. \eqref{dxdt3PN}, and we recover

\small
\begin{eqnarray}
 b_{5/2} & = & \pi \left( -\frac{4159}{672} - \frac{189}{8} \nu \right) + \frac{1}{Gm^2} \left[ \left( -\frac{40127}{1008} \right. \right. \nonumber \\ 
        & & + \left. \left. \frac{1465}{28} \nu \right) S_l + \left( -\frac{583}{42} + \frac{3049}{168} \nu\right) \frac{\delta m}{m} \Sigma_l \right].
\end{eqnarray}
\normalsize

\subsection{3PN without spin terms from Blanchet et al. 2002}

In Luc Blanchet's living review \cite{blanchet2006} (see also \cite{blanchetiyerjoguet2002, blanchetdamouresposito2004, blanchetdamourespositoiyer2005}), the 3PN 
expression for the total energy of non-spinning compact binaries can be found to be

\begin{eqnarray}
 E & = & -\frac{1}{2} \mu c^2 x \left\{ 1 + \left( -\frac{3}{4} -\frac{1}{12} \nu \right) x + \left( -\frac{27}{8} + \frac{19}{8} \nu \right. \right. \nonumber \\  
    &  &   \left. -\frac{1}{24} \nu^2\right) x^2 + \left( -\frac{675}{64} + \left[ \frac{34445}{576} - \frac{205}{96} \pi^2 \right] \nu \right. \nonumber \\
    &  & \left. \left. - \frac{155}{96} \nu^2 - \frac{35}{5184} \nu^3\right) x^3 \right\} \text{ ,}\nonumber \\
\end{eqnarray}
and the energy flux is

\begin{eqnarray}
 \frac{dE}{dt} & = & -\frac{32c^5}{5G} \nu^2 x^5 \left\{ 1 + \left( -\frac{1247}{336} -\frac{35}{12} \nu \right) x + 4\pi x^{3/2} \right. \nonumber \\
  & & + \left( - \frac{44711}{9072} + \frac{9271}{504} \nu + \frac{65}{18} \nu^2\right) x^2 + \left( -\frac{8191}{672} \right. \nonumber \\
 & & - \left. \frac{535}{24} \nu\right) \pi x^{5/2} + \left( \frac{6643739519}{69854400} + \frac{16}{3} \pi^2 - \frac{1712}{105} C \right. \nonumber \\
  & & - \frac{856}{105} \log(16x) + \left[ -\frac{134543}{7776} + \frac{41}{48} \pi^2 \right] \nu - \frac{94403}{3024} \nu^2 \nonumber \\
  & & - \left. \frac{775}{324} \nu^3 \right) x^3 + \left( -\frac{16285}{504} + \frac{176419}{1512} \nu + \frac{19897}{378} \nu^2 \right) \nonumber \\
  & & \times \left. \pi x^{7/2} \right\}. \nonumber \\
\end{eqnarray}

\noindent Here $\nu$ is the symmetric mass ratio and $C = 0.577..$ is the Euler constant. The logarithm in $dE/dt$ will lead to a logarithmic term in the 3PN 
expansion. The PN coefficients $b_i$ can be recovered by computing the frequency evolution as a series in the dimensionless frequency $x$ in the adiabatic approximation:

\begin{eqnarray}
 \frac{dx}{dt} & = & \frac{dE}{dt} \text{ } \left(\frac{dE}{dx}\right)^{-1} \nonumber \\
   & = & \frac{64\nu}{5} \frac{c^3}{Gm} x^5 \left[ b_1 x + b_{3/2} x^{3/2} + b_2 x^2 + b^{5/2} x^{5/2} \right. \nonumber \\
   & &  \left. + b_3 x^3 + b_{3\text{,log}} x^3 \log(x)\right], \nonumber \\
\end{eqnarray}

\noindent with

\begin{eqnarray}
 b_3 & = & \frac{16447322263}{139708800} - \frac{1712 \gamma_e}{105} + \frac{16\pi^2}{3} - \frac{56198689\nu}{217728} \nonumber \\
     &   & + \frac{451\pi^2 \nu}{48} + \frac{541 \nu^2}{896} - \frac{5605\nu^3}{2592} - \frac{856}{105} \log(16), \nonumber \\
 & & \nonumber \\
b_{3,\text{log}} & = & - \frac{856}{105}. \nonumber \\ 
\end{eqnarray}

\clearpage
\begin{widetext}
\section{\label{sec:tables}Tables}

\begin{table*}[!Hb]
 \begin{center}
  \caption{\label{table:m1_21} \small Median, 5\% and 95\% quantiles of the estimated measurement errors on $m_1$ for different mass configurations at redshift $z=1$ with alternative theory parameters included.}
  \footnotesize
  \begin{ruledtabular}
\begin{tabular}{cccccccc}
   $m_1[\Msun]$  & $m_2[\Msun]$ & \multicolumn{6}{c}{$\Delta m_1 / m_1$ \scriptsize with corrections \normalsize} \\
   & &  \multicolumn{2}{c}{5\%-quantile} & \multicolumn{2}{c}{Median} & \multicolumn{2}{c}{95\%-quantile} \\
   & & RWF & FWF & RWF & FWF & RWF & FWF \\
   \hline
   $3 \times 10^{5}$ & $3 \times 10^{4}$ & $5.2 \times 10^{-4}$ & $4.0 \times 10^{-4}$ & $1.4 \times 10^{-3}$ & $1.1 \times 10^{-3}$ & $7.1 \times 10^{-3}$ & $2.7 \times 10^{-3}$ \\
   $3 \times 10^{5}$ & $1 \times 10^{5}$ & $5.1 \times 10^{-4}$ & $4.0 \times 10^{-4}$ & $1.6 \times 10^{-3}$ & $8.8 \times 10^{-4}$ & $1.1 \times 10^{-2}$ & $2.3 \times 10^{-3}$ \\
   $3 \times 10^{5}$ & $3 \times 10^{5}$ & $6.4 \times 10^{-4}$ & $4.1 \times 10^{-4}$ & $2.4 \times 10^{-3}$ & $1.0 \times 10^{-3}$ & $1.5 \times 10^{-2}$ & $2.7 \times 10^{-3}$ \\
   $1 \times 10^{6}$ & $1 \times 10^{5}$ & $1.2 \times 10^{-3}$ & $7.9 \times 10^{-4}$ & $2.8 \times 10^{-3}$ & $1.8 \times 10^{-3}$ & $1.4 \times 10^{-2}$ & $4.4 \times 10^{-3}$ \\
   $1 \times 10^{6}$ & $3 \times 10^{5}$ & $1.1 \times 10^{-3}$ & $7.5 \times 10^{-4}$ & $3.3 \times 10^{-3}$ & $1.7 \times 10^{-3}$ & $1.9 \times 10^{-2}$ & $4.0 \times 10^{-3}$ \\
\end{tabular}
  \end{ruledtabular}
  \normalsize
 \end{center}
\end{table*}

\begin{table*}[!Hb]
 \begin{center}
  \caption{\label{table:m1_15} \small Median, 5\% and 95\% quantiles of the estimated measurement errors on $m_1$ for different mass configurations at redshift $z=1$ without considering alternative theory parameters.}
  \footnotesize
  \begin{ruledtabular}
  \begin{tabular}{cccccccc}
   $m_1[\Msun]$  & $m_2[\Msun]$ & \multicolumn{6}{c}{$\Delta m_1 / m_1$ \scriptsize without corrections \normalsize} \\
   & &  \multicolumn{2}{c}{5\%-quantile} & \multicolumn{2}{c}{Median} & \multicolumn{2}{c}{95\%-quantile} \\
   & & RWF & FWF & RWF & FWF & RWF & FWF \\
   \hline
   $3 \times 10^{5}$ & $3 \times 10^{4}$ & $1.1 \times 10^{-4}$ & $8.2 \times 10^{-5}$ & $3.2 \times 10^{-4}$ & $2.4 \times 10^{-4}$ & $2.5 \times 10^{-3}$ & $2.5 \times 10^{-3}$ \\
   $3 \times 10^{5}$ & $1 \times 10^{5}$ & $1.8 \times 10^{-4}$ & $1.4 \times 10^{-4}$ & $7.7 \times 10^{-4}$ & $4.5 \times 10^{-4}$ & $8.2 \times 10^{-3}$ & $8.2 \times 10^{-3}$ \\
   $3 \times 10^{5}$ & $3 \times 10^{5}$ & $2.1 \times 10^{-4}$ & $1.4 \times 10^{-4}$ & $1.2 \times 10^{-3}$ & $2.9 \times 10^{-4}$ & $9.1 \times 10^{-3}$ & $9.1 \times 10^{-3}$ \\
   $1 \times 10^{6}$ & $1 \times 10^{5}$ & $2.2 \times 10^{-4}$ & $1.5 \times 10^{-4}$ & $7.0 \times 10^{-4}$ & $4.2 \times 10^{-4}$ & $5.1 \times 10^{-3}$ & $5.1 \times 10^{-3}$ \\
   $1 \times 10^{6}$ & $3 \times 10^{5}$ & $3.9 \times 10^{-4}$ & $2.7 \times 10^{-4}$ & $1.4 \times 10^{-3}$ & $8.6 \times 10^{-4}$ & $1.3 \times 10^{-2}$ & $1.3 \times 10^{-2}$ \\
\end{tabular}
  \end{ruledtabular}
  \normalsize
 \end{center}
\end{table*}

\begin{table*}[!Hb]
 \begin{center}
  \caption{\label{table:m2_21} \small Median, 5\% and 95\% quantiles of the estimated measurement errors on $m_2$ for different mass configurations at redshift $z=1$ with alternative theory parameters included.}
  \footnotesize
  \begin{ruledtabular}
  \begin{tabular}{cccccccc}
   $m_1[\Msun]$  & $m_2[\Msun]$ & \multicolumn{6}{c}{$\Delta m_2 / m_2$ \scriptsize with corrections \normalsize} \\
   & &  \multicolumn{2}{c}{5\%-quantile} & \multicolumn{2}{c}{Median} & \multicolumn{2}{c}{95\%-quantile} \\
   & & RWF & FWF & RWF & FWF & RWF & FWF \\
   \hline
   $3 \times 10^{5}$ & $3 \times 10^{4}$ & $1.6 \times 10^{-4}$ & $1.2 \times 10^{-4}$ & $5.6 \times 10^{-4}$ & $4.0 \times 10^{-4}$ & $4.8 \times 10^{-3}$ & $2.2 \times 10^{-3}$ \\
   $3 \times 10^{5}$ & $1 \times 10^{5}$ & $3.2 \times 10^{-4}$ & $2.2 \times 10^{-4}$ & $1.1 \times 10^{-3}$ & $7.6 \times 10^{-4}$ & $9.2 \times 10^{-3}$ & $2.3 \times 10^{-3}$ \\
   $3 \times 10^{5}$ & $3 \times 10^{5}$ & $6.4 \times 10^{-4}$ & $4.2 \times 10^{-4}$ & $2.4 \times 10^{-3}$ & $1.0 \times 10^{-3}$ & $1.6 \times 10^{-2}$ & $2.6 \times 10^{-3}$ \\
   $1 \times 10^{6}$ & $1 \times 10^{5}$ & $3.4 \times 10^{-4}$ & $2.3 \times 10^{-4}$ & $1.3 \times 10^{-3}$ & $8.7 \times 10^{-4}$ & $1.3 \times 10^{-2}$ & $4.5 \times 10^{-3}$ \\
   $1 \times 10^{6}$ & $3 \times 10^{5}$ & $7.1 \times 10^{-4}$ & $4.9 \times 10^{-4}$ & $2.3 \times 10^{-3}$ & $1.5 \times 10^{-3}$ & $1.7 \times 10^{-2}$ & $4.1 \times 10^{-3}$ \\
\end{tabular}
  \end{ruledtabular}
  \normalsize
 \end{center}
\end{table*}

\begin{table*}[!Hb]
 \begin{center}
  \caption{\label{table:m2_15} \small Median, 5\% and 95\% quantiles of the estimated measurement errors on $m_2$ for different mass configurations at redshift $z=1$ without considering alternative theory parameters.}
  \footnotesize
  \begin{ruledtabular}
  \begin{tabular}{cccccccc}
   $m_1[\Msun]$  & $m_2[\Msun]$ & \multicolumn{6}{c}{$\Delta m_2 / m_2$ \scriptsize without corrections \normalsize} \\
   & &  \multicolumn{2}{c}{5\%-quantile} & \multicolumn{2}{c}{Median} & \multicolumn{2}{c}{95\%-quantile} \\
   & & RWF & FWF & RWF & FWF & RWF & FWF \\
   \hline
   $3 \times 10^{5}$ & $3 \times 10^{4}$ & $8.0 \times 10^{-5}$ & $5.9 \times 10^{-5}$ & $2.3 \times 10^{-4}$ & $1.7 \times 10^{-4}$ & $1.7 \times 10^{-3}$ & $1.7 \times 10^{-3}$ \\
   $3 \times 10^{5}$ & $1 \times 10^{5}$ & $1.5 \times 10^{-4}$ & $1.2 \times 10^{-4}$ & $6.3 \times 10^{-4}$ & $3.7 \times 10^{-4}$ & $6.7 \times 10^{-3}$ & $6.7 \times 10^{-3}$ \\
   $3 \times 10^{5}$ & $3 \times 10^{5}$ & $2.1 \times 10^{-4}$ & $1.3 \times 10^{-4}$ & $1.2 \times 10^{-3}$ & $2.9 \times 10^{-4}$ & $9.2 \times 10^{-3}$ & $9.2 \times 10^{-3}$ \\
   $1 \times 10^{6}$ & $1 \times 10^{5}$ & $1.6 \times 10^{-4}$ & $1.1 \times 10^{-4}$ & $5.0 \times 10^{-4}$ & $3.0 \times 10^{-4}$ & $3.6 \times 10^{-3}$ & $3.6 \times 10^{-3}$ \\
   $1 \times 10^{6}$ & $3 \times 10^{5}$ & $3.1 \times 10^{-4}$ & $2.2 \times 10^{-4}$ & $1.1 \times 10^{-3}$ & $6.9 \times 10^{-4}$ & $1.1 \times 10^{-2}$ & $1.1 \times 10^{-2}$ \\
\end{tabular}
  \end{ruledtabular}
  \normalsize
 \end{center}
\end{table*}

\begin{table*}[!Hb]
 \begin{center}
  \caption{\label{table:ch1_21} \small Median, 5\% and 95\% quantiles of the estimated measurement errors on $\chi_1$ for different mass configurations at redshift $z=1$ with alternative theory parameters included.}
  \footnotesize
  \begin{ruledtabular}
  \begin{tabular}{cccccccc}
   $m_1[\Msun]$  & $m_2[\Msun]$ & \multicolumn{6}{c}{$\Delta\chi_1$ \scriptsize with corrections \normalsize} \\
   & &  \multicolumn{2}{c}{5\%-quantile} & \multicolumn{2}{c}{Median} & \multicolumn{2}{c}{95\%-quantile} \\
   & & RWF & FWF & RWF & FWF & RWF & FWF \\
   \hline
   $3 \times 10^{5}$ & $3 \times 10^{4}$ & $3.3 \times 10^{-4}$ & $2.2 \times 10^{-4}$ & $9.7 \times 10^{-4}$ & $6.7 \times 10^{-4}$ & $4.3 \times 10^{-3}$ & $2.6 \times 10^{-3}$ \\
   $3 \times 10^{5}$ & $1 \times 10^{5}$ & $8.3 \times 10^{-4}$ & $5.6 \times 10^{-4}$ & $3.1 \times 10^{-3}$ & $2.0 \times 10^{-3}$ & $1.6 \times 10^{-2}$ & $1.0 \times 10^{-2}$ \\
   $3 \times 10^{5}$ & $3 \times 10^{5}$ & $2.1 \times 10^{-3}$ & $1.6 \times 10^{-3}$ & $1.3 \times 10^{-2}$ & $8.9 \times 10^{-3}$ & $0.24$ & $0.14$ \\
   $1 \times 10^{6}$ & $1 \times 10^{5}$ & $6.6 \times 10^{-4}$ & $4.3 \times 10^{-4}$ & $1.9 \times 10^{-3}$ & $1.2 \times 10^{-3}$ & $7.9 \times 10^{-3}$ & $4.4 \times 10^{-3}$ \\
   $1 \times 10^{6}$ & $3 \times 10^{5}$ & $1.6 \times 10^{-3}$ & $1.0 \times 10^{-3}$ & $4.9 \times 10^{-3}$ & $3.0 \times 10^{-3}$ & $2.3 \times 10^{-2}$ & $1.2 \times 10^{-2}$ \\
\end{tabular}
  \end{ruledtabular}
  \normalsize
 \end{center}
\end{table*}

\begin{table*}[!Hb]
 \begin{center}
  \caption{\label{table:ch1_15} \small Median, 5\% and 95\% quantiles of the estimated measurement errors on $\chi_1$ for different mass configurations at redshift $z=1$ without considering alternative theory parameters.}
  \footnotesize
  \begin{ruledtabular}
  \begin{tabular}{cccccccc}
   $m_1[\Msun]$  & $m_2[\Msun]$ & \multicolumn{6}{c}{$\Delta\chi_1$ \scriptsize without corrections \normalsize} \\
   & &  \multicolumn{2}{c}{5\%-quantile} & \multicolumn{2}{c}{Median} & \multicolumn{2}{c}{95\%-quantile} \\
   & & RWF & FWF & RWF & FWF & RWF & FWF \\
   \hline
   $3 \times 10^{5}$ & $3 \times 10^{4}$ & $1.8 \times 10^{-4}$ & $1.1 \times 10^{-4}$ & $4.2 \times 10^{-4}$ & $2.5 \times 10^{-4}$ & $1.6 \times 10^{-3}$ & $1.6 \times 10^{-3}$ \\
   $3 \times 10^{5}$ & $1 \times 10^{5}$ & $4.6 \times 10^{-4}$ & $2.8 \times 10^{-4}$ & $1.4 \times 10^{-3}$ & $8.1 \times 10^{-4}$ & $9.3 \times 10^{-3}$ & $9.3 \times 10^{-3}$ \\
   $3 \times 10^{5}$ & $3 \times 10^{5}$ & $1.0 \times 10^{-3}$ & $7.8 \times 10^{-4}$ & $6.0 \times 10^{-3}$ & $3.6 \times 10^{-3}$ & $8.8 \times 10^{-2}$ & $8.8 \times 10^{-2}$ \\
   $1 \times 10^{6}$ & $1 \times 10^{5}$ & $3.6 \times 10^{-4}$ & $1.9 \times 10^{-4}$ & $8.7 \times 10^{-4}$ & $4.4 \times 10^{-4}$ & $2.9 \times 10^{-3}$ & $2.9 \times 10^{-3}$ \\
   $1 \times 10^{6}$ & $3 \times 10^{5}$ & $8.6 \times 10^{-4}$ & $4.8 \times 10^{-4}$ & $2.2 \times 10^{-3}$ & $1.2 \times 10^{-3}$ & $1.2 \times 10^{-2}$ & $1.2 \times 10^{-2}$ \\
\end{tabular}
  \end{ruledtabular}
  \normalsize
 \end{center}
\end{table*}

\begin{table*}[!Hb]
 \begin{center}
  \caption{\label{table:ch2_21} \small Median, 5\% and 95\% quantiles of the estimated measurement errors on $\chi_2$ for different mass configurations at redshift $z=1$ with alternative theory parameters included.}
  \footnotesize
  \begin{ruledtabular}
  \begin{tabular}{cccccccc}
   $m_1[\Msun]$  & $m_2[\Msun]$ & \multicolumn{6}{c}{$\Delta\chi_2$ \scriptsize with corrections \normalsize} \\
   & &  \multicolumn{2}{c}{5\%-quantile} & \multicolumn{2}{c}{Median} & \multicolumn{2}{c}{95\%-quantile} \\
   & & RWF & FWF & RWF & FWF & RWF & FWF \\
   \hline
   $3 \times 10^{5}$ & $3 \times 10^{4}$ & $1.4 \times 10^{-3}$ & $8.9 \times 10^{-4}$ & $1.3 \times 10^{-2}$ & $9.0 \times 10^{-3}$ & $0.13$ & $7.2 \times 10^{-2}$ \\
   $3 \times 10^{5}$ & $1 \times 10^{5}$ & $2.0 \times 10^{-3}$ & $1.3 \times 10^{-3}$ & $9.7 \times 10^{-3}$ & $6.7 \times 10^{-3}$ & $8.9 \times 10^{-2}$ & $5.0 \times 10^{-2}$ \\
   $3 \times 10^{5}$ & $3 \times 10^{5}$ & $2.0 \times 10^{-3}$ & $1.5 \times 10^{-3}$ & $1.4 \times 10^{-2}$ & $9.0 \times 10^{-3}$ & $0.25$ & $0.13$ \\
   $1 \times 10^{6}$ & $1 \times 10^{5}$ & $3.3 \times 10^{-3}$ & $1.9 \times 10^{-3}$ & $2.7 \times 10^{-2}$ & $1.7 \times 10^{-2}$ & $0.32$ & $0.15$ \\
   $1 \times 10^{6}$ & $3 \times 10^{5}$ & $3.4 \times 10^{-3}$ & $2.2 \times 10^{-3}$ & $1.8 \times 10^{-2}$ & $1.2 \times 10^{-2}$ & $0.13$ & $5.7 \times 10^{-2}$ \\
\end{tabular}
  \end{ruledtabular}
  \normalsize
 \end{center}
\end{table*}

\begin{table*}[!Hb]
 \begin{center}
  \caption{\label{table:ch2_15} \small Median, 5\% and 95\% quantiles of the estimated measurement errors on $\chi_2$ for different mass configurations at redshift $z=1$ without considering alternative theory parameters.}
  \footnotesize
  \begin{ruledtabular}
  \begin{tabular}{cccccccc}
   $m_1[\Msun]$  & $m_2[\Msun]$ & \multicolumn{6}{c}{$\Delta\chi_2$ \scriptsize without corrections \normalsize} \\
   & &  \multicolumn{2}{c}{5\%-quantile} & \multicolumn{2}{c}{Median} & \multicolumn{2}{c}{95\%-quantile} \\
   & & RWF & FWF & RWF & FWF & RWF & FWF \\
   \hline
   $3 \times 10^{5}$ & $3 \times 10^{4}$ & $7.6 \times 10^{-4}$ & $4.5 \times 10^{-4}$ & $2.7 \times 10^{-3}$ & $1.6 \times 10^{-3}$ & $1.5 \times 10^{-2}$ & $1.5 \times 10^{-2}$ \\
   $3 \times 10^{5}$ & $1 \times 10^{5}$ & $8.2 \times 10^{-4}$ & $5.3 \times 10^{-4}$ & $3.3 \times 10^{-3}$ & $1.9 \times 10^{-3}$ & $1.9 \times 10^{-2}$ & $1.9 \times 10^{-2}$ \\
   $3 \times 10^{5}$ & $3 \times 10^{5}$ & $1.0 \times 10^{-3}$ & $8.2 \times 10^{-4}$ & $6.3 \times 10^{-3}$ & $3.7 \times 10^{-3}$ & $7.9 \times 10^{-2}$ & $7.9 \times 10^{-2}$ \\
   $1 \times 10^{6}$ & $1 \times 10^{5}$ & $1.6 \times 10^{-3}$ & $7.7 \times 10^{-4}$ & $5.6 \times 10^{-3}$ & $2.9 \times 10^{-3}$ & $3.3 \times 10^{-2}$ & $3.3 \times 10^{-2}$ \\
   $1 \times 10^{6}$ & $3 \times 10^{5}$ & $1.4 \times 10^{-3}$ & $8.6 \times 10^{-4}$ & $5.0 \times 10^{-3}$ & $2.9 \times 10^{-3}$ & $2.9 \times 10^{-2}$ & $2.9 \times 10^{-2}$ \\
\end{tabular}
  \end{ruledtabular}
  \normalsize
 \end{center}
\end{table*}

\begin{table*}[!Hb]
 \begin{center}
  \caption{\label{table:2a_21} \small Median, 5\% and 95\% quantiles of the estimated measurement errors on $2a$ for different mass configurations at redshift $z=1$ with alternative theory parameters included.}
  \footnotesize
  \begin{ruledtabular}
  \begin{tabular}{cccccccc}
   $m_1[\Msun]$  & $m_2[\Msun]$ & \multicolumn{6}{c}{$2a [']$ \scriptsize with corrections \normalsize} \\
   & &  \multicolumn{2}{c}{5\%-quantile} & \multicolumn{2}{c}{Median} & \multicolumn{2}{c}{95\%-quantile} \\
   & & RWF & FWF & RWF & FWF & RWF & FWF \\
   \hline
   $3 \times 10^{5}$ & $3 \times 10^{4}$ & $7.5$ & $4.5$ & $21$ & $13$ & $83$ & $67$ \\
   $3 \times 10^{5}$ & $1 \times 10^{5}$ & $5.3$ & $3.3$ & $24$ & $15$ & $99$ & $81$ \\
   $3 \times 10^{5}$ & $3 \times 10^{5}$ & $6.9$ & $4.3$ & $29$ & $21$ & $110$ & $100$ \\
   $1 \times 10^{6}$ & $1 \times 10^{5}$ & $12$ & $8.2$ & $37$ & $22$ & $130$ & $96$ \\
   $1 \times 10^{6}$ & $3 \times 10^{5}$ & $11$ & $7.2$ & $36$ & $23$ & $140$ & $100$ \\
\end{tabular}
  \end{ruledtabular}
  \normalsize
 \end{center}
\end{table*}

\begin{table*}[!Hb]
 \begin{center}
  \caption{\label{table:2a_15} \small Median, 5\% and 95\% quantiles of the estimated measurement errors on $2a$ for different mass configurations at redshift $z=1$ without considering alternative theory parameters.}
  \footnotesize
  \begin{ruledtabular}
  \begin{tabular}{cccccccc}
   $m_1[\Msun]$  & $m_2[\Msun]$ & \multicolumn{6}{c}{$2a [']$ \scriptsize without corrections \normalsize} \\
   & &  \multicolumn{2}{c}{5\%-quantile} & \multicolumn{2}{c}{Median} & \multicolumn{2}{c}{95\%-quantile} \\
   & & RWF & FWF & RWF & FWF & RWF & FWF \\
   \hline
   $3 \times 10^{5}$ & $3 \times 10^{4}$ & $7.2$ & $4.4$ & $20$ & $13$ & $77$ & $77$ \\
   $3 \times 10^{5}$ & $1 \times 10^{5}$ & $5.0$ & $3.1$ & $21$ & $14$ & $91$ & $91$ \\
   $3 \times 10^{5}$ & $3 \times 10^{5}$ & $6.0$ & $3.7$ & $26$ & $18$ & $100$ & $100$ \\
   $1 \times 10^{6}$ & $1 \times 10^{5}$ & $8.5$ & $7.6$ & $35$ & $20$ & $120$ & $120$ \\
   $1 \times 10^{6}$ & $3 \times 10^{5}$ & $8.5$ & $6.2$ & $33$ & $21$ & $120$ & $120$ \\
\end{tabular}
  \end{ruledtabular}
  \normalsize
 \end{center}
\end{table*}

\begin{table*}[!Hb]
 \begin{center}
  \caption{\label{table:2b_21} \small Median, 5\% and 95\% quantiles of the estimated measurement errors on $2b$ for different mass configurations at redshift $z=1$ with alternative theory parameters included.}
  \footnotesize
  \begin{ruledtabular}
  \begin{tabular}{cccccccc}
   $m_1[\Msun]$  & $m_2[\Msun]$ & \multicolumn{6}{c}{$2b [']$ \scriptsize with corrections \normalsize} \\
   & &  \multicolumn{2}{c}{5\%-quantile} & \multicolumn{2}{c}{Median} & \multicolumn{2}{c}{95\%-quantile} \\
   & & RWF & FWF & RWF & FWF & RWF & FWF \\
   \hline
   $3 \times 10^{5}$ & $3 \times 10^{4}$ & $0.98$ & $0.57$ & $4.7$ & $2.7$ & $13$ & $7.3$ \\
   $3 \times 10^{5}$ & $1 \times 10^{5}$ & $0.9$ & $0.52$ & $4.1$ & $2.4$ & $17$ & $9.3$ \\
   $3 \times 10^{5}$ & $3 \times 10^{5}$ & $1.5$ & $0.89$ & $5.8$ & $3.4$ & $22$ & $15$ \\
   $1 \times 10^{6}$ & $1 \times 10^{5}$ & $2.3$ & $1.0$ & $11$ & $5.0$ & $28$ & $13$ \\
   $1 \times 10^{6}$ & $3 \times 10^{5}$ & $1.8$ & $1.1$ & $9.0$ & $4.8$ & $30$ & $17$ \\
   $1 \times 10^{6}$ & $1 \times 10^{6}$ & $2.0$ & $1.1$ & $9.7$ & $5.8$ & $39$ & $27$ \\
   $3 \times 10^{6}$ & $3 \times 10^{5}$ & $2.4$ & $1.2$ & $11$ & $5.4$ & $31$ & $14$ \\
   $3 \times 10^{6}$ & $1 \times 10^{6}$ & $2.1$ & $1.0$ & $10$ & $4.9$ & $34$ & $16$ \\
   $3 \times 10^{6}$ & $3 \times 10^{6}$ & $2.2$ & $1.1$ & $12$ & $5.7$ & $56$ & $30$ \\
   $1 \times 10^{7}$ & $1 \times 10^{6}$ & $3.4$ & $1.1$ & $17$ & $5.6$ & $66$ & $16$ \\
   $1 \times 10^{7}$ & $3 \times 10^{6}$ & $3.9$ & $1.4$ & $19$ & $6.1$ & $87$ & $25$ \\
   $1 \times 10^{7}$ & $1 \times 10^{7}$ & $14$ & $4.3$ & $64$ & $18$ & $310$ & $93$ \\
   $3 \times 10^{7}$ & $3 \times 10^{6}$ & $29$ & $3.8$ & $110$ & $15$ & $919$ & $56$ \\
   $3 \times 10^{7}$ & $1 \times 10^{7}$ & $88$ & $10$ & $600$ & $34$ & $6.1 \times 10^{3}$ & $140$ \\
   $3 \times 10^{7}$ & $3 \times 10^{7}$ & $1.2 \times 10^{3}$ & $93$ & $2.9 \times 10^{4}$ & $530$ & $6.3 \times 10^{5}$ & $3.0 \times 10^{3}$ \\
   $1 \times 10^{8}$ & $1 \times 10^{7}$ & $\infty$ & $490$ & $\infty$ & $1.8 \times 10^{3}$ & $\infty$ & $8.2 \times 10^{3}$ \\
   $1 \times 10^{8}$ & $3 \times 10^{7}$ & $\infty$ & $3.2 \times 10^{3}$ & $\infty$ & $1.3 \times 10^{4}$ & $\infty$ & $9.2 \times 10^{4}$ \\
  \end{tabular}
  \end{ruledtabular}
  \normalsize
 \end{center}
\end{table*}

\begin{table*}[!Hb]
 \begin{center}
  \caption{\label{table:2b_15} \small Median, 5\% and 95\% quantiles of the estimated measurement errors on $2b$ for different mass configurations at redshift $z=1$ without considering alternative theory parameters.}
  \footnotesize
  \begin{ruledtabular}
  \begin{tabular}{cccccccc}
   $m_1[\Msun]$  & $m_2[\Msun]$ & \multicolumn{6}{c}{$2b [']$ \scriptsize without corrections \normalsize} \\
   & &  \multicolumn{2}{c}{5\%-quantile} & \multicolumn{2}{c}{Median} & \multicolumn{2}{c}{95\%-quantile} \\
   & & RWF & FWF & RWF & FWF & RWF & FWF \\
   \hline
   $3 \times 10^{5}$ & $3 \times 10^{4}$ & $0.95$ & $0.55$ & $4.6$ & $2.6$ & $12$ & $12$ \\
   $3 \times 10^{5}$ & $1 \times 10^{5}$ & $0.82$ & $0.44$ & $3.8$ & $2.1$ & $15$ & $15$ \\
   $3 \times 10^{5}$ & $3 \times 10^{5}$ & $1.2$ & $0.69$ & $5.1$ & $3.0$ & $19$ & $19$ \\
   $1 \times 10^{6}$ & $1 \times 10^{5}$ & $2.1$ & $0.91$ & $10$ & $4.7$ & $25$ & $25$ \\
   $1 \times 10^{6}$ & $3 \times 10^{5}$ & $1.5$ & $0.92$ & $8.4$ & $4.4$ & $27$ & $27$ \\
   $1 \times 10^{6}$ & $1 \times 10^{6}$ & $1.6$ & $0.97$ & $8.7$ & $5.1$ & $34$ & $34$ \\
   $3 \times 10^{6}$ & $3 \times 10^{5}$ & $2.3$ & $1.1$ & $11$ & $5.1$ & $27$ & $27$ \\
   $3 \times 10^{6}$ & $1 \times 10^{6}$ & $1.9$ & $0.89$ & $9.0$ & $4.3$ & $29$ & $29$ \\
   $3 \times 10^{6}$ & $3 \times 10^{6}$ & $1.7$ & $0.83$ & $9.7$ & $4.6$ & $38$ & $38$ \\
   $1 \times 10^{7}$ & $1 \times 10^{6}$ & $2.8$ & $0.91$ & $15$ & $4.9$ & $41$ & $41$ \\
   $1 \times 10^{7}$ & $3 \times 10^{6}$ & $2.7$ & $0.99$ & $14$ & $4.9$ & $54$ & $54$ \\
   $1 \times 10^{7}$ & $1 \times 10^{7}$ & $6.1$ & $2.3$ & $27$ & $10$ & $130$ & $130$ \\
   $3 \times 10^{7}$ & $3 \times 10^{6}$ & $15$ & $2.5$ & $60$ & $12$ & $260$ & $260$ \\
   $3 \times 10^{7}$ & $1 \times 10^{7}$ & $43$ & $4.8$ & $170$ & $18$ & $919$ & $919$ \\
   $3 \times 10^{7}$ & $3 \times 10^{7}$ & $550$ & $49$ & $4.2 \times 10^{3}$ & $250$ & $7.1 \times 10^{4}$ & $7.1 \times 10^{4}$ \\
   $1 \times 10^{8}$ & $1 \times 10^{7}$ & $\infty$ & $270$ & $\infty$ & $890$ & $\infty$ & $\infty$ \\
   $1 \times 10^{8}$ & $3 \times 10^{7}$ & $\infty$ & $1.4 \times 10^{3}$ & $\infty$ & $4.9 \times 10^{3}$ & $\infty$ & $\infty$ \\
  \end{tabular}
  \end{ruledtabular}
  \normalsize
 \end{center}
\end{table*}

\begin{table*}[!Hb]
 \begin{center}
  \caption{\label{table:dl_21} \small Median, 5\% and 95\% quantiles of the estimated measurement errors on $d_L$ for different mass configurations at redshift $z=1$ with alternative theory parameters included.}
  \footnotesize
  \begin{ruledtabular}
  \begin{tabular}{cccccccc}
   $m_1[\Msun]$  & $m_2[\Msun]$ & \multicolumn{6}{c}{$\Delta d_L / d_L$ \scriptsize with corrections \normalsize} \\
   & &  \multicolumn{2}{c}{5\%-quantile} & \multicolumn{2}{c}{Median} & \multicolumn{2}{c}{95\%-quantile} \\
   & & RWF & FWF & RWF & FWF & RWF & FWF \\
   \hline
   $3 \times 10^{5}$ & $3 \times 10^{4}$ & $2.4 \times 10^{-3}$ & $1.5 \times 10^{-3}$ & $4.8 \times 10^{-3}$ & $3.0 \times 10^{-3}$ & $1.4 \times 10^{-2}$ & $8.8 \times 10^{-3}$ \\
   $3 \times 10^{5}$ & $1 \times 10^{5}$ & $2.3 \times 10^{-3}$ & $1.5 \times 10^{-3}$ & $4.9 \times 10^{-3}$ & $3.3 \times 10^{-3}$ & $1.8 \times 10^{-2}$ & $1.2 \times 10^{-2}$ \\
   $3 \times 10^{5}$ & $3 \times 10^{5}$ & $2.7 \times 10^{-3}$ & $2.0 \times 10^{-3}$ & $7.0 \times 10^{-3}$ & $5.1 \times 10^{-3}$ & $2.1 \times 10^{-2}$ & $1.7 \times 10^{-2}$ \\
   $1 \times 10^{6}$ & $1 \times 10^{5}$ & $3.7 \times 10^{-3}$ & $2.5 \times 10^{-3}$ & $8.0 \times 10^{-3}$ & $5.0 \times 10^{-3}$ & $2.2 \times 10^{-2}$ & $1.4 \times 10^{-2}$ \\
   $1 \times 10^{6}$ & $3 \times 10^{5}$ & $3.0 \times 10^{-3}$ & $2.3 \times 10^{-3}$ & $7.6 \times 10^{-3}$ & $5.1 \times 10^{-3}$ & $2.4 \times 10^{-2}$ & $1.6 \times 10^{-2}$ \\
\end{tabular}
  \end{ruledtabular}
  \normalsize
 \end{center}
\end{table*}

\begin{table*}[!Hb]
 \begin{center}
  \caption{\label{table:dl_15} \small Median, 5\% and 95\% quantiles of the estimated measurement errors on $d_L$ for different mass configurations at redshift $z=1$ without considering alternative theory parameters.}
  \footnotesize
  \begin{ruledtabular}
  \begin{tabular}{cccccccc}
   $m_1[\Msun]$  & $m_2[\Msun]$ & \multicolumn{6}{c}{$\Delta d_L / d_L$ \scriptsize without corrections \normalsize} \\
   & &  \multicolumn{2}{c}{5\%-quantile} & \multicolumn{2}{c}{Median} & \multicolumn{2}{c}{95\%-quantile} \\
   & & RWF & FWF & RWF & FWF & RWF & FWF \\
   \hline
   $3 \times 10^{5}$ & $3 \times 10^{4}$ & $1.1 \times 10^{-3}$ & $7.8 \times 10^{-4}$ & $2.5 \times 10^{-3}$ & $1.5 \times 10^{-3}$ & $9.4 \times 10^{-3}$ & $9.4 \times 10^{-3}$ \\
   $3 \times 10^{5}$ & $1 \times 10^{5}$ & $9.2 \times 10^{-4}$ & $6.1 \times 10^{-4}$ & $2.8 \times 10^{-3}$ & $1.7 \times 10^{-3}$ & $1.4 \times 10^{-2}$ & $1.4 \times 10^{-2}$ \\
   $3 \times 10^{5}$ & $3 \times 10^{5}$ & $1.2 \times 10^{-3}$ & $7.7 \times 10^{-4}$ & $4.3 \times 10^{-3}$ & $2.8 \times 10^{-3}$ & $1.8 \times 10^{-2}$ & $1.8 \times 10^{-2}$ \\
   $1 \times 10^{6}$ & $1 \times 10^{5}$ & $1.9 \times 10^{-3}$ & $9.8 \times 10^{-4}$ & $5.0 \times 10^{-3}$ & $2.5 \times 10^{-3}$ & $1.8 \times 10^{-2}$ & $1.8 \times 10^{-2}$ \\
   $1 \times 10^{6}$ & $3 \times 10^{5}$ & $1.7 \times 10^{-3}$ & $1.1 \times 10^{-3}$ & $4.9 \times 10^{-3}$ & $2.8 \times 10^{-3}$ & $2.0 \times 10^{-2}$ & $2.0 \times 10^{-2}$ \\
\end{tabular}
  \end{ruledtabular}
  \normalsize
 \end{center}
\end{table*}

\begin{table*}[!Hb]
 \begin{center}
  \caption{\label{table:Psim1_21} \small Median, 5\% and 95\% quantiles of the estimated measurement errors on $\Psi_{-1}$ for different mass configurations at redshift $z=1$}
  \footnotesize
  \begin{ruledtabular}
  \begin{tabular}{cccccccc}
   $m_1[\Msun]$  & $m_2[\Msun]$ & \multicolumn{6}{c}{$\Delta \Psi_{-1}$} \\
   & &  \multicolumn{2}{c}{5\%-quantile} & \multicolumn{2}{c}{Median} & \multicolumn{2}{c}{95\%-quantile} \\
   & & RWF & FWF & RWF & FWF & RWF & FWF \\
   \hline
   $3 \times 10^{5}$ & $3 \times 10^{4}$ & $3.0 \times 10^{-6}$ & $1.8 \times 10^{-6}$ & $7.3 \times 10^{-6}$ & $5.4 \times 10^{-6}$ & $7.0 \times 10^{-5}$ & $4.9 \times 10^{-5}$ \\
   $3 \times 10^{5}$ & $1 \times 10^{5}$ & $2.8 \times 10^{-6}$ & $1.8 \times 10^{-6}$ & $7.3 \times 10^{-6}$ & $6.0 \times 10^{-6}$ & $5.2 \times 10^{-5}$ & $4.6 \times 10^{-5}$ \\
   $3 \times 10^{5}$ & $3 \times 10^{5}$ & $5.3 \times 10^{-6}$ & $3.2 \times 10^{-6}$ & $1.5 \times 10^{-5}$ & $1.1 \times 10^{-5}$ & $1.2 \times 10^{-4}$ & $8.2 \times 10^{-5}$ \\
   $1 \times 10^{6}$ & $1 \times 10^{5}$ & $1.4 \times 10^{-5}$ & $5.9 \times 10^{-6}$ & $3.3 \times 10^{-5}$ & $2.2 \times 10^{-5}$ & $2.9 \times 10^{-4}$ & $2.2 \times 10^{-4}$ \\
   $1 \times 10^{6}$ & $3 \times 10^{5}$ & $1.4 \times 10^{-5}$ & $7.8 \times 10^{-6}$ & $3.6 \times 10^{-5}$ & $2.6 \times 10^{-5}$ & $2.3 \times 10^{-4}$ & $1.8 \times 10^{-4}$ \\
\end{tabular}
  \end{ruledtabular}
  \normalsize
 \end{center}
\end{table*}

\begin{table*}[!Hb]
 \begin{center}
  \caption{\label{table:Psi0_21} \small Median, 5\% and 95\% quantiles of the estimated measurement errors on $\Psi_0$ for different mass configurations at redshift $z=1$}
  \footnotesize
  \begin{ruledtabular}
  \begin{tabular}{cccccccc}
   $m_1[\Msun]$  & $m_2[\Msun]$ & \multicolumn{6}{c}{$\Delta \Psi_0$} \\
   & &  \multicolumn{2}{c}{5\%-quantile} & \multicolumn{2}{c}{Median} & \multicolumn{2}{c}{95\%-quantile} \\
   & & RWF & FWF & RWF & FWF & RWF & FWF \\
   \hline
   $3 \times 10^{5}$ & $3 \times 10^{4}$ & $1.4 \times 10^{-3}$ & $7.0 \times 10^{-4}$ & $3.1 \times 10^{-3}$ & $2.1 \times 10^{-3}$ & $1.7 \times 10^{-2}$ & $1.3 \times 10^{-2}$ \\
   $3 \times 10^{5}$ & $1 \times 10^{5}$ & $1.3 \times 10^{-3}$ & $8.1 \times 10^{-4}$ & $3.2 \times 10^{-3}$ & $2.4 \times 10^{-3}$ & $1.4 \times 10^{-2}$ & $1.3 \times 10^{-2}$ \\
   $3 \times 10^{5}$ & $3 \times 10^{5}$ & $2.2 \times 10^{-3}$ & $1.2 \times 10^{-3}$ & $5.9 \times 10^{-3}$ & $4.1 \times 10^{-3}$ & $3.1 \times 10^{-2}$ & $2.3 \times 10^{-2}$ \\
   $1 \times 10^{6}$ & $1 \times 10^{5}$ & $4.7 \times 10^{-3}$ & $1.7 \times 10^{-3}$ & $1.0 \times 10^{-2}$ & $6.2 \times 10^{-3}$ & $6.3 \times 10^{-2}$ & $4.7 \times 10^{-2}$ \\
   $1 \times 10^{6}$ & $3 \times 10^{5}$ & $4.6 \times 10^{-3}$ & $2.4 \times 10^{-3}$ & $1.2 \times 10^{-2}$ & $7.9 \times 10^{-3}$ & $5.3 \times 10^{-2}$ & $3.9 \times 10^{-2}$ \\
\end{tabular}
  \end{ruledtabular}
  \normalsize
 \end{center}
\end{table*}

\begin{table*}[!Hb]
 \begin{center}
  \caption{\label{table:Psi12_21} \small Median, 5\% and 95\% quantiles of the estimated measurement errors on $\Psi_{1/2}$ for different mass configurations at redshift $z=1$}
  \footnotesize
  \begin{ruledtabular}
  \begin{tabular}{cccccccc}
   $m_1[\Msun]$  & $m_2[\Msun]$ & \multicolumn{6}{c}{$\Delta \Psi_{1/2}$} \\
   & &  \multicolumn{2}{c}{5\%-quantile} & \multicolumn{2}{c}{Median} & \multicolumn{2}{c}{95\%-quantile} \\
   & & RWF & FWF & RWF & FWF & RWF & FWF \\
   \hline
   $3 \times 10^{5}$ & $3 \times 10^{4}$ & $1.5 \times 10^{-2}$ & $5.7 \times 10^{-3}$ & $3.6 \times 10^{-2}$ & $2.0 \times 10^{-2}$ & $0.13$ & $9.6 \times 10^{-2}$ \\
   $3 \times 10^{5}$ & $1 \times 10^{5}$ & $1.4 \times 10^{-2}$ & $7.0 \times 10^{-3}$ & $3.3 \times 10^{-2}$ & $2.3 \times 10^{-2}$ & $0.13$ & $0.11$ \\
   $3 \times 10^{5}$ & $3 \times 10^{5}$ & $1.8 \times 10^{-2}$ & $1.0 \times 10^{-2}$ & $5.0 \times 10^{-2}$ & $3.6 \times 10^{-2}$ & $0.24$ & $0.18$ \\
   $1 \times 10^{6}$ & $1 \times 10^{5}$ & $3.9 \times 10^{-2}$ & $1.5 \times 10^{-2}$ & $8.7 \times 10^{-2}$ & $4.8 \times 10^{-2}$ & $0.4$ & $0.3$ \\
   $1 \times 10^{6}$ & $3 \times 10^{5}$ & $3.5 \times 10^{-2}$ & $2.0 \times 10^{-2}$ & $9.1 \times 10^{-2}$ & $6.2 \times 10^{-2}$ & $0.35$ & $0.27$ \\
\end{tabular}
  \end{ruledtabular}
  \normalsize
 \end{center}
\end{table*}

\begin{table*}[!Hb]
 \begin{center}
  \caption{\label{table:Psi1_21} \small Median, 5\% and 95\% quantiles of the estimated measurement errors on $\Psi_1$ for different mass configurations at redshift $z=1$}
  \footnotesize
  \begin{ruledtabular}
  \begin{tabular}{cccccccc}
   $m_1[\Msun]$  & $m_2[\Msun]$ & \multicolumn{6}{c}{$\Delta \Psi_1$} \\
   & &  \multicolumn{2}{c}{5\%-quantile} & \multicolumn{2}{c}{Median} & \multicolumn{2}{c}{95\%-quantile} \\
   & & RWF & FWF & RWF & FWF & RWF & FWF \\
   \hline
   $3 \times 10^{5}$ & $3 \times 10^{4}$ & $0.14$ & $3.6 \times 10^{-2}$ & $0.32$ & $0.16$ & $0.91$ & $0.61$ \\
   $3 \times 10^{5}$ & $1 \times 10^{5}$ & $0.15$ & $4.5 \times 10^{-2}$ & $0.33$ & $0.21$ & $1.0$ & $0.74$ \\
   $3 \times 10^{5}$ & $3 \times 10^{5}$ & $0.19$ & $6.0 \times 10^{-2}$ & $0.48$ & $0.29$ & $1.6$ & $1.4$ \\
   $1 \times 10^{6}$ & $1 \times 10^{5}$ & $0.3$ & $7.5 \times 10^{-2}$ & $0.65$ & $0.34$ & $2.0$ & $1.3$ \\
   $1 \times 10^{6}$ & $3 \times 10^{5}$ & $0.27$ & $0.11$ & $0.67$ & $0.4$ & $1.8$ & $1.2$ \\
\end{tabular}
  \end{ruledtabular}
  \normalsize
 \end{center}
\end{table*}

\begin{table*}[!Hb]
 \begin{center}
  \caption{\label{table:Psi32_21} \small Median, 5\% and 95\% quantiles of the estimated measurement errors on $\Psi_{3/2}$ for different mass configurations at redshift $z=1$}
  \footnotesize
  \begin{ruledtabular}
  \begin{tabular}{cccccccc}
   $m_1[\Msun]$  & $m_2[\Msun]$ & \multicolumn{6}{c}{$\Delta \Psi_{3/2}$} \\
   & &  \multicolumn{2}{c}{5\%-quantile} & \multicolumn{2}{c}{Median} & \multicolumn{2}{c}{95\%-quantile} \\
   & & RWF & FWF & RWF & FWF & RWF & FWF \\
   \hline
   $3 \times 10^{5}$ & $3 \times 10^{4}$ & $1.1$ & $0.13$ & $2.5$ & $1.4$ & $8.2$ & $6.3$ \\
   $3 \times 10^{5}$ & $1 \times 10^{5}$ & $1.3$ & $0.16$ & $3.0$ & $1.9$ & $8.7$ & $7.3$ \\
   $3 \times 10^{5}$ & $3 \times 10^{5}$ & $2.0$ & $0.22$ & $5.1$ & $2.9$ & $16$ & $14$ \\
   $1 \times 10^{6}$ & $1 \times 10^{5}$ & $2.8$ & $0.24$ & $6.1$ & $3.4$ & $23$ & $15$ \\
   $1 \times 10^{6}$ & $3 \times 10^{5}$ & $3.0$ & $0.63$ & $7.5$ & $4.5$ & $21$ & $14$ \\
\end{tabular}
  \end{ruledtabular}
  \normalsize
 \end{center}
\end{table*}

\begin{table*}[!Hb]
 \begin{center}
  \caption{\label{table:Psi2_21} \small Median, 5\% and 95\% quantiles of the estimated measurement errors on $\Psi_2$ for different mass configurations at redshift $z=1$}
  \footnotesize
  \begin{ruledtabular}
  \begin{tabular}{cccccccc}
   $m_1[\Msun]$  & $m_2[\Msun]$ & \multicolumn{6}{c}{$\Delta \Psi_2$} \\
   & &  \multicolumn{2}{c}{5\%-quantile} & \multicolumn{2}{c}{Median} & \multicolumn{2}{c}{95\%-quantile} \\
   & & RWF & FWF & RWF & FWF & RWF & FWF \\
   \hline
   $3 \times 10^{5}$ & $3 \times 10^{4}$ & $4.0$ & $0.35$ & $8.9$ & $5.3$ & $30$ & $23$ \\
   $3 \times 10^{5}$ & $1 \times 10^{5}$ & $5.1$ & $0.6$ & $12$ & $7.6$ & $34$ & $28$ \\
   $3 \times 10^{5}$ & $3 \times 10^{5}$ & $8.2$ & $0.83$ & $21$ & $12$ & $63$ & $52$ \\
   $1 \times 10^{6}$ & $1 \times 10^{5}$ & $10$ & $0.76$ & $23$ & $13$ & $86$ & $55$ \\
   $1 \times 10^{6}$ & $3 \times 10^{5}$ & $12$ & $2.1$ & $30$ & $18$ & $85$ & $54$ \\
\end{tabular}
  \end{ruledtabular}
  \normalsize
 \end{center}
\end{table*}

\end{widetext}

\clearpage

\bibliography{bibliography.bib}

\end{document}